\documentclass[11pt]{article}
\usepackage[utf8]{inputenc}

\usepackage{putex}
\usepackage{amsmath}    
\usepackage[pdftex]{graphicx}   
\usepackage{epstopdf}   
\usepackage{verbatim}   
\usepackage{color}      
\usepackage{subfigure}  
\usepackage{hyperref}   
\usepackage{amssymb}
\usepackage{slashed}
\usepackage{bbm}
\usepackage{psfrag}
\usepackage{array}

\makeatletter
\newcommand{\thickhline}{%
    \noalign {\ifnum 0=`}\fi \hrule height 1pt
    \futurelet \reserved@a \@xhline
}
\newcolumntype{"}{@{\hskip\tabcolsep\vrule width 1pt\hskip\tabcolsep}}
\makeatother

\begin{document}
\preprint{CCTP-2014-27\\ 
CCQCN-2014-53\\
COLO-HEP-585}

\institution{CU}{${}^1$Department of Physics, 390 UCB, University of Colorado, Boulder, CO 80309, USA}

\institution{CCTP}{${}^3$Crete Center for Theoretical Physics, University of
Crete, 71003 Heraklion, Greece}

\title{Fermi surface behavior in the ABJM M2-brane theory}

\authors{Oliver DeWolfe,${}^\CU$ Oscar Henriksson,${}^\CU$ and Christopher Rosen${}^\CCTP$}

\abstract{We calculate fermionic Green's functions for states of the three-dimensional ABJM M2-brane theory at large N using the gauge-gravity correspondence. We embed extremal black brane solutions in four-dimensional maximally supersymmetric gauged supergravity, obtain the linearized Dirac equations for each spin-1/2 mode that cannot mix with a gravitino, and solve these equations with infalling boundary conditions to calculate retarded Green's functions. For generic values of the chemical potentials, we find Fermi surfaces with universally non-Fermi liquid behavior, matching the situation for four-dimensional ${\cal N}=4$ Super-Yang-Mills. Fermi surface singularities appear and disappear discontinuously at the point with all chemical potentials equal, reminiscent of a quantum critical point. One limit of parameter space has zero entropy at zero temperature, and fermionic fluctuations are perfectly stable inside an energy region around the Fermi surface. An ambiguity in the quantization of the fermions is resolved by supersymmetry. 
}

\date{October 2014}

\maketitle



\section{Introduction and Summary}

\subsection{Holographic realizations of non-Fermi liquids}

Many systems of interacting fermions, including most metals, behave as Landau-Fermi liquids,
where the interactions dress the fermions into quasiparticles whose fluctuations around a Fermi surface are asymptotically stable at low energies. However, a number of interesting strongly coupled systems --- notably cuprate superconductors \cite{Varma, Anderson} and heavy fermion systems \cite{Gegenwart} --- display ``strange metal" behavior which deviates from the Fermi liquid paradigm. In such systems, a Fermi surface is evident, but the fluctuations are not stable, and transport properties are correspondingly different. It is of interest to develeop theoretical mechanisms to study such ``non-Fermi liquids".

The gauge-gravity correspondence, or AdS/CFT correspondence \cite{Maldacena:1997re,Gubser:1998bc,Witten:1998qj}, has become a valuable tool for exploring strongly coupled systems that lack a straightforward quasiparticle description. Systems at zero temperature and finite density  are described holographically by charged, extremal, asymptotically anti-de Sitter black hole geometries living in one higher dimension
\cite{Lee:2008xf}-
\cite{Iqbal:2011ae}.
Normal modes of fermions in such backgrounds compute fermionic Green's functions, whose zero energy, finite momentum poles may be interpreted as Fermi surface singularities, with near-pole behavior determining the dispersion of nearby excitations.

Such systems were considered first from a ``bottom-up" perspective, where simple Dirac equations were postulated and studied in Reissner-Nordstr\"om black brane backgrounds \cite{Lee:2008xf,Liu:2009dm, Cubrovic:2009ye,Faulkner:2009wj}.  These studies showed that holographic Fermi surfaces could indeed exist, and depending on the charge and mass parameters of the fermion, could manifest either Fermi liquid behavior, with asymptotically stable quasiparticles, or non-Fermi liquid behavior, where the decay width of excitations typically remains of the same size as the energy. Thus gravity duals to systems having non-Fermi liquid behavior were shown to be possible, albeit in systems whose precise field theory dual is not known. Initial studies included (constant) masses and gauge couplings; Pauli couplings were added in  \cite{Edalati:2010ge, Edalati:2010ww}.

A natural next step is to study ``top-down" constructions, where the black brane backgrounds and fluctuating fermions are part of a known supergravity theory descending from string theory, and hence have a precisely known field theory dual. Natural candidate theories for such a study are ${\cal N}=4$ Super-Yang-Mills (SYM) theory in four spatial dimensions, and the ${\cal N}=8$ supersymmetric Aharony-Bergman-Jafferis-Maldacena (ABJM) theory in three dimensions; these theories are maximally superconformal and are the most symmetric avatars of four-dimensional non-Abelian gauge theory and three-dimensional Chern-Simons-matter theory, describing the dynamics of stacks of D3-branes and M2-branes, respectively. The finite-density behavior of these theories is interesting in its own right, and the gauge-gravity correspondence provides an opportunity to study them at strong coupling and large N.

Fluctuations of the gravitino field in supergravity were studied in \cite{Gauntlett:2011mf, Belliard:2011qq, Gauntlett:2011wm}, but no Fermi surface singularities were found. The first Fermi surfaces were identified in \cite{DeWolfe:2011aa}, where one fermion in one particular background for each of ${\cal N}=4$ SYM and ABJM theories was shown to have a Fermi surface with non-Fermi liquid behavior. A systematic study of the ${\cal N}=4$ Super-Yang-Mills case was carried out in \cite{DeWolfe:2012uv}, where the Dirac equation of every spin-1/2 fermion not mixing with the gravitino was solved across a one-dimensional slice of the two-dimensional parameter space defined by ratios of the three SO(6) chemical potentials. The Dirac equations were more complicated than the typical bottom-up examples, featuring mass and Pauli terms that depend on scalar fields that generically vary in the background. Every value of the chemical potentials showed at least one fermion with a Fermi surface, and in all cases, the excitations near the Fermi surfaces displayed non-Fermi liquid behavior. As the chemical potentials varied, in general Fermi momenta vary but the existence of a Fermi surface persists, except when the Fermi momentum enters a so-called oscillatory region, where the Green's function displays log oscillatory behavior and the Fermi surface singularity cannot exist. One class of fermion asymptoted to the case separating Fermi and non-Fermi liquid behaviors, the marginal Fermi liquid (MFL) which was proposed as a description of the optimally doped cuprates \cite{Varma}, as it approached the edge of the parameter space.

For generic values of the chemical potentials, the extremal black brane backgrounds possess a regular event horizon, which implies a nonzero entropy at zero temperature. Such a feature is shared by the Reissner-Nordstr\"om backgrounds studied in many bottom-up models, but is somewhat unusual from the field theory point of view. In \cite{Iqbal:2011in}, it was suggested that such states should be understood not as the true ground state of the dual gauge theory, but instead as states in a semi-local quantum liquid (SLQL) phase characterized by scaling at intermediate energies, before a true ground state phase emerges due to the condensation of instabilities or the manifestation of subleading N effects.
There are exceptions to this behavior, however, at the edges of the ${\cal N}=4$ SYM chemical potential parameter space. When two of the three charges are set to zero, the ``extremal" geometry loses its horizon, becoming a non-thermodynamic renormalization group (RG) flow geometry previously studied in \cite{Brandhuber:1999jr, Freedman:1999gk, Bianchi:2000sm, Bianchi:2001de, Bianchi:2001kw}. 
Perhaps more interestingly, when one of the three charges is set to zero, the geometry becomes singular at the horizon, and the  entropy at zero-temperature correspondingly vanishes \cite{Gubser:2009qt, Iizuka:2011hg}. This case was studied in detail in \cite{DeWolfe:2013uba}, where it was shown how a lift to six dimensions resolves the singularity as well as providing constraints on consistent parameters for fermion fields. It was found that there is a region in energy around the Fermi surface where the fermionic fluctuations are perfectly stable, before 
returning to non-Fermi liquid behavior outside. An interpretation of this is a gap developing in another sector, removing a large number of degrees of freedom and depriving the fermions of the catalyst for their decay; the removal of many but not all degrees of freedom from the region is reminiscent  of a pseudogap phase. It also shares features with the semi-local quantum liquid resolutions described in \cite{Iqbal:2011in}, as the non-Fermi liquid behavior exists at intermediate energies, while the true ground state is controlled by a Fermi surface with Fermi liquid-like excitations and vanishing entropy.

Hence it has been demonstrated that non-Fermi liquid behavior exists in nonzero-density gauge theories at strong coupling, and studied in great detail for four-dimensional ${\cal N}=4$ Super-Yang-Mills. Given the associations to cuprate superconductors and other strongly correlated systems in two spatial dimensions, it is natural to extend the thorough, systematic study of \cite{DeWolfe:2011aa, DeWolfe:2013uba} to the case of the maximally supersymmetric ABJM M2-brane theory, both for its potential application to realistic systems and for its inherent interest as one of the maximally superconformal theories. This is the goal of the present work.

\subsection{Fermionic response in the M2-brane theory}

The M2-brane theory has an SO(8) R-symmetry, and hence 4 distinct chemical potentials. The dual description is M-theory on $AdS_4 \times S^7$, which reduces to four-dimensional ${\cal N}=8$ gauged supergravity. Finite density black brane solutions corresponding to rotating M2-brane systems are known in a truncated theory of the metric, gauge fields and three scalars, but no fermions \cite{Duff:1999gh}. We use the known embedding \cite{Cvetic:1999xp} of this truncated theory to lift the solutions to the full ${\cal N}=8$ gauged supergravity, and we use these backgrounds to derive the 
corresponding Dirac equations for all spin-1/2 fluctuations with quantum numbers forbidding mixing with the gravitino. We then solve these Dirac equations in the black brane backgrounds with the infalling boundary conditions at the horizon that calculate retarded Green's functions in the dual field theory. Because the mass of the fermions approaches zero at the boundary, there is an ambiguity between which terms in the near-boundary expansion to identify as the source, and which as the response;  the mass functions are nonzero away from the boundary, however, so the choice has physical content. We demonstrate how to use supersymmetry to resolve this ambiguity, producing a unique prescription for the dual Green's functions.

We calculate Green's functions over two one-dimensional cuts through the three-dimensional space of chemical potential ratios, one where three charges are set equal, and one where the charges are set equal in pairs; these cuts meet at the point where all four charges are equal. Results over this parameter space are strongly in accord with the ${\cal N}=4$ SYM case. In particular, Fermi surface singularities are common and are in all cases associated to non-Fermi liquid behavior. In particular, one class of excitations, the {\em net-charged} fermions, are qualitatively identical to the higher-dimensional case; Fermi surfaces persist as the chemical potentials are varied unless the Fermi momentum falls into an oscillatory region. One such fermion again approaches marginal Fermi liquid behavior at a limit of the parameter space.  The other class of excitations, so-called {\em net-neutral} fermions, shows novel behavior: while all Fermi surface singularities still show non-Fermi liquid behavior, there are no oscillatory regions, and yet a Fermi surface can discontinuously appear or disappear at a nonzero value of the Fermi momentum as one tunes the chemical potentials past the four-charge black hole. This abrupt change in the spectrum at zero temperature as a dimensionless parameter is varied is reminiscent in aspects of a quantum phase transition; however, no singularities in the susceptibilities are visible in the thermodynamics. 

Some interest has appeared recently in identifying zeros of a fermionic Green's function as a sign of Mott insulator behavior, and a duality between zeros and poles in certain bottom-up models has been noted \cite{Alsup:2014uca, Vanacore:2014hka}. We also obtain the zeros of the Green's function, which are also of interest as in the alternate quantization of the fermions --- which would correspond to an alternate theory breaking supersymmetry --- they exchange roles with the poles. In this alternate quantization, ordinary Fermi liquid behavior would appear for certain excitations, while the true ABJM theory has only non-Fermi liquid excitations. We note that the zero/pole duality of \cite{Alsup:2014uca, Vanacore:2014hka} is a consequence of the symmetry of the Dirac equation under a flip of chirality, and does not obtain for our models where the mass and Pauli couplings are nonzero.

As for ${\cal N}=4$ SYM, the ABJM theory again has exceptional cases at the limits of parameter space. When one of the four charges is set to zero, we encounter again a naively singular geometry, with vanishing entropy at zero temperature. An analysis closely following \cite{DeWolfe:2013uba} holds, again demonstrating pseudogap-like behavior, with a region in energy around the Fermi surface where the fermionic fluctuations are perfectly stable. As in \cite{DeWolfe:2013uba}, there is a lift to a higher dimension resolving the singularity, which also results in a constraint between the mass, charge and Pauli couplings of consistent fermions, which are obeyed by all the cases in the maximal gauged supergravity. When two or three charges are set to zero, we find renormalization group flow solutions, with only a running scalar modifying the geometry. While these backgrounds are non-thermodynamic, they may be of interest both in their relation to the nonzero temperature backgrounds with the same charge, and as RG flow geometries in their own right. In these cases we are able to solve for the fluctuations of fermions, and find the corresponding Green's functions, exactly.

Overall a similar picture has emerged for the ABJM case as for the ${\cal N}=4$ case: the bulk of the parameter space, with all charges nonzero, leads to regular black holes dual to zero temperature states with nonzero entropy showing non-Fermi liquid behavior. Limits of the parameter space  either lack horizons, or are naively singular, resulting in zero entropy states with an energy pseudogap around the Fermi surface where the fermionic fluctuations are stable. For the ABJM case, moreover, Fermi surfaces appear and disappear discontinuously around the most symmetric point in the parameter space, suggestive of a quantum phase transition.

In section~\ref{GaugedSUGRASec}, we recap the M2-brane theory, describe the reduction of four-dimensional maximally supersymmetric gauged supergravity to the truncated model, present the general black brane solution with four charges, and derive the Dirac equations of the theory's fermions in these backgrounds.
In section~\ref{GreenFunctionSec}, we review methods for solving these Dirac equations and generating Green's functions, presenting discrete symmetries of the equations and using supersymmetry to resolve the apparent ambiguity in quantization for the fermions in asymptotically anti-de Sitter space.
In section~\ref{RegularSec}, we review the properties of such equations in the background of regular extremal black holes, present the solutions with three charges set equal and with charges set equal in pairs, and numerically obtain Fermi surface singularities and their corresponding Fermi momenta, as well as oscillatory regions and the locations of zeros of the Green's function, for each fermion.
In section~\ref{ThreeChargeSec}, we consider the special case with three charges equal and one charge zero, and demonstrate the existence of an energy gap wherein the fluctuations are exactly stable, and solve for the dispersion relations for each fermion throughout this region. We match these results on to the limit of the regular black holes.
In section~\ref{RGFlowSec}, we exactly solve the Dirac equations in the backgrounds where two and three charges are zero, with the rest set equal, and again match the results onto the limit of the regular sequence.
Certain details of the gauged supergravity analysis, and of the lift of the three-charge geometry to five dimensions, are presented in appendices.

\section{Black branes and fermions in maximal gauged supergravity}
\label{GaugedSUGRASec}

We begin this section by reviewing a few  features of the M2-brane theory and its gravity dual. We then show how four-dimensional maximally supersymmetric gauged supergravity reduces to a truncated bosonic theory and present its black brane solutions, and finally derive the Dirac fluctuation equations for the set of spin-1/2 fields not mixing with the gravitino.

\subsection{The M2-brane theory and its gravity dual}

The three-dimensional exactly superconformal field theory living on a stack of $N$ M2-branes is of great interest, both in its own right as one of the three fundamental maximally superconformal field theories (the others being four-dimensional ${\cal N}=4$ Super-Yang-Mills theory and six-dimensional $(2, 0)$ theory) and as a potential source of insight into strongly correlated theories in two spatial dimensions. It can be described as
 a Chern-Simons gauge theory with $U(N) \times U(N)$ gauge group at levels $(1, -1)$ coupled to bifundamental matter called the ABJM theory \cite{Bagger:2006sk, Gustavsson:2007vu, Bagger:2007jr, Bagger:2007vi, Aharony:2008ug}; for reviews see  \cite{Klebanov:2009sg, Bagger:2012jb}. The R-symmetry group is $SO(8)$, leading to four independent chemical potentials. We are interested in studying fermionic response in this theory at finite density, that is, with some combination of the chemical potentials turned on. 
 
While the Chern-Simons-matter theory is the proper description, when identifying gauge-invariant operators it is often sufficient to think simply about taking the 8 scalar fields $X$ and 8 Majorana fermions $\lambda$ that describe the case for a single M2-brane ($N=1$), and generalizing these to $N \times N$ matrices; this is an oversimplified way of describing the theory, but allows us to simply describe the operators we are interested in. The scalar $X$ and fermion $\lambda$
transform in the ${\bf 8}_{\rm v}$ and ${\bf 8}_{\rm c}$ representations of $SO(8)$, respectively, 
and
the supersymmetry transformation $\delta X \sim \epsilon  \Gamma \lambda$ 
together with the product rule ${\bf 8}_i \otimes {\bf 8}_j = {\bf 8}_k + {\bf 56}_k$ for $i$, $j$, $k$ different 
 implies the supercharges are in the ${\bf 8}_{\rm s}$.
The ${\bf 8}_{\rm v}$ scalars $X$  may be arranged into complex combinations, each of which has charge $\pm 1$ under precisely one of the $SO(8)$ Cartan generators. The  ${\bf 8}_{\rm c}$ fermions $\lambda$ are each charged under all four generators, with charges $\pm (-\frac{1}{2}, \frac{1}{2}, \frac{1}{2}, \frac{1}{2})$ + permutations. Turning on the chemical potential for each Cartan generator will thus affect only two bosons, but all eight fermions.

Gauge/gravity duality calculates correlation functions of gauge-invariant operators, and the chiral primaries and their descendants are the operators accessible to gravity calculations. In the $X$, $\lambda$ notation, the chiral primary operators of the theory have the form Tr~$\!X^k$, $k = 2, 3, \ldots$, with dimensions $\Delta = k/2$. The lowest-dimension chiral primary Tr~$\!X^2$\ transforms in the ${\bf 35}_{\rm v}$ of $SO(8)$, and its first descendent is
the lowest-dimension gauge-invariant fermionic operator Tr~$\!X \lambda$, which has $\Delta = 3/2$ and sits in the ${\bf 56}_{\rm s}$. This is the operator for which we will calculate Green's functions and investigate Fermi surface behavior. We will also have cause to mention the second descendant, the bosonic operator Tr~$\!\lambda^2$ with $\Delta = 2$ in the ${\bf 35}_{\rm c}$. A table of operators for this theory may be found in \cite{D'Hoker:2000vb}.

The AdS/CFT dual of the M2-brane theory is given by the near-horizon limit of a stack of M2-branes, which is M-theory on an $AdS_4 \times S^7$ background with $N$ units of 4-form flux on $AdS_4$ \cite{Maldacena:1997re, Aharony:2008ug}. The $SO(8)$ R-symmetry is realized as the isometry group of the seven-sphere. In the large-$N$ limit, M-theory reduces to eleven-dimensional supergravity. The Kaluza-Klein reduction of eleven-dimensional supergravity on $S^7$ \cite{Biran:1983iy, Castellani:1984vv, Casher:1984ym} includes an infinite tower of supersymmetry multiplets; the theory of the modes sharing the multiplet of the four-dimensional massless graviton is four-dimensional ${\cal N}=8$ (maximal) gauged supergravity \cite{deWit:1981eq, deWit:1982ig}, which represents a consistent truncation of the higher-dimensional theory \cite{deWit:1986iy}. %

Maximal gauged supergravity in four dimensions consists of the metric $g_{\mu\nu}$, eight gravitino fields $\psi_\mu^i$  in the ${\bf 8}_{\rm s}$ of $SO(8)$, 28 gauge fields $A_\mu^{IJ} \equiv A_\mu^{[IJ]}$ filling out the adjoint of $SO(8)$, 56 spin-1/2 fermions $\chi_{ijk} \equiv \chi_{[ijk]}$ in the ${\bf 56}_{\rm s}$, and 70 scalars parameterizing an $E_{7(7)}/SU(8)$ coset transforming in the ${\bf 35}_{\rm v} \oplus {\bf 35}_{\rm c}$, with the two sets parity-even and parity-odd respectively. Here $I, J = 1 \ldots 8$ are $SO(8)$ indices, and $i, j = 1 \ldots 8$ are $SU(8)$ indices. The scalars in the ${\bf 35}_{\rm v}$ are dual to the lowest chiral primary Tr~$\!X^2$, and the remaining modes are dual to descendants, as summarized in the table:

\medskip
\begin{center}
\begin{tabular}{|c|cccccc|}\hline
SUGRA mode & $g_{\mu\nu}$ & $\psi_\mu^i$ & $A_\mu^{IJ}$ & $\chi_{ijk}$ & Re $\phi_{ijkl}$ & Im $\phi_{ijkl}$\\ \hline
Dual operator & $T^{\mu\nu}$ & ${\cal S}^{\mu i}$ & $J_R^{\mu IJ}$ & Tr~$X \lambda$ & Tr~$X^2$ & Tr~$\lambda^2$ \\ \hline
Conformal dimension & 3 & 5/2 & 2 & 3/2 & 1 & 2 \\ \hline 
$SO(8)$ rep & ${\bf 1}$& ${\bf 8}_{\rm s}$&${\bf 28}$ & ${\bf 56}_{\rm s}$ & ${\bf 35}_{\rm v}$ & ${\bf 35}_{\rm c}$ \\ \hline
\end{tabular}
\end{center}
\medskip
\noindent
Note the $SO(8)$ triality frame is forced on us by the identification of the gravitini as dual to the supercurrents ${\cal S}^{\mu i}$ in the ${\bf 8}_{\rm s}$. 
We will be interested in the fermionic fluctuations of the $\chi_{ijk}$, dual to the operators Tr~$X \lambda$. The set of charge vectors of the ${\bf 56}_{\rm s}$ contains three copies of the ${\bf 8}_{\rm s}$, each vector with norm 1, as well as 32 unique charge vectors with norm $\sqrt{3}$. The former may mix with the gravitini, and so to avoid this complication it is the latter we will consider.

In the next subsection, we will relate the bosonic sector of this theory to a truncated theory consisting of the metric, four gauge fields and three scalars, and discuss the black brane solutions of this truncated theory. In the subsection following, we will derive the Dirac equation for  linearized fluctuations of the $\chi_{ijk}$ in these backgrounds.

\subsection{Bosonic sector of maximal gauged supergravity}

The coset representative containing the scalars is written in the form of a sechsundf\"unfzigbein
(56-bein) \cite{deWit:1982ig}:
\begin{equation}
\label{bein}
 \mathcal{V}=
\begin{pmatrix} 
  u_{ij}^{\ \ IJ}     & v_{ijKL}\\ 
  v^{klIJ} & u^{kl}_{\ \ KL} 
\end{pmatrix} \,.
\end{equation}
Here each pair $IJ$ or $ij$ is antisymmetric, and thus may be thought of as a single composite index running from 1 to 28, decomposing  the $56 \times 56$ coset representative into $28 \times 28$ blocks corresponding to $u$ and $v$.
Everything in the Lagrangian involving the scalars, including the potential and interactions, can then be written in terms of the $u$- and $v$-tensors thus defined. Important objects are the T-tensor,
\begin{equation}
\label{t-tensor}
 T_i^{\ jkl}=(u^{kl}_{\ \ IJ}+v^{klIJ})(u_{im}^{\ \ JK}u^{jm}_{\ \ KI}-v_{imJK}v^{jmKI}) \,,
\end{equation}
 the A-tensors derived from it,
\begin{equation}
\label{Atensors}
 A^1_{ij}=\frac{4}{21}T^k_{\ ikj} \,, \ \ \ A^2_{ijkl}=-\frac{4}{3}T_{i[jkl]} \,,
\end{equation}
and the S-tensor, which can be defined in terms of the equation
\begin{equation}
\label{Stensor}
 (u^{ij}_{\ \ IJ} + v^{ijIJ})S^{IJ,KL} = u^{ij}_{\ \ KL} \,.
\end{equation}
A derivative of the scalar is represented as ${\cal A}_\mu^{ijkl}$,
\eqn{}{
{\cal A}_\mu^{ijkl} \equiv - 2 \sqrt{2} \left( u^{ij}_{\;\;\,IJ} \partial_\mu v^{klIJ} - v^{ijIJ}\partial_\mu u^{kl}_{\;\;\;IJ} \right) \,.
}
The  parts of the ${\cal N}=8$ Lagrangian involving the metric, scalars and gauge fields are 
\eqn{MaximalLagrangian}{
e^{-1} {\cal L} = R - {1 \over 48} {\cal A}_\mu^{ijkl} {\cal A}^\mu_{ijkl} - {1 \over 4} \left( F^+_{\mu\nu IJ} \left( 2 S^{IJ,KL} - \delta^{IJ}_{KL} \right) F^{+ \mu\nu}_{\;\;\;\;\;\;\;KL} + {\rm h.c.} \right) - V(\phi)\,,
}
with the self-dual and anti-self-dual parts of the field strength and the generalized Kronecker delta defined as $F^\pm_{\mu\nu} \equiv \frac{1}{2}(F_{\mu\nu}\pm \frac{i}{2} \epsilon_{\mu\nu\rho\sigma} F^{\rho\sigma})$ and $\delta^{IJ}_{KL} \equiv \frac{1}{2} ( \delta^I_K \delta^J_L - \delta^I_L \delta^K_J)$, 
and the scalar potential given by
\begin{equation}
\label{potential}
 V=-2g^2 \left( \frac{3}{4}|A^1_{ij}|^2-\frac{1}{24}|A^2_{ijkl}|^2\right) \,.
\end{equation}
We will use the so-called symmetric gauge \cite{deWit:1982ig}, where the 56-bein reduces to 
\begin{equation}
\label{gauge_bein}
 \mathcal{V}=\exp \left[-\frac{1}{2 \sqrt{2}} 
\begin{pmatrix} 
  0  & \phi_{ijkl}\\ 
  \phi^{mnpq} & 0 
\end{pmatrix} \right] \,,
\end{equation}
with $\phi_{ijkl}$ obeying the self-duality relation $\phi_{ijkl}=\frac{1}{24}\epsilon_{ijklmnpq}\phi^{mnpq}$.
In this gauge the scalar kinetic function reduces to
\eqn{ScalarKinetic}{
{\cal A}_\mu^{\;\;ijkl} = \partial_\mu \phi^{ijkl} \,.
}
Following Duff and Liu \cite{Duff:1999gh}, we can reduce to a truncated theory including only the metric, the four Cartan gauge fields, and three scalars $\phi_A$ using the ansatz
\begin{equation}
\label{ansatz}
\phi_{ijkl}=\frac{1}{\sqrt{2}}[\phi_1(\epsilon^{12}+\epsilon^{34})_{ijkl}+\phi_2(\epsilon^{13}+\epsilon^{24})_{ijkl}+\phi_3(\epsilon^{14}+\epsilon^{23})_{ijkl}] \,.
\end{equation}
Here the special Levi-Civita symbols $\epsilon^{\alpha \beta}_{ijkl}$ are non-zero only when the indices $i,j,k,l$ take values within the index pairs specified by the superscripts, where $\alpha=1,...,4$ runs over the SO(8) index pairs $\{12,34,56,78\}$. For example, $\epsilon^{13}_{ijkl}=1 (-1)$ when ${i,j,k,l}$ is an even (odd) permutation of ${1,2,5,6}$. 
One can then see using \eno{MaximalLagrangian} and \eno{ScalarKinetic} that $\phi_1$, $\phi_2$, $\phi_3$ have canonical kinetic terms.

One may now calculate the $u$ and $v$ tensors in terms of this scalar ansatz, and from them the $T$-, $A$- and $S$-tensors. We present the results in the appendix. One  then finds the potential
\begin{equation}
 V= -4g^2[\cosh\phi_1+\cosh\phi_2+\cosh\phi_3] \,.
\end{equation}
Finally, we define the gauge fields $A_\mu^a$,  $A_\mu^b$, $A_\mu^c$, $A_\mu^d$ in terms of the Cartan generators $A_\mu^{IJ}$ as 
\begin{equation}
 \begin{pmatrix} A^{12}_\mu \\ A^{34}_\mu \\ A^{56}_\mu \\ A^{78}_\mu \end{pmatrix} \equiv {1 \over 2 \sqrt{2}} 
 \begin{pmatrix} 1 & 1 & 1 & 1 \\ 1 & 1 & -1 & -1 \\ 1 & -1 & 1 & -1 \\ 1 & -1 & -1 & 1 \end{pmatrix}
 \begin{pmatrix} A^a_\mu \\ A^b_\mu \\ A^c_\mu \\ A^d_\mu \end{pmatrix} \,,
\end{equation}
where the factor $1/2\sqrt{2}$ is for obtaining canonical gauge kinetic terms, and the matrix may be thought of as an $SO(8)$ triality rotation \cite{Duff:1999gh}, which diagonalizes the couplings to the scalars. The Lagrangian for this restricted set of fields is then
\begin{equation}\label{lagrangian}
 e^{-1}\mathcal{L} = R-\frac{1}{2}(\partial \vec\phi)^2 + \frac{2}{L^2}(\cosh\phi_1+\cosh\phi_2+\cosh\phi_3) - \frac{1}{4}\sum_{i=a,b,c,d} e^{-\lambda_i} F_i^2 \,,
\end{equation}
where 
\eqn{lambdas}{
\lambda_a\equiv-\phi_1-\phi_2-\phi_3 \,,\quad
\lambda_b\equiv-\phi_1+\phi_2+\phi_3 \,, \quad
\lambda_c\equiv \phi_1-\phi_2+\phi_3 \,,\quad
\lambda_d\equiv\phi_1+\phi_2-\phi_3  \,,
}
and where we have defined\footnote{Our normalization of $g$ is from \cite{deWit:1982ig} and matches \cite{Duff:1999gh}; \cite{Cvetic:1999xp} uses a $g$ smaller by $1/\sqrt{2}$.} $L$,
\eqn{gResult}{
g= {1 \over \sqrt{2}L}\,.
}
Families of  black brane solutions are known in this truncated theory \cite{Duff:1999gh, Cvetic:1999xp}.
The black branes asymptote to the Poincar\'e patch of four-dimensional anti-de Sitter space. In general the three scalars of the truncated theory run with the radial coordinate, and the electric potentials of the four gauge fields are turned on as well, which will be associated with the nonzero chemical potentials. The solutions are of the form \cite{Cvetic:1999xp},
\eqn{}{
 ds^2_4 = e^{2A(r)} (-h(r) dt^2 + d\vec x^2_2) + \frac{e^{2B(r)}}{h(r)}dr^2 \,, \quad\quad
 A_i = \Phi_i(r) dt \,, \quad \quad \phi_A = \phi_A(r) \,.
}
They are characterized by four charges $Q_i$ and a mass parameter, the latter of which we may trade for a horizon radius $r_H$. It is convenient for us to take $Q_i > 0$, and separate out the signs of the gauge fields  $\eta_i \equiv \pm 1$. Then in terms of the functions
\begin{equation}
 H_i = 1+\frac{Q_i}{r} \,,
\end{equation}
the solutions are
\begin{equation}
  A(r)=- B(r) = \log\frac{r}{L}+\frac{1}{4}\sum_i \log H_i \,,
\end{equation}
\begin{equation}
 h(r) = 1-\frac{r(r_H+Q_a)(r_H+Q_b)(r_H+Q_c)(r_H+Q_d)}{r_H(r+Q_a)(r+Q_b)(r+Q_c)(r+Q_d)} \,,
\end{equation}
\begin{equation}
 \phi_1 = \frac{1}{2}\log\left(\frac{H_a H_b}{H_c H_d}\right) \ ,\ \  \phi_2 = \frac{1}{2}\log\left(\frac{H_a H_c}{H_b H_d}\right) \ ,\ \ \phi_3 = \frac{1}{2}\log\left(\frac{H_a H_d}{H_b H_c}\right) \,,
\end{equation}
\begin{equation}
 \Phi_i = \frac{\eta_i}{L}\sqrt{\frac{Q_i}{r_H}}\frac{\sqrt{(r_H+Q_a)(r_H+Q_b)(r_H+Q_c)(r_H+Q_d)}}{r_H+Q_i}\left(1-\frac{r_H+Q_i}{r+Q_i}\right) \,.
\end{equation}
The horizon $ r= r_H$ is the largest zero of the horizon function $h(r)$. These solutions are asymptotically anti-de Sitter at large $r$,
\eqn{}{
A(r \to \infty) &= - B(r \to \infty) \to \log {r \over L} \,, \cr h(r \to \infty) &\to 1\,, \quad\quad \phi_A(r \to \infty) \to 0 \,,\quad\quad \Phi_i (r \to \infty) \to {\rm const} \,,
}
with $AdS$ radius $L$.
These black brane solutions, when lifted to 11D, have the interpretation as rotating M2-brane configurations, with the conserved charges corresponding to conserved angular momenta in the eight directions transverse to the branes; this is analogous to the five-dimensional solutions studied in \cite{DeWolfe:2012uv}, corresponding to rotating D3-branes.

The thermodynamics may be calculated from standard formulas, with the temperature $T$ and entropy density $s$ determined by the metric,
\begin{equation}
 T=\frac{1}{4\pi}h'(r_H)e^{A(r_H)-B(r_H)} \ , \ \ s=\frac{1}{4G}e^{2A(r_H)} \,,
\end{equation}
and the chemical potentials $\mu_i$ and charge densities $\rho_i$ for the conserved charges  from the near-boundary expansion of the gauge fields,
\eqn{MuRho}{
\Phi_i(r \to \infty) \to \mu L - {8 \pi G L \rho \over r} + \ldots
}
The results are
\eqn{TEqn}{
 T=\frac{\sqrt{(r_H+Q_a)(r_H+Q_b)(r_H+Q_c)(r_H+Q_d)}}{4 \pi L^2}\left(-\frac{1}{r_H}+\frac{1}{r_H+Q_a}+\frac{1}{r_H+Q_b}+\frac{1}{r_H+Q_c}+\frac{1}{r_H+Q_d}\right)
}
\begin{equation}
s=\frac{1}{4GL^2}\sqrt{(r_H+Q_a)(r_H+Q_b)(r_H+Q_c)(r_H+Q_d)} \,,
\end{equation}
\begin{equation}
 \mu_i = \frac{\eta_i}{L^2}\sqrt{\frac{Q_i}{r_H}}\frac{\sqrt{(r_H+Q_1)(r_H+Q_2)(r_H+Q_3)(r_H+Q_4)}}{r_H+Q_i} \ .
\end{equation}
\eqn{RhoEqn}{
\rho_i = {\eta_i \over 2 \pi} \sqrt{Q_i \over r_H} s\,.
}
Extremal black holes have $T=0$, and generically display a double pole in $h(r \to r_H)$. It will be extremal solutions that we will focus on.

The simplest special case is the so-called four-charge black hole 
 (4QBH) where $Q_a = Q_b = Q_c = Q_d$; here the scalars all vanish and we are left with a Reissner-Nordstr\"om black brane. If $A_a = A_b = A_c = A_d \equiv \Phi_4(r) dt$, then the 4QBH solution is
\eqn{4QBH}{
 A(r)=-B(r)=\log\frac{r}{L}+\log\left(1+\frac{Q_4}{r}\right)  \,, \quad h(r) = 1-\frac{r(r_H+Q_4)^4}{r_H(r+Q_4)^4} \,,
 }
 \eqn{}{
 \Phi_4(r)= \frac{\eta_4}{L}\sqrt{\frac{Q_4}{r_H}}(r_H+Q_4)\left(1-\frac{r_H+Q_4}{r+Q_4}\right) \,.
}
Other simplifications can be chosen where two or three charges are set equal, which we will discuss in later sections.
Interesting special cases arise when one or more charges $Q_i$ vanishes, which we will explore in turn. For now, we turn to the fermionic Lagrangian. In what follows, we take all the signs of the charges to be positive, $\eta_i  = +1$.

\subsection{Fermionic action}

We are interested in the quadratic action for spin-1/2 fields. In general spin-1/2 fields may mix with the gravitini. The ${\bf 56}_{\rm s}$ representation consists of 32 unique weight vectors, along with three copies of the weights of the ${\bf 8}_{\rm s}$. Because the bosonic fields turned on in the background are all neutral under the Cartan gauge fields $U(1)_a \times U(1)_b \times U(1)_c \times U(1)_d$, and the action must respect this gauge symmetry, fermi fields can only mix in the quadratic action if they have the same weight vector. Thus the 32 spin-1/2 fields that have unique weight vectors cannot mix with the gravitini or each other. We will therefore consider these fields, and drop the couplings to the gravitini.

The quadratic fermion Lagrangian for these fields  has the form \cite{deWit:1982ig}:
\begin{equation}
\begin{split}
\label{FermiLagrangian}
 e^{-1} \mathcal{L}= \frac{i}{12}(\bar \chi^{ijk} \gamma^{\mu} D_{\mu} \chi_{ijk}-\bar \chi^{ijk} \overleftarrow{D}_{\mu} \gamma^{\mu} \chi_{ijk}) -\frac{1}{2}(F^+_{\mu\nu IJ}S^{IJ,KL}O^{+\mu\nu KL} + h.c.) & \\+ \left(\frac{\sqrt{2}}{144} g\, \epsilon^{ijklmnpq}A^2_{\;\;rlmn}\bar\chi_{ijk}\chi_{pq}^{\;\;\;\;r}+ h.c.\right) \,.
\end{split}
\end{equation}
Here $S^{IJ,KL}$ and $A^2_{\;\;rlmn}$ are the scalar tensors defined previously \eno{Atensors}, \eno{Stensor},
and $O^{+\mu\nu IJ}$ is a tensor quadratic in fermion fields and dependent on the scalars, given by (dropping gravitino terms)
\begin{equation}
\label{Otensor}
 u^{ij}_{\ IJ} O_{\mu\nu}^{+\ IJ}= \frac{\sqrt{2}}{144}\,\epsilon^{ijklmnpq}\bar\chi_{klm}\sigma_{\mu\nu}\chi_{npq} \,.
\end{equation}
The covariant derivative acting on the fermion is
\begin{equation}
 D_{\mu}\chi_{ijk}= \nabla_{\mu}\chi_{ijk}-\frac{1}{2}\mathcal{B}_{\mu\;\, i}^{\;\;\,l}\chi_{ljk}-\frac{1}{2}\mathcal{B}_{\mu\;\, j}^{\;\;\,l}\chi_{ilk}-\frac{1}{2}\mathcal{B}_{\mu\;\, k}^{\;\;\,l}\chi_{ijl} \,,
\end{equation}
with $\nabla_\mu$ containing the spin connection,
\eqn{}{
\nabla_\mu \equiv \partial_\mu - {1 \over 4} \omega_{\hat{a} \hat{b} \mu} \gamma^{\hat{a} \hat{b}} \,,
}
and where the composite connection is
\eqn{}{
{\cal B}_{\mu\;\,j}^{\;\;\,i} \equiv {2 \over 3} \left( u^{ik}_{\;\;\;IJ} \partial_\mu u_{jk}^{\;\;\;\,IJ} - v^{ikIJ} \partial_\mu v_{jkIL} \right)\,,
}
which for the ansatz \eno{ansatz} evaluates simply to
\eqn{}{
{\cal B}_{\mu\;\,j}^{\;\;\,i} = - 2 g A_{\mu\;\,j}^{\;\;\,i} \,.
}
The covariant derivative thus becomes
\begin{equation}
 D_{\mu}\chi_{ijk}= \nabla_{\mu}\chi_{ijk}+g A_{\mu \;\,i}^{\;\;\,l}\chi_{ljk}+gA_{\mu\;\, j}^{\;\;\,l}\chi_{ilk}+g A_{\mu \;\,k}^{\;\;\,l}\chi_{ijl} \,.
\end{equation}
Note that in this background, $SU(8)$ and $SO(8)$ indices are freely mixed together. One must still be careful to include factors of $u$ to translate between the two index types in appropriate places, as in  \eno{Otensor}. 

In dealing with the hermitian conjugates in \eno{FermiLagrangian}, we note that in our background $S^{IJKL}$ and $A^i_{2\, jkl}$ are real, and that $(O^{+\ IJ}_{\mu\nu})^{\dagger}=O^{+\ IJ}_{\mu\nu}$. Thus in the Pauli term the only thing that is different in the conjugate term is $F^+ \rightarrow F^-$. In the mass term the conjugate flips the $\chi$ and the $\bar\chi$, but since the fermions are Majorana, we have  $\bar\lambda \chi=\bar\chi \lambda$ and this just adds a factor of two. Integrating the kinetic term by parts and substituting \eno{gResult} for $g$, 
we can rewrite the fermionic Lagrangian as
\begin{equation}
\begin{split}
 e^{-1} \mathcal{L}= & \frac{i}{6}\bar \chi^{ijk} \gamma^{\mu} \nabla_{\mu} \chi_{ijk} +\frac{i}{8L}\bar \chi^{ijk} \gamma^{\mu} A_{\mu\;\, i}^{\;\;\, l}\chi_{ljk}    + \frac{1}{72L}\epsilon^{ijklmnpq}A^2_{\;\;rlmn}\bar\chi_{ijk}\chi_{pq}^{\;\;\;\;r}\\ & -\frac{1}{576}F_{\mu\nu ij}S^{ijkl}(u^{-1})_{klmn}\epsilon^{mnpqrstu}\bar\chi_{pqr}\sigma^{\mu\nu}\chi_{stu} \,.
\end{split}
\end{equation}
Thinking of the $\chi_{ijk}$ as a 56-component vector $\vec\chi$, this Lagrangian has the form 
\begin{equation}
 e^{-1}\mathcal{L}={1 \over 2} \vec{\bar{\chi}} ( i\gamma^{\mu}\nabla_{\mu}\mathbf{\mathbf{1}} +\mathbf{Q} +\mathbf{M} +\mathbf{P})\vec\chi
\end{equation}
where ${\bf 1}$, ${\bf Q}$, ${\bf M}$ and ${\bf P}$ are $56\times 56$ matrices for the kinetic, gauge, mass and Pauli-type terms, respectively.
We then diagonalize these matrices to find eigenvectors and eigenvalues.
Diagonalizing first the gauge term, we find 32 eigenvectors with distinct, non-degenerate eigenvalues, and 24 eigenvectors that are degenerate in groups of three, as expected. The latter contain some additional mixing to the gravitini which we have ignored, and therefore we set them aside. The remaining 32 cannot mix thanks to gauge invariance, and are therefore also eigenvectors of the mass and Pauli terms.

In general the eigenvectors are complex linear combinations of the form $\chi  = \chi_1 + i \chi_2$ where $\chi_1$ and $\chi_2$ are two of the $\chi_{ijk}$, and are hence Dirac spinors; 16 are then conjugates of the other 16. The Dirac equation for these eigenvectors takes the form
\eqn{GeneralDirac}{
 \Big[& i\gamma^{\mu}\nabla_{\mu} +\frac{1}{4L}\, \sum_{i=a,b,c,d} m_i e^{\lambda_i/2}+\frac{1}{4L} \gamma^{\mu} \sum_{i=a,b,c,d}  q_i A_{\mu}^i +\frac{i}{8}\sigma^{\mu\nu} \sum_{i=a,b,c,d} \left( p_i e^{-\lambda_i/2} F_{\mu\nu}^i\right) \Big] \chi   =0 \ \,.
}
Here, $m_i$, $q_i$, and $p_i$, $i=a,b,c,d$ are integer numbers characterizing each fermion. The $\lambda_i$ are combinations of the scalars given in (\ref{lambdas}). The table of 16 independent eigenvectors is
\begin{center}
\begin{tabular}{|l"l"l|l|l|l"l|l|l|l"l|l|l|l|}\hline
 $\chi^{(q_a,q_b,q_c,q_d)}$ & Operator &$m_a$ & $m_b$ & $m_c$ & $m_d$ & $q_a$ & $q_b$ & $q_c$ & $q_d$ & $p_a$ & $p_b$ & $p_c$ & $p_d$\\\thickhline
 
 $\chi^{(+3,-1,+1,+1)}$& Tr~$Z_1 \Lambda_2$& $-3$ & $+1$ & $+1$ & $+1$ & $+3$ & $-1$ & $+1$ & $+1$ & $-1$ & $-1$ & $+1$ & $+1$\\ 
$\chi^{(+3,+1,-1,+1)}$& Tr~$Z_1 \Lambda_3$& $-3$ & $+1$ & $+1$ & $+1$ & $+3$ & $+1$ & $-1$ & $+1$ & $-1$ & $+1$ & $-1$ & $+1$\\
$\chi^{(+3,+1,+1,-1)}$& Tr~$Z_1 \Lambda_4$& $-3$ & $+1$ & $+1$ & $+1$ & $+3$ & $+1$ & $+1$ & $-1$ & $-1$ & $+1$ & $+1$ & $-1$\\
 
 $\chi^{(-1,+3,+1,+1)}$&Tr~$Z_2 \Lambda_1$ & $+1$ & $-3$ & $+1$ & $+1$ & $-1$ & $+3$ & $+1$ & $+1$ & $-1$ & $-1$ & $+1$ & $+1$\\
 $\chi^{(+1,+3,-1,+1)}$&Tr~$Z_2 \Lambda_3$ & $+1$ & $-3$ & $+1$ & $+1$ & $+1$ & $+3$ & $-1$ & $+1$ & $+1$ & $-1$ & $-1$ & $+1$\\
 $\chi^{(+1,+3,+1,-1)}$& Tr~$Z_2 \Lambda_4$& $+1$ & $-3$ & $+1$ & $+1$ & $+1$ & $+3$ & $+1$ & $-1$ & $+1$ & $-1$ & $+1$ & $-1$\\

$\chi^{(-1,+1,+3,+1)}$& Tr~$Z_3 \Lambda_1$& $+1$ & $+1$ & $-3$ & $+1$ & $-1$ & $+1$ & $+3$ & $+1$ & $-1$ & $+1$ & $-1$ & $+1$\\
 $\chi^{(+1,-1,+3,+1)}$&Tr~$Z_3 \Lambda_2$ & $+1$ & $+1$ & $-3$ & $+1$ & $+1$ & $-1$ & $+3$ & $+1$ & $+1$ & $-1$ & $-1$ & $+1$\\
 $\chi^{(+1,+1,+3,-1)}$&Tr~$Z_3 \Lambda_4$ & $+1$ & $+1$ & $-3$ & $+1$ & $+1$ & $+1$ & $+3$ & $-1$ & $+1$ & $+1$ & $-1$ & $-1$\\

 $\chi^{(-1,+1,+1,+3)}$&Tr~$Z_4 \Lambda_1$ & $+1$ & $+1$ & $+1$ & $-3$ & $-1$ & $+1$ & $+1$ & $+3$ & $-1$ & $+1$ & $+1$ & $-1$\\
$\chi^{(+1,-1,+1,+3)}$&Tr~$Z_4 \Lambda_2$ & $+1$ & $+1$ & $+1$ & $-3$ & $+1$ & $-1$ & $+1$ & $+3$ & $+1$ & $-1$ & $+1$ & $-1$\\
 $\chi^{(+1,+1,-1,+3)}$&Tr~$Z_4 \Lambda_3$ & $+1$ & $+1$ & $+1$ & $-3$ & $+1$ & $+1$ & $-1$ & $+3$ & $+1$ & $+1$ & $-1$ & $-1$\\
\hline
 $\chi^{(+3,-1,-1,-1)}$&Tr~$Z_1 \bar\Lambda_1$ & $-3$ & $+1$ & $+1$ & $+1$ & $+3$ & $-1$ & $-1$ & $-1$ & $-1$ & $-1$ & $-1$ & $-1$\\
$\chi^{(-1,+3,-1,-1)}$& Tr~$Z_2 \bar\Lambda_2$& $+1$ & $-3$ & $+1$ & $+1$ & $-1$ & $+3$ & $-1$ & $-1$ & $-1$ & $-1$ & $-1$ & $-1$\\
$\chi^{(-1,-1,+3,-1)}$&Tr~$Z_3 \bar\Lambda_3$ & $+1$ & $+1$ & $-3$ & $+1$ & $-1$ & $-1$ & $+3$ & $-1$ & $-1$ & $-1$ & $-1$ & $-1$\\
 $\chi^{(-1,-1,-1,+3)}$&Tr~$Z_4 \bar\Lambda_4$ & $+1$ & $+1$ & $+1$ & $-3$ & $-1$ & $-1$ & $-1$ & $+3$ & $-1$ & $-1$ & $-1$ & $-1$\\\hline
\end{tabular}
\end{center}

\bigskip
\noindent
The 16 conjugate fermions simply have $(m_i, q_i, p_i) \to (m_i, -q_i, -p_i)$.
The $q_i$ in the table are proportional to the 32 weight vectors of  the ${\bf 56}_{\rm s}$ representation of $SO(8)$ with norm $\sqrt{3}$, as expected, and can be characterized as follows: one of the four $q_i$ is $q_i = 3$. Of the remaining three $q_j$, an odd number (either one or all three) are $-1$, with the remaining charges, if any, equal to $+1$. There are 16 such combinations.  We identify the dual operators in the table, where the complex scalars $Z_j \equiv X_{2j - 1} + i X_{2j}$, $j = 1,2,3,4$ have weight vectors proportional to $(+1,0,0,0)$ and permutations, and $\Lambda_j\equiv \lambda_{2j - 1} + i \lambda_{2j}$, $j = 1,2,3,4$ are complex combinations of the eight spinors with weight vectors proportional to $(-\frac{1}{2}, \frac{1}{2}, \frac{1}{2}, \frac{1}{2})$ and permutations.

The $m_i$ are then determined by the $q_i$: if $|q_i| = 3$, $m_i = -3$, while if $|q_i| = 1$, $m_i = 1$. Finally the $p_i$ are all $\pm 1$, and are simply the ratios
\eqn{pRatio}{
p_i = {m_i \over q_i} \,,
}
for each $i$. 
Thus the four charges completely characterize the Dirac equation. We note that for each fermion the $m_i$ satisfy
\eqn{}{
m_a + m_b + m_c + m_d = 0 \,.
}
We find it useful to sort these fermions into two categories: the first 12 are {\em net-charged} fermions, for which only one $q_i$ is $-1$, and for which $\sum_i q_i = +4$ and $\sum_i p_i = 0$, while the final four are the {\em net-neutral} fermions, for which three $q_i$ are $-1$, and for which $\sum_i q_i = 0$ and $\sum_i p_i = -4$.

\section{Fermionic Green's functions}
\label{GreenFunctionSec}

In this section, we discuss how to solve the Dirac equation obtained in the previous section, and review how the retarded Green's function may be obtained from such a solution. There is an apparent ambiguity in how to treat the quantization of this fermionic fluctuation, and we discuss how this ambiguity is resolved by supersymmetry.

\subsection{Solving the Dirac equation}

Solutions to Dirac equations of the form \eno{GeneralDirac} were discussed in \cite{Faulkner:2009wj} for constant mass and gauge couplings, and Pauli couplings were added in \cite{Edalati:2010ge, Edalati:2010ww}. Further development, including cases with scalar-dependent couplings, was carried out in \cite{DeWolfe:2012uv, Gubser:2012yb, DeWolfe:2013uba}. We begin by Fourier transforming the $t$, $\vec{x}$ directions and rescaling the spinor $\chi$,
\begin{equation}
\label{Fourier}
\chi \equiv (e^{6A} h)^{-1/4} e^{-i\omega t+ikx} \psi \ ,
\end{equation}
where $\omega$ is the frequency and $k$ is the spatial momentum (chosen to lie in the $x$-direction) of the fermion mode. The factor of $(e^{6A} h)^{-1/4}$ is chosen so as to exactly cancel the spin connection term coming from $\nabla_{\mu}$ in the Dirac equations above. Next we choose a Clifford basis where the relevant matrices are block diagonal,
\eqn{}{
\gamma^{\hat{r}} =  \begin{pmatrix} i \sigma_3 & 0 \\  0 & i \sigma_3 \end{pmatrix} \,, \quad
\gamma^{\hat{t}} =  \begin{pmatrix} \sigma_1 & 0 \\  0 & \sigma_1\end{pmatrix} \,, \quad
\gamma^{\hat{i}} =  \begin{pmatrix} i \sigma_2  & 0 \\  0 & -i \sigma_2\end{pmatrix} \,.
}
We can characterize the four components of the spinor as
\eqn{}{
\psi_{\alpha \pm} \equiv \Pi_\alpha P_\pm \psi \,.
}
with $\alpha = 1,2$, in terms of the projectors
\eqn{Projectors}{
\Pi_\alpha \equiv {1 \over 2} \left(1 - (-1)^{\alpha} i \gamma^{\hat{r}} \gamma^{\hat{t}} \gamma^{\hat{i}} \right) \,,\quad \quad
P_\pm \equiv {1 \over 2}\left ( 1 \pm i \gamma^{\hat{r}} \right) \,.
}
The two-component objects $\psi_+$ and $\psi_-$ (each with both values of $\alpha$) transform as three-dimensional Dirac spinors. However, it is in terms of the two-component objects $\psi_\alpha$,
\begin{equation}
\psi_{\alpha}= \left(
\begin{array}{c}
\psi_{\alpha-} \\
\psi_{\alpha+} \end{array} \right)
\end{equation}
(which are not lower-dimensional spinors) that the Dirac equation decomposes into two decoupled pairs of equations:
\eqn{FirstOrderDirac}{
(\partial_r + X \sigma_3 + Y i \sigma_2 + Z \sigma_1) \psi_\alpha = 0 \,,
}
where
\begin{equation}
X= -{e^B \over 4 L \sqrt{h}} \sum_i m_i e^{\lambda_i / 2} \,, 
\qquad
Y= -\frac{e^{B-A}}{\sqrt{h}} u \,,
\qquad
Z= -\frac{e^{B-A}}{\sqrt{h}}\big[ (-1)^{\alpha} k - v \big] \,,
\end{equation}
with
\begin{equation}
\label{uandv}
u= \frac{1}{\sqrt{h}}\big[\omega  +\frac{1}{4L} \sum_i q_i \Phi_i\big]
\,, \qquad
v= \frac{e^{-B}}{4} \sum_i p_i e^{-\lambda_i /2} \partial_r \Phi_i \,.
\end{equation}
We note that the solutions for $\psi_{\alpha = 1}$ and $\psi_{\alpha = 2}$ are related to each other simply by $k \to -k$.

We may turn the coupled first-order equations \eno{FirstOrderDirac} into decoupled second-order equations for each component,
\eqn{SecondOrderDirac}{
 \psi''_{ \alpha \pm} - F_\pm  \psi'_{ \alpha \pm}  + \left( \mp X' - X^2 + Y^2- Z^2 \pm X F_\pm \right)\psi_{\alpha \pm}=0\,,
}
with $F_\pm \equiv \partial_r \log \left(\mp Y+Z \right)$,
where we keep in mind that \eno{FirstOrderDirac} keeps the solutions for different components from being independent. The form \eno{SecondOrderDirac} is convenient for an analysis at $r \to \infty$, but for $r \to r_H$
it is convenient to define the combinations \cite{Gubser:2012yb},
\begin{equation}
U_{\pm} \equiv \psi_{-} \pm i\psi_{+}
\end{equation}
in terms of which the Dirac equations become
\eqn{}{
U'_{-}+i Y U_{-} = (-X+i Z)U_{+} \quad \quad
U'_{+}-i Y U_{+} = (-X-i Z)U_{-} \,.
}
From here one can derive the uncoupled second-order equations
\eqn{}{
U''_{-}+pU'_{-}+(iY'-X^2+Y^2-Z^2+iYp) U_{-} &= 0 \cr
U''_{+}+\bar p U'_{+}+(-iY'-X^2+Y^2-Z^2-iY \bar p) U_{+} &= 0 \ .
}
with $p\equiv -\partial_r\log(-X+i Z)$.

We note there are two independent discrete transformations acting on the Dirac equation. Conjugation is implemented by showing that if $\chi$ satisfies the Dirac equation with parameters $\{m_i, q_i, p_i \}$, then $\gamma^{\hat{r}} \chi^*$ satisfies it with parameters $\{ m_i, -q_i, -p_i\}$. Conjugation of \eno{Fourier} also exchanges the signs of $k$ and $\omega$, so the net transformation is
\eqn{Conjugation}{
{\rm Conjugation:} \quad \quad \quad  q \to -q \,, \quad \quad p \to -p \,, \quad \quad \omega \to - \omega \,, \quad \quad k \to -k \,,
}
which is equivalent to $Y \to -Y$, $Z \to -Z$; one can see the $\psi_\pm$ second-order equations respect this symmetry, while it exchanges the equations for $U_+$ and $U_-$. Meanwhile one can also show that if $\chi$ satisfies the Dirac equation with parameters $\{m_i, q_i, p_i \}$, then $\gamma_5 \chi$ satisfies it with parameters $\{ -m_i, q_i, -p_i\}$. The chirality matrix exchanges both $\psi_+$ and $\psi_-$ and the two values of $\alpha$; since the latter is equivalent to flipping the sign of $k$, we have
\eqn{ChiralityFlip}{
{\rm Chirality \ flip:} \quad \quad \quad m \to -m \,, \quad \quad p \to -p \,, \quad \quad k \to - k \,, \quad \quad \psi_+ \leftrightarrow \psi_- \,,
}
which is $X \to -X$, $Z \to -Z$; while this exchanges $\psi_+ \leftrightarrow \psi_-$, it is a symmetry of the $U_\pm$ equations.

\subsection{Quantization of Fermi fields and Green's functions}

To define any field in an asymptotically anti-de Sitter space, one must impose appropriate boundary conditions at $r \to \infty$. The functions appearing in the Dirac equation have the asymptotic behavior
\eqn{BdySeries}{
X \to { m_0 L \over r} \,, \quad \quad
Y \to  -{\tilde\omega L^2 \over r^2} \,, \quad \quad
Z \to -{k L^2 \over r^2} \,,
}
with $m_0$ the value of $m(\phi)$ at infinity, and
\eqn{}{
\tilde\omega \equiv \omega + q A_0(r \to \infty) \,.
}
We discuss this first for the case of general $m_0$, discussed in \cite{Iqbal:2009fd}, and then specialize to our case, where $m_0 = 0$. 
The behavior \eno{BdySeries} leads to the near-boundary second-order equation,
\eqn{}{
 \psi''_{\alpha \pm}  + {2 \over r}  \psi'_{ \alpha\pm}  - {m_0^2 L^2 \pm m_0 L \over r^2}\psi_{\alpha \pm}\,,
}
The asymptotic solutions are then,
\eqn{}{
\psi_{\alpha+} \sim A_{\alpha +}(\omega, k) r^{m_0L} + B_{\alpha +}(\omega, k) r^{-m_0L - 1} \,, \quad \quad
\psi_{\alpha-} \sim A_{\alpha -}(\omega, k) r^{-m_0L} + B_{\alpha -}(\omega, k) r^{m_0L-1} \,,
}
or in terms of the original spinor,
\eqn{BoundaryScalings}{
\chi_{\alpha+} &\sim A_{\alpha +}(\omega, k) r^{- d/2+ m_0L} + B_{\alpha +}(\omega, k) r^{-d/2-m_0L - 1} \,, \cr
\chi_{\alpha-} &\sim  A_{\alpha -}(\omega, k) r^{-d/2-m_0L}  +B_{\alpha -}(\omega, k) r^{-d/2+m_0L-1} \,.
}
Using the full Dirac equation on the asymptotic solution \eno{BoundaryScalings}, one finds that
the $B_\pm$ are not independent of the $A_\mp$, but rather are derivatives of them:
\eqn{BRelations}{
B_{\alpha \mp} = {L^2 (\tilde\omega \pm (-1)^\alpha k) \over 2m_0L\mp 1} A_{\alpha \pm} \,, \quad \quad
}
One must choose whether $A_+$ or $A_-$ is the mode that one imposes boundary conditions on; this is only allowed when the mode is normalizable, which depends on the value of $m_0$. The chosen mode is then interpreted as the response (vev) of the dual operator, while the other mode is interpreted as the source.  The retarded Greens function is then given by the ratio of the response over the source, for a solution of the Dirac equation for which infalling boundary conditions have been imposed at the black hole horizon.

The $A_-$ quantization is allowed for $m_0 L > -1/2$ and corresponds to a dual operator with $\Delta = d/2 + m_0 L$, with Green's function 
\eqn{AminusGreen}{
\quad  \quad \quad \quad \quad \quad \quad \quad \quad \quad \quad \quad \quad G_{R, \alpha} = {A_{\alpha -} \over A_{\alpha +}} \,, \quad \quad \quad\quad \quad (A_- \ {\rm quantization})
}
while the $A_+$ quantization is allowed for $m_0 L < 1/2$ and corresponds to a dual operator with $\Delta = d/2- m_0 L$, with Green's function
\eqn{AplusGreen}{
\quad  \quad \quad \quad \quad \quad \quad \quad \quad \quad \quad \quad \quad G_{R, \alpha} = {A_{\alpha +} \over A_{\alpha -}} \,. \quad \quad \quad\quad \quad (A_+ \ {\rm quantization})
}
For the range $-1/2 < m_0 L < 1/2$, both quantizations are possible. Note that the Green's function is diagonal on the space of two-component spinors $\alpha = 1, 2$;  in what follows we will pick a single component $\alpha=2$ for convenience, knowing that $G_{R, 1}(k) = G_{R, 2}(-k)$.

Now for our special case $m_0 = 0$, we find $\psi_+$ and $\psi_-$ have the same scaling in $r$: 
\eqn{}{
\psi_{\alpha+} \sim A_{\alpha +}(\omega, k) + {B_{\alpha +}(\omega, k) \over r} \,, \quad \quad
\psi_{\alpha-} \sim  A_{\alpha -}(\omega, k)+ {B_{\alpha -}(\omega, k)  \over r}  \,.
}
The leading term in $X$ is now
\eqn{}{
X = {m_1 \over r^2} + \ldots \,,
}
and the relations between the $B_\mp$ and $A_\pm$  from the first-order equations are modified to
\eqn{}{
B_{\alpha \mp} = \mp L^2 (\tilde\omega \pm (-1)^\alpha k)  A_{\alpha \pm}  \pm m_1 A_{\alpha \mp} \,.
}
There is now an ambiguity in the identification of the source and the response: both $A_+$ and $A_-$ appear in symmetric fashion with the same scaling in $r$, appropriate to the situation where $\Delta = 3/2 = d/2$, and the operator and its source have the same conformal dimension. 
In the case of scalar fluctuations in AdS/CFT, the $\Delta = d/2$ case involves a term of the form $r^{-d/2}$ and a term of the form $r^{d/2} \log r$, and there is only one conformally invariant choice of quantization \cite{Klebanov:1999tb}. For spinors, however, there is no log and two choices of quantization are possible. 

In the simple case of a fermion with $m=p=0$ exactly, the chirality flip \eno{ChiralityFlip} implies that the two quantizations are equivalent up to $k \to -k$, so there is no loss of generality to simply picking one; this is the case usually discussed in the literature, for example \cite{Faulkner:2009wj}. However in our case, both $p$ and $m$ are nonzero (though $m$ is asymptotically zero) and depend on $r$. In this more general situation, the two different choices of quantization lead to distinct physics; in particular, for us, they will exchange poles of the fermionic Green's function with zeros. This exchange in the $m=0$ case was noted in \cite{Alsup:2014uca}. Thus we must find a way to resolve the ambiguity to correctly identify the fermionic response.

To resolve the issue, we will use supersymmetry. The 70 scalars of maximal gauged supergravity are divided into ${\bf 35_{\rm v}}$ parity-even scalars with $\Delta =1$, and ${\bf 35_{\rm c}}$ pseudoscalars with $\Delta = 2$. All the scalar modes, however, asymptotically have $m^2 L^2 = -2$. The well-known analog of \eno{BoundaryScalings} for scalars is
\eqn{AsymptoticScalar}{
\phi = A_-(\vec{x}, t) \,r^{-\Delta_-} + \ldots + A_+(\vec{x}, t) \,r^{-\Delta_+} + \ldots \,, \quad \quad \Delta_\pm = {d \over 2} \pm \sqrt{{d^2 \over 4} + m^2 L^2} \,.
}
which for the case at hand gives $\Delta_- = 1$, $\Delta_+ = 2$. Again there is a choice of quantization \cite{Klebanov:1999tb}, and to match the dual field theory, we must place the ${\bf 35}_{\rm v}$ scalars in the alternate quantization to get $\Delta = 1$, and the ${\bf 35}_{\rm c}$ pseudoscalars in the regular quantization to obtain $\Delta = 2$. We now show how supersymmetry relates this choice of scalar quantization to a definite choice of spinor quantization. These results were discussed in pre-AdS/CFT language in \cite{Breitenlohner:1982bm, Breitenlohner:1982jf}; for a related discussion see \cite{Freedman:2013oja}.

It is sufficient to consider a single ${\cal N}=1$ supersymmetry, under which a scalar $\phi$, a pseudoscalar ${\cal P}$ and a Majorana spinor $\chi$ assemble into a single chiral multiplet. The action for such a multiplet is 
\eqn{ChiralMultAction}{
S = {1 \over 2} \int d^4x \sqrt{-g} \Big( -g^{\mu\nu} \partial_\mu \phi \partial_\nu \phi -g^{\mu\nu} \partial_\mu {\cal P} \partial_\nu {\cal P} + i \bar\chi \gamma^\mu \nabla_\mu \chi - m_\phi^2 \phi^2 -  m_{\cal P}^2 {\cal P}^2 - m \bar\chi \chi \Big) \,,
}
where the scalars have masses
\eqn{}{
m_\phi^2 \equiv  \left(m^2-{m \over L}- {2 \over L^2}\right)  \,, \quad \quad
m_{\cal P}^2 \equiv  \left(m^2+{m \over L}- {2 \over L^2}\right) \,.
}
Being in anti-de Sitter space has split the three masses of the multiplet, but all masses are determined by the single fermion mass parameter $m$. It is straightforward to see that as $m$ varies from $m=-\infty$ to $m = \infty$, the scalar mass-squareds go down from infinity, reach a minimum at the Breitenlohner-Freedman bound $m_{\rm BF}^2 L^2 = -{9 \over 4}$, and go back to infinity. The action \eno{ChiralMultAction} is invariant under the transformations
\eqn{}{
\delta \phi = \bar\varepsilon \chi \,, \quad \quad
\delta {\cal P} = i \bar\varepsilon \gamma_5 \chi \,, \quad \quad
\delta \chi = - \left[i \gamma^\mu \partial_\mu (\phi + i \gamma_5 {\cal P}) + {1 \over L} (\phi-i\gamma_5{\cal P}) + m (\phi + i \gamma_5 {\cal P}) \right] \varepsilon \,.
}
Our strategy is to use a Killing spinor of the AdS background to generate a near-boundary solution for the scalars from a near-boundary solution of the spinor; this will match the fluctuations on which boundary conditions are imposed between the scalar and spinor sectors, which will allow us to choose our spinor quantization.
A Killing spinor is obtained by requiring that the gravitino supersymmetry variation \cite{deWit:1982ig},
\eqn{MaximalGravitinoVariation}{
\delta \psi^i_\mu = 2 \nabla_\mu \varepsilon^i -i \sqrt{2} g A_1^{ji} \gamma_\mu \varepsilon_j + \ldots \,,
}
vanishes. In AdS space where $A_1^{ij} = \delta^{ij}$ this becomes
\eqn{}{
\delta \psi^i_\mu = 2 \nabla_\mu \varepsilon^i -{1 \over L}  \gamma_\mu \varepsilon^i =0\,.
}
The $r$-dependent Killing spinor solution for any $i$ is:
\eqn{KillingSpinor}{
\varepsilon(r) = r^{1/2} \varepsilon^{(0)}_+ \,,
}
with $\gamma^{\hat{r}}$-chirality $\varepsilon_+ \equiv P_+ \varepsilon_+$ as in \eno{Projectors}. The supersymmetry variations of the scalars with this Killing spinor as supersymmetry parameter then each involve only one of $\chi_\pm$,
\eqn{}{
\delta \phi = \bar\varepsilon \chi_- \,, \quad \quad \delta {\cal P} = i\bar\varepsilon \gamma_5 \chi_+ \,.
}
Consider the $A_-$ quantization of $\chi$; this is permitted for $mL \geq -1/2$, and has
\eqn{}{
\Delta_\chi = {3 \over 2} + mL \,. 
}
We consider a fluctuation of $\chi$ with no ``source" term; thus only $A_-$ and $B_+$ are turned on:
\eqn{}{
\chi_+ =  B_+ r^{-5/2-mL } \,, \quad \quad
\chi_- =  A_- r^{-3/2-mL}   \,.
}
We then find the corresponding scalar fluctuations,
\eqn{}{
\delta \phi = r^{-mL - 1} \left( {1 \over \sqrt{2} }\bar\varepsilon^{(0)}_+ A_- \right) \,, \quad \quad
\delta {\cal P} = r^{-mL - 2} \left( {i \over \sqrt{2} }\bar\varepsilon^{(0)}_+ \gamma_5 B_+ \right) \,.
}
Thus supersymmetry requires we pick the quantizations of the scalars giving the operator dimensions
\eqn{ABDeltas}{
\Delta_\phi = 1 + mL \,, \quad \quad \Delta_{\cal P} = 2 + mL \,.
}
Analogously, the $A_+$ quantization would lead to $\Delta_\phi = 2 - mL$, $\Delta_{\cal P} = 1- mL$. For us, we require $\Delta_\phi = 1$, $\Delta_{\cal P} = 2$, which obtains for $m=0$ in the $A_-$ quantization, with the response in $\chi_-$ and the source in $\chi_+$.  Thus we have resolved the ambiguity, and \eno{AminusGreen} will be our expression for the fermionic Green's function.

The $A_+$ quantization would place the scalars in the regular quantization and the pseudoscalars in the alternate quantization, contrary to maximal gauged supergravity. In principle this represents some other non-supersymmetric AdS/CFT dual pair. Since the Green's functions for the two quantizations \eno{AminusGreen} and \eno{AplusGreen} are reciprocals, the poles and zeros of the Green's function are exchanged between the two. We will indicate the zeros of the Green's function in many of our backgrounds; one may give them the alternate interpretation as Fermi surface singularities for the non-supersymmetric theory of the other quantization.

\section{Regular black holes and non-Fermi liquids}
\label{RegularSec}

We turn now to solving the Dirac equation to obtain retarded Green's functions for different fermions at zero temperature and various values of the chemical potentials, obtaining information about the fermionic response over the parameter space of the ABJM theory.

\begin{figure}
\begin{center}
\includegraphics[scale=0.5]{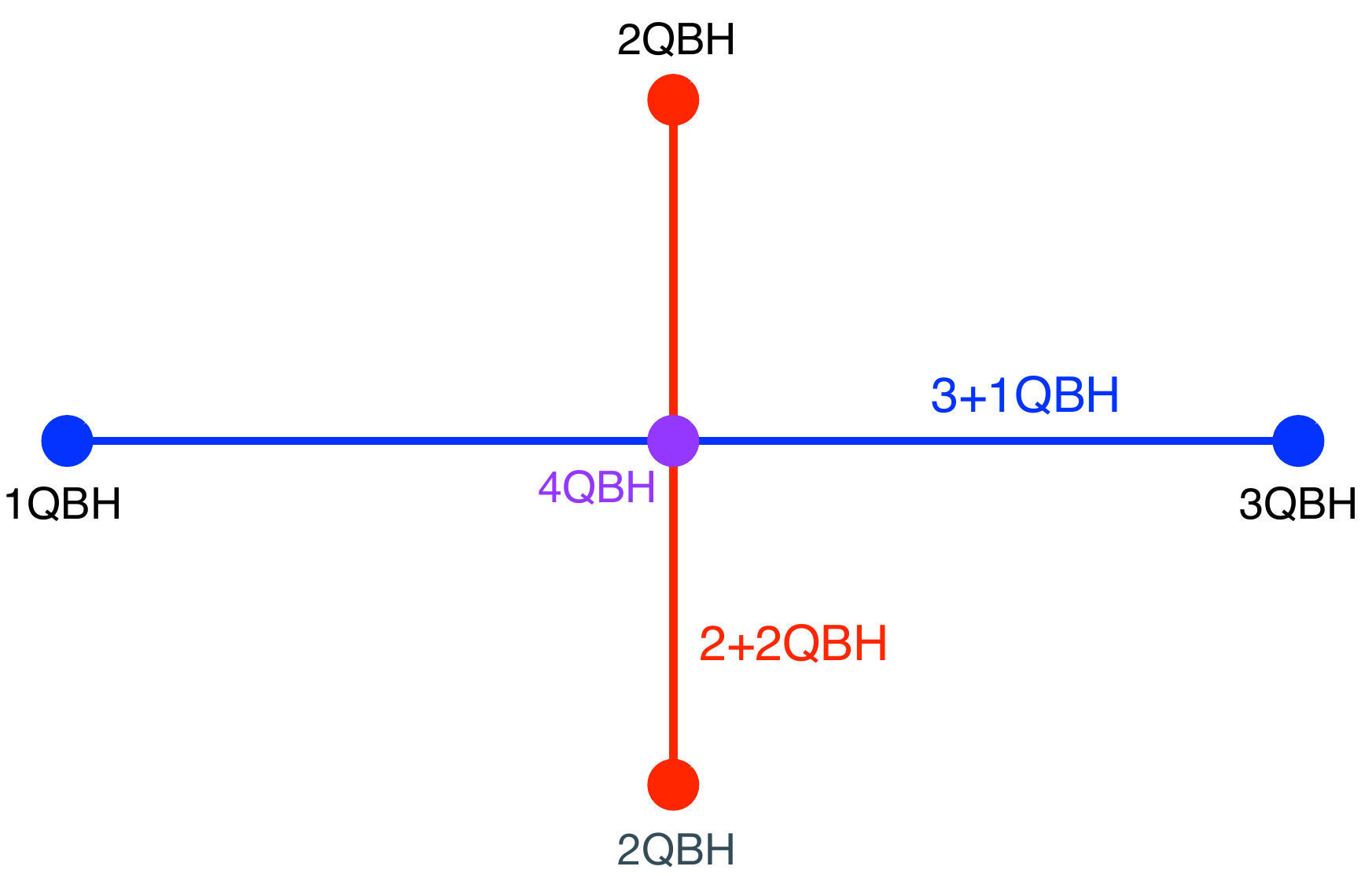}
\caption{A cartoon of the parameter space of black holes we consider.
\label{BHFig}}
\end{center}
\end{figure}

In principle, one could study the entire black hole parameter space of four independent charges of black holes. However, dealing with four charges can be somewhat tedious. To simplify matters, we will consider a truncated parameter space, examining two classes of simplified black holes: one class with three charges set equal, and the other distinct (the ``3+1-charge black hole")
and one
with the four charges set to two values in pairs (the ``2+2-charge black hole"). As we will see, each class simplifies the solutions to consist of two gauge fields and a single scalar. Since only the ratio of charges matters, the parameter space consists of two one-dimensional segments that intersect at the point where all four charges are equal, the four-charge black hole, which has vanishing scalars and is simply a Reissner-Nordstr\"om black brane. 
A cartoon of the parameter space is displayed in Figure~\ref{BHFig}.

Generic black branes with all four charges nonzero are ``regular", with a regular horizon, and display qualitatively similar behavior; these will be explored in this section.
Novel phenomena occur when one or more charges vanish. These interesting special cases occur at the boundaries of our parameter space, and we will investigate them in more detail in future sections. While we do not cover the entire parameter space of four charges, we expect that the unexplored areas are qualitatively similar to corresponding regions in our explored space with the same number of nonzero charges.

\subsection{Regular black holes and non-Fermi liquids}

Regular black holes are characterized by a regular horizon; for the extremal case there is a double pole in the horizon function, $h(r) \sim (r - r_H)^2$, but the horizon remains of nonzero size. Since the entropy density of the dual field theory is simply proportional to the area of the  horizon, these systems have a nonzero entropy density even as the temperature goes to zero. Zero entropy at zero temperature will require a singular event horizon, as we review in a later section.

The fermionic response of regular black holes was considered in \cite{Faulkner:2009wj} for fermions with constant masses, and Pauli couplings were added in \cite{Edalati:2010ww, Edalati:2010ge}. In \cite{DeWolfe:2012uv}, it was shown that top-down supergravities in five dimensions generically are of this type, albeit with masses and Pauli couplings depending on the radial coordinate; this did not change the overall structure.

Let us review how the Green's function may be calculated in this case. One must solve the Dirac equation for a fermionic fluctuation with infalling boundary conditions at the horizon, and then calculate the ratio \eno{AminusGreen} of the components near the boundary. For the general case of $\omega \neq 0$, this can be done straightforwardly. Near $\omega =0 $ there is a subtlety \cite{Faulkner:2009wj}.
For Dirac equations in the background of regular extremal black holes, the near-horizon ($r \to r_H$) limit has the structure
\eqn{}{
U'' + \left( {1 \over r-r_H} + \ldots \right) U' + \left( {\# L^4 \omega^2  \over  (r -r_H)^4} + {\#' L^2 \omega  \over (r-r_H)^3} - {\nu^2 \over (r-r_H)^2 }+ \ldots \right) U = 0 \,,
}
where $\#$ and $\#'$ are constants we are not interested in and $\nu^2$ is a constant we are interested in, and we have neglected both higher-order terms in $1/(r-r_H)$ and in $\omega$. Because the near-horizon and small-frequency limits do not commute, to study dynamics at low energy one must define an inner region with $\omega \to 0$, $r \to r_H$, $\omega/(r-r_H)$ fixed, where the infalling boundary condition is imposed; this is then matched to an outer region with $\omega =0$ strictly, and the result may be extended to small $\omega$. The inner (IR) region for black branes in $AdS_{d+1}$ has the geometry $AdS_2 \times \mathbb{R}^{d-1}$, and this region governs the low energy properties of the dual gauge theory. An infalling solution at the horizon translates to a solution bridging the gap between inner and outer regions with the form
\eqn{AdS2Bdy}{
U \sim (r - r_H)^{- {1 \over 2} + \nu} + {\cal G}(\omega) (r - r_H)^{- {1 \over 2} - \nu} \,,
}
where the relative weighting ${\cal G}(\omega)$ between the two solutions depends on $k$ and the other parameters as well. Thinking of the two terms in \eno{AdS2Bdy} as the source and response in the near-boundary region of an $AdS_2$ fluctuation, we may interpret ${\cal G}(\omega)$ as an $AdS_2$ Green's function. 

The exponent $\nu$ takes the form
\eqn{Nu}{
\nu^2 = \nu_{\rm m}^2 + \nu_{\rm k}^2 - \nu_{\rm q}^2 \,,
}
where $\nu_{\rm m}$ depends on the mass parameters $m_i$, $\nu_{\rm k}$ depends on the momentum $k$ and the Pauli couplings $p_i$, and $\nu_{\rm q}$ depends on the charges $q_i$; all three depend on the ratios of chemical potentials encoding where we are in the parameter space.  The term $\nu_{\rm k}^2$ depends on $k$ and the $p_i$ only in the combination $\tilde{k}^2$, where
\eqn{ktilde}{
\tilde{k} \equiv k + \sum_i \alpha_i \, p_i \, \mu_i \,,
}
where $\alpha_i$ are some constants and $\mu_i$ is the corresponding chemical potential, so the effect of the Pauli couplings is to shift the momentum.
In \cite{Iqbal:2011in}, the combination of the terms $\nu_{\rm m}^2 - \nu_{\rm q}^2$ was identified as being proportional to the inverse correlation length squared,
\eqn{CorrLength}{
\nu\sim \sqrt{\tilde{k}^2 + {1 \over \xi^2} } \,,
}
where for our more general case we have replaced $k^2$ with $\tilde{k}^2$.

For regions where the contribution of the charge to \eno{Nu} is not too strong, $\nu^2$ is positive and one may find Fermi surface singularities where the retarded Green's function $G_R$ diverges at $\omega = 0$ for some $k = k_F$, corresponding to the vanishing of the source term $A_-$. Negative values of $k_F$ correspond to Fermi surfaces for the antiparticles associated to our (Dirac) fermionic operators.
One may then determine the properties of excitations near the Fermi surface using ${\cal G}(\omega)$.
The full form of ${\cal G}(\omega)$ is recorded in \cite{Faulkner:2009wj};
for small $\omega$ it scales as a power law,
\eqn{RegularCalG}{
{\cal G}(\omega) = |c(k)| e^{i \gamma_k} (2\omega)^{2 \nu} \,,
}
with real quantities $|c(k)|$ and $\gamma_k$.  The phase $\gamma_k$ can be written as
\eqn{}{
\gamma_k \equiv \arg \left( \Gamma(-2 \nu) \left( e^{-2 \pi i \nu} - e^{-2 \pi  \nu_{\rm q}} \right) \right)\,.
}
The retarded Green's function near the Fermi surface for small $\omega$  takes the form
\eqn{GreensFunction}{
G_R(k, \omega) \sim {h_1 \over k_\perp - {1 \over v_F} \omega - h_2 e^{i \gamma_{k_F}} (2 \omega)^{2 \nu_{k_F}}} \,,
}
with $h_1$, $h_2$ positive constants and $k_\perp \equiv k - k_F$.

While $h_1$ and $h_2$ depend on the details of the UV physics, certain properties are determined solely by the IR $AdS_2$ region \cite{Faulkner:2009wj}. The denominator of \eno{GreensFunction} determines the dispersion relation of fluctuations near the Fermi surface.
The nature of the dispersion relation depends crucially on $\nu_{k_F}$. For $\nu_{k_F} > 1/2$, the leading imaginary part comes from ${\cal G}(\omega)$, but the leading real part comes from the generic ${\cal O}(\omega)$ corrections given by the $1/v_F$ term. In this case the ratio of excitation width $\Gamma$ to excitation energy $\omega_*$ goes to zero as one approaches the Fermi surface; the excitations are true quasiparticles and the system behaves as a Fermi liquid. The leading dispersion relation is $\omega_* \sim v_F k_\perp$, and the
residue $Z$ quantifying the overlap between the state created by the fermionic operator and the quasiparticle excitation approaches a nonzero constant proportional to $v_F$.

On the other hand, if $\nu_{k_F} < 1/2$, both the leading real and imaginary parts of the dispersion relation come from ${\cal G}(\omega)$, and they are of the same order; we can ignore the Fermi velocity $v_F$ term as subleading.
The ratio of the excitation width to its energy then approaches a constant, given by
\eqn{Width}{
{\Gamma \over \omega_*} &= \tan \left( \gamma_{k_F} \over 2 \nu_{k_F} \right) \,, \quad \quad k_\perp>0 \,, \cr
&= \tan \left({ \gamma_{k_F} \over 2 \nu_{k_F} } - \pi z \right) \,, \quad \quad k_\perp<0 \,,
}
where the exponent is 
\eqn{}{
z \equiv {1 \over 2 \nu_{k_F}} \,.
}
In this case the excitations remain unstable as one approaches the Fermi surface; this behavior is similar to what one expects in a non-Fermi liquid.
The dispersion relation between the excitation energy $\omega_*$ and the momentum $k_\perp$ is then
\eqn{}{
\omega_* \sim (k_\perp)^z \,.
}
Furthermore the residue $Z$ vanishes at the Fermi surface like
\eqn{}{
Z \sim (k_\perp)^{z-1} \,,
}
another property characteristic of a non-Fermi liquid. The intermediate case of $\nu_{k_F} = 1/2$ is the so-called marginal Fermi liquid, where the ratio $\Gamma/\omega_*$ and the residue $Z$ vanish logarithmically in $\omega$ as the Fermi surface is approached.

If the charge contribution to \eno{Nu} is sufficiently strong, $\nu$ will become imaginary. This has been interpreted as the $AdS_2$ region developing an instability to pair creation of charged excitations \cite{Pioline:2005pf}. 
The range of $k$ for which this is the case is called an oscillatory region, as the retarded Green's function displays periodic behavior in log~$\omega$  \cite{Liu:2009dm, Faulkner:2009wj}.
In this case the boundary condition \eno{AdS2Bdy} acquires a complex exponent, and in general one cannot have Im~$G_R^{-1} = 0$ even when Re~$G_R^{-1} = 0$; thus there are no Fermi surface singularities, as the width of would-be excitations persists even as the energy goes to zero, washing out the Fermi surface. We will find lines of Fermi surfaces as we vary the chemical potentials that terminate at an oscillatory region.

While bottom-up models can easily show both Fermi and non-Fermi liquid behavior, ${\cal N}=4$ Super-Yang-Mills at finite density was found at strong coupling to exclusively have excitations behaving as a non-Fermi liquid \cite{DeWolfe:2012uv}. A primary result of this work is that the ABJM theory is the same: only non-Fermi liquid behavior is found. It is interesting to note that for the alternate non-supersymmetric quantization this is no longer the case.

\subsection{The 3+1-charge black hole}\label{3plus1_section}
The $3+1$-charge black hole solutions (3+1QBH) are defined by setting three of the charges equal, while allowing the fourth to vary independently:
\begin{eqnarray}
Q_1 \equiv Q_a \,, \quad \quad Q_3 \equiv Q_b = Q_c = Q_d \,.
\end{eqnarray}
The corresponding gauge fields turned on in the bulk are
\begin{eqnarray}
a \equiv A_a  \equiv \Phi_1(r) dt \,, \quad \quad
A \equiv A_b = A_c = A_d \equiv \Phi_3(r) dt \,,
\end{eqnarray}
with field strengths $f \equiv da$, and $F \equiv dA$.
This simplification also relates the three active scalars to one another,
\begin{eqnarray}
\phi \equiv- \phi_1 = -\phi_2 = - \phi_3 \,,
\end{eqnarray}
where the minus sign is for later convenience.
The simplified Lagrangian then becomes
\begin{equation}\label{3plus1_lagrangian}
 e^{-1}\mathcal{L} = R-\frac{3}{2}(\partial \phi)^2 + \frac{6}{L^2}\cosh\phi - \frac{3}{4} e^{\phi} F^2 - \frac{1}{4} e^{-3\phi} f^2 \,.
\end{equation}
The (3+1)QBH solutions are
\eqn{}{
& A(r)=-B(r)=\log\frac{r}{L}+ \frac{1}{4}\log\left(1+\frac{Q_1}{r}\right)+\frac{3}{4}\log\left(1+\frac{Q_3}{r}\right)  \cr
 &h(r) = 1-\frac{r(r_H+Q_1)(r_H+Q_3)^3}{r_H(r+Q_1)(r+Q_3)^3} \,, \quad \quad
 \phi = \frac{1}{2}\log\left(1+\frac{Q_3}{r}\right)-\frac{1}{2}\log\left(1+\frac{Q_1}{r}\right) \cr
&\Phi_1(r) = \frac{\eta_1}{L}\sqrt{\frac{Q_1}{r_H}}\frac{(r_H+Q_3)^{3/2}}{(r_H+Q_1)^{1/2}}\left(1-\frac{r_H+Q_1}{r+Q_1}\right)\cr
 &\Phi_3(r) = \frac{\eta_3}{L}\sqrt{\frac{Q_3}{r_H}(r_H+Q_3)(r_H+Q_1)}\left(1-\frac{r_H+Q_3}{r+Q_3}\right) \,,
}
with temperature and entropy density
\begin{align}\label{eq:sT3p1}
 T=\frac{3r_H^2+2 Q_1 r_H-Q_1 Q_3}{4\pi L^2 r_H}\sqrt{\frac{r_H+Q_3}{r_H+Q_1}} \ \ , \ \ \ \ s = \frac{1}{4G L^2}(r_H+Q_3)^{3/2} (r_H+Q_1)^{1/2} \,,
\end{align}
and the chemical potentials and charge densities
\eqn{}{
\mu_1 = \frac{\eta_1}{L^2}\sqrt{\frac{Q_1}{r_H}}\frac{(r_H+Q_3)^{3/2}}{(r_H+Q_1)^{1/2}} \,, \quad \quad
\mu_3 = \frac{\eta_3}{L^2}\sqrt{\frac{3\, Q_3}{r_H}(r_H+Q_3)(r_H+Q_1)} \,,
}
\eqn{}{
\rho_1 = {\eta_1 \over 2 \pi} \sqrt{Q_1 \over r_H} s \,, \quad \quad
\rho_3 = {\eta_3 \over 2 \pi} \sqrt{3Q_3 \over r_H} s\,,
}
where the factor of $\sqrt{3}$ comes from defining $\mu_3$ and $\rho$ relative to a canonically normalized gauge field $\sqrt{3} A$.

\begin{figure}
\begin{center}
\includegraphics[scale=0.4]{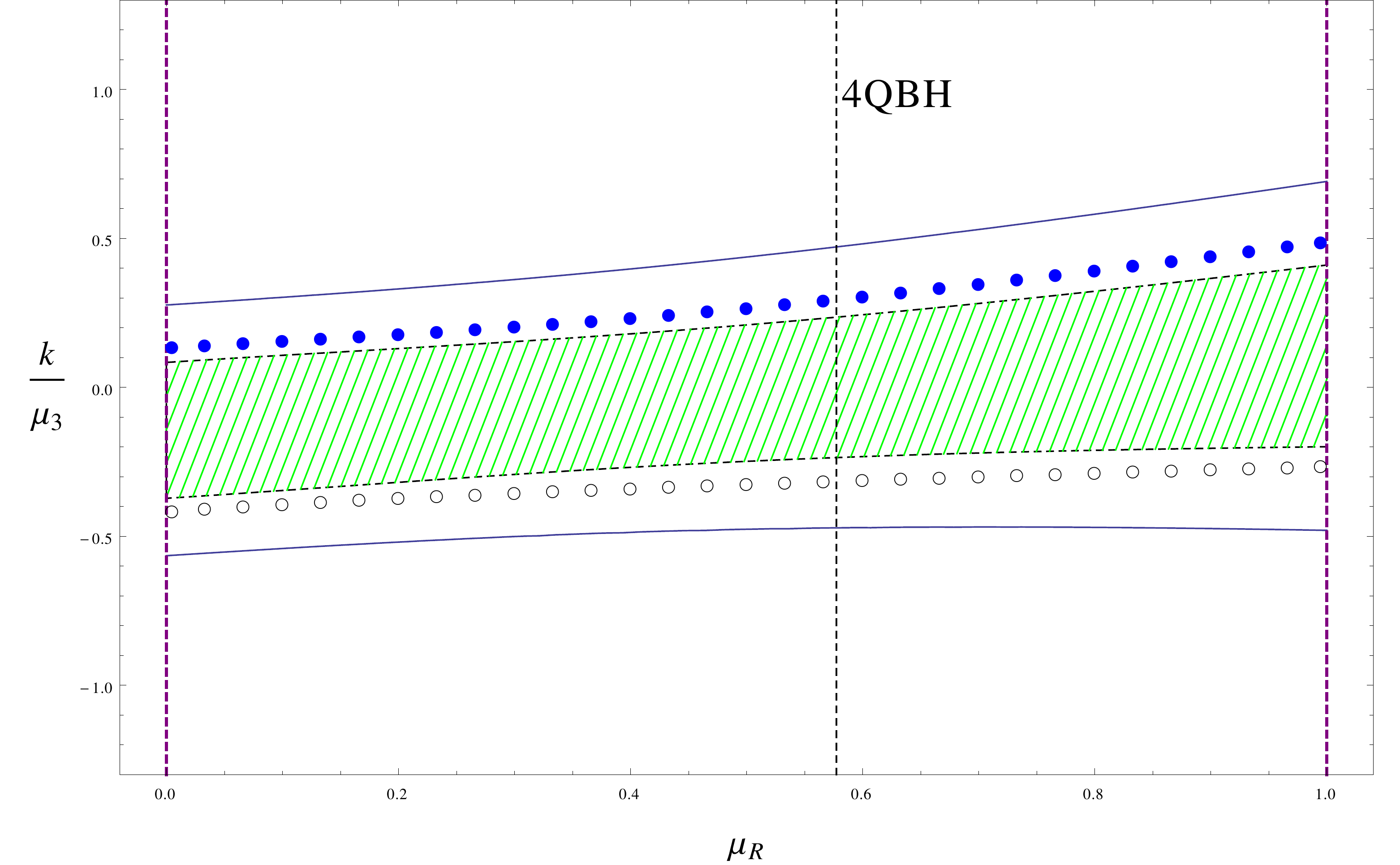}
\caption{Class 1 fermions for the (3+1)QBH. Fermi surface singularities are shown as blue dots, while zeroes are marked by empty circles. The green hatched region is the ``oscillatory region" characteristic of an infrared instability towards pair production in the bulk. The solid blue contours bound the region of Fermi surfaces with non-Fermi liquid-like excitations.
\label{3plus1-1Fig}}
\end{center}
\end{figure}

We will be interested in extremal black holes, which satisfy
\eqn{ExtremalThreeOne}{
3r_H^2+2 Q_1 r_H-Q_1 Q_3 =0  \quad\quad\quad {\rm (extremal \ (3+1)QBH)} \,.
}
To solve \eno{ExtremalThreeOne} it is generally most convenient to eliminate $Q_1$ in favor of $Q_3$ and $r_H$,
\eqn{x3p1}{
Q_1 = {3 r_H^2 \over Q_3-2r_H}
\quad\quad\quad {\rm (extremal \ (3+1)QBH)} \,,
}
indicating $Q_3 \geq 2 r_H$ for extremal solutions (recall we have taken the $Q_i$ positive).  These black holes all have a nonsingular event horizon at $r=r_H$, and are thus regular.
 Correspondingly, the entropy density (which is just the area density of the event horizon) is nonzero, even at zero temperature. 

The parameter space is naively two-dimensional, but since the underlying field theory is conformal, only the ratio of dimensionful quantities matters; hence there is a one-parameter space of extremal solutions, given by $r_H /Q_3$, or equivalently by the ratio of the chemical potentials:
\eqn{}{
\mu_R \equiv {\mu_1 \over \mu_3} = \sqrt{1 - {2 r_H \over Q_3}} 
\quad\quad\quad {\rm (extremal \ (3+1)QBH)} \,.
}
This runs over values $0 \leq \mu_R \leq 1$. 
The endpoints of the parameter range are not regular black holes:
the limit $\mu_R \to 1$ connects to the three-charge black hole, to be discussed in section~\ref{ThreeChargeSec}, while the opposite limit $\mu_R \to 0$ connects to the one-charge black hole, discussed in section~\ref{RGFlowSec}.  At $\mu_R = 1/\sqrt{3}$, we obtain the Reissner-Nordstr\"om four-charge black hole \eno{4QBH}.

\begin{figure}
\begin{center}
\includegraphics[scale=0.4]{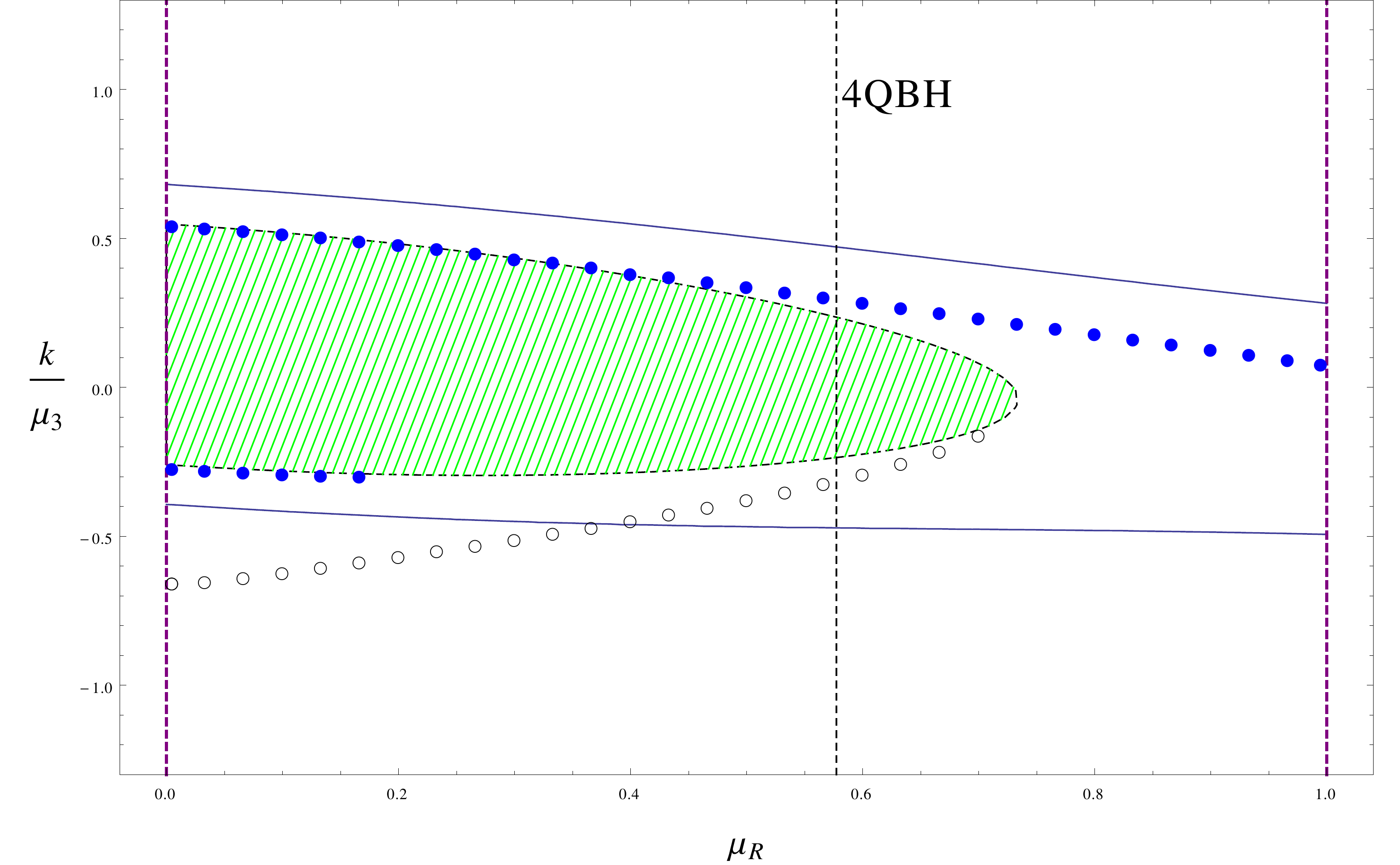}
\caption{Class 2 fermions for the (3+1)QBH. These modes are unique in that they exhibit multiple Fermi surfaces for small $\mu_R$.
\label{3plus1-2Fig}}
\end{center}
\end{figure}
The Dirac equation \eno{GeneralDirac} in the (3+1)QBH backgrounds is
\begin{equation}\label{3QBH_Lagrangian}
 \Big[i\gamma^{\mu}\nabla_{\mu} +\frac{m}{4L}(e^{-\phi/2}-e^{3\phi/2})+\frac{q_1}{4L} \gamma^{\mu} a_{\mu}  +\frac{q_3}{4L} \gamma^{\mu}A_{\mu} +\frac{i}{8}\sigma^{\mu\nu}\Big(p_1 e^{-3\phi/2} f_{\mu\nu}+ p_3 e^{\phi/2} F_{\mu\nu}  \Big) \Big] \chi  =0 \,.
\end{equation}
The quantities $m$, $q_1$, $q_3$, $p_1$, $p_3$ are combinations of the $m_i, q_i, p_i$ characterizing each fermion:
\eqn{31Charges}{
m \equiv - m_a =&\  m_b + m_c +m_d  \,, \cr  q_1 \equiv q_a \,, \quad q_3 \equiv q_b + q_c + q_d \,,\quad &
\quad p_1 \equiv p_a \,, \quad p_3 \equiv p_b + p_c + p_d \,.
}
In these backgrounds the functions $X$, $u$ and $v$ are
\eqn{}{
X= \frac{m\, (e^{3\phi/2}-e^{-\phi/2})\  e^B }{4L\sqrt{h}} \,, \quad
u= \frac{1}{\sqrt{h}}\big[\omega  +\frac{q_1}{4L} \Phi_1+\frac{q_3}{4L} \Phi_3 \big]
\,,\quad
v= \frac{e^{-B}}{4}\big[p_1 e^{-3\phi/2} \Phi_1' + p_3 e^{\phi/2} \Phi_3'  \big] \,.
}
Several fermions that have distinct charges in general backgrounds  satisfy the same Dirac equation when restricted to the (3+1)QBH backgrounds. We find that the 16 fermions given in the previous section organize into five distinct (3+1)QBH equations, which we label as classes 1-5:
\begin{center}
\begin{tabular}{|c|l|l|l|l|l|l|}\hline
Class & $\chi^{(q_a,q_b,q_c,q_d)}$ & $m$ & $q_3$ & $q_1$ & $p_3$ & $p_1$\\\hline
1& \begin{tabular}{l} $\chi^{(+1,+3,-1,+1)}$, $\chi^{(+1,-1,+3,+1)}$, $\chi^{(+1,+1,+3,-1)}$, \cr \quad $\chi^{(+1,+3,+1,-1)}$, $\chi^{(+1,+1,-1,+3)}$, $\chi^{(+1,-1,+1,+3)}$  \end{tabular}& $-1$ & 3 & 1 & $-1$ & $1$\\
2& $\chi^{(-1,+1,+1,+3)}$, $\chi^{(-1,+3,+1,+1)}$, $\chi^{(-1,+1,+3,+1)}$ & $-1$ & 5 & $-1$ & $1$ & $-1$\\
3& $\chi^{(+3,-1,+1,+1)}$, $\chi^{(+3,+1,-1,+1)}$, $\chi^{(+3,+1,+1,-1)}$ & $3$ & 1 & 3 & $1$ & $-1$\\
 \hline
4&  $\chi^{(-1,+3,-1,-1)}$, $\chi^{(-1,-1,+3,-1)}$, $\chi^{(-1,-1,-1,+3)}$  & $-1$ & 1 & $-1$ & $-3$ & $-1$\\
 5&$\chi^{(+3,-1,-1,-1)}$ & $3$ & $-3$& 3 & $-3$ & $-1$\\
 \hline
\end{tabular}
\end{center}

\begin{figure}
\begin{center}
\includegraphics[scale=0.4]{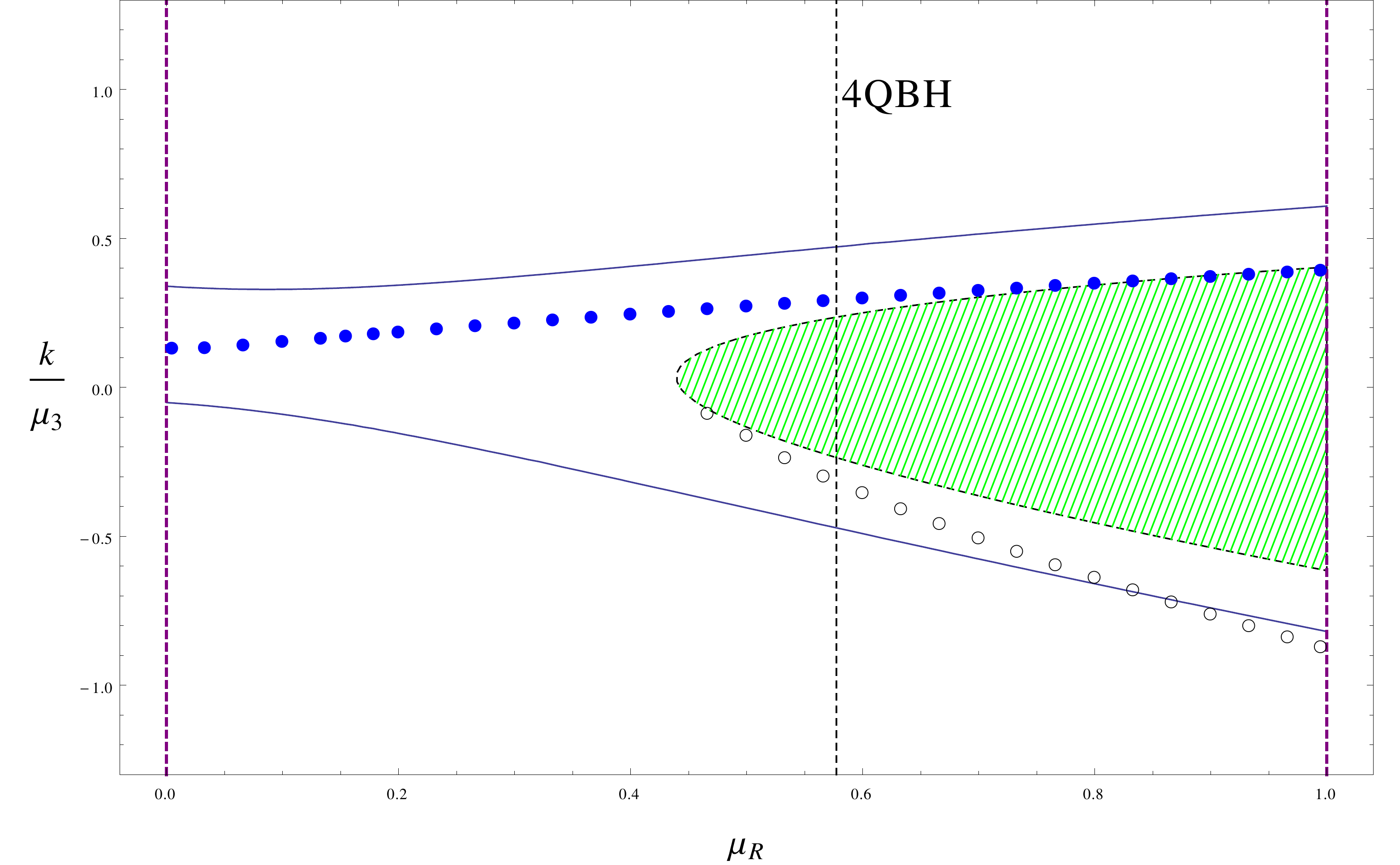}
\caption{Class 3 fermions for the (3+1)QBH. The poles end at the oscillatory region just before $\mu_R = 1$.
\label{3plus1-3Fig}}
\end{center}
\end{figure}

\bigskip
\noindent
Classes 1-3 are net-charged fermions, while classes 4 and 5 are net-neutral. We note that at the 4QBH point, the vanishing scalar makes the mass function vanish, while the gauge and scalar couplings depend only on $q_1 + q_3 = \sum_i q_i$ and $p_1 + p_3 + \sum_i p_i$; thus all net-charged fermions have the same Dirac equation at the 4QBH point, with a gauge coupling only, and all net-neutral fermions have the same Dirac equation at the 4QBH point, with a Pauli coupling only.

\begin{figure}
\begin{center}
\includegraphics[scale=0.4]{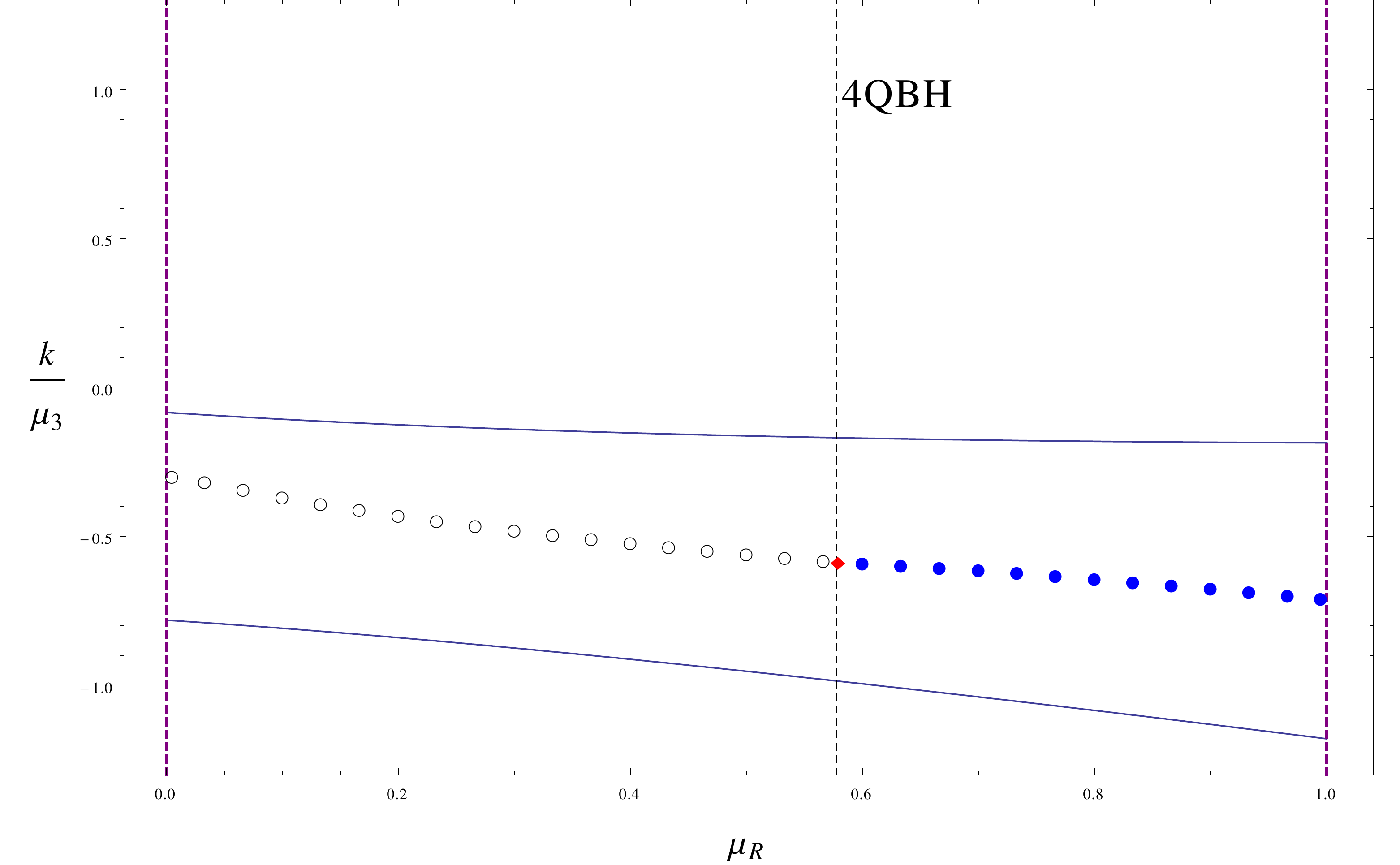}
\caption{Class 4 fermions for the (3+1)QBH. For the net-neutral modes, there is a novel transition at the 4QBH state from Fermi surface singularities to zeroes.
\label{3plus1-4Fig}}
\end{center}
\end{figure}

\begin{figure}
\begin{center}
\includegraphics[scale=0.4]{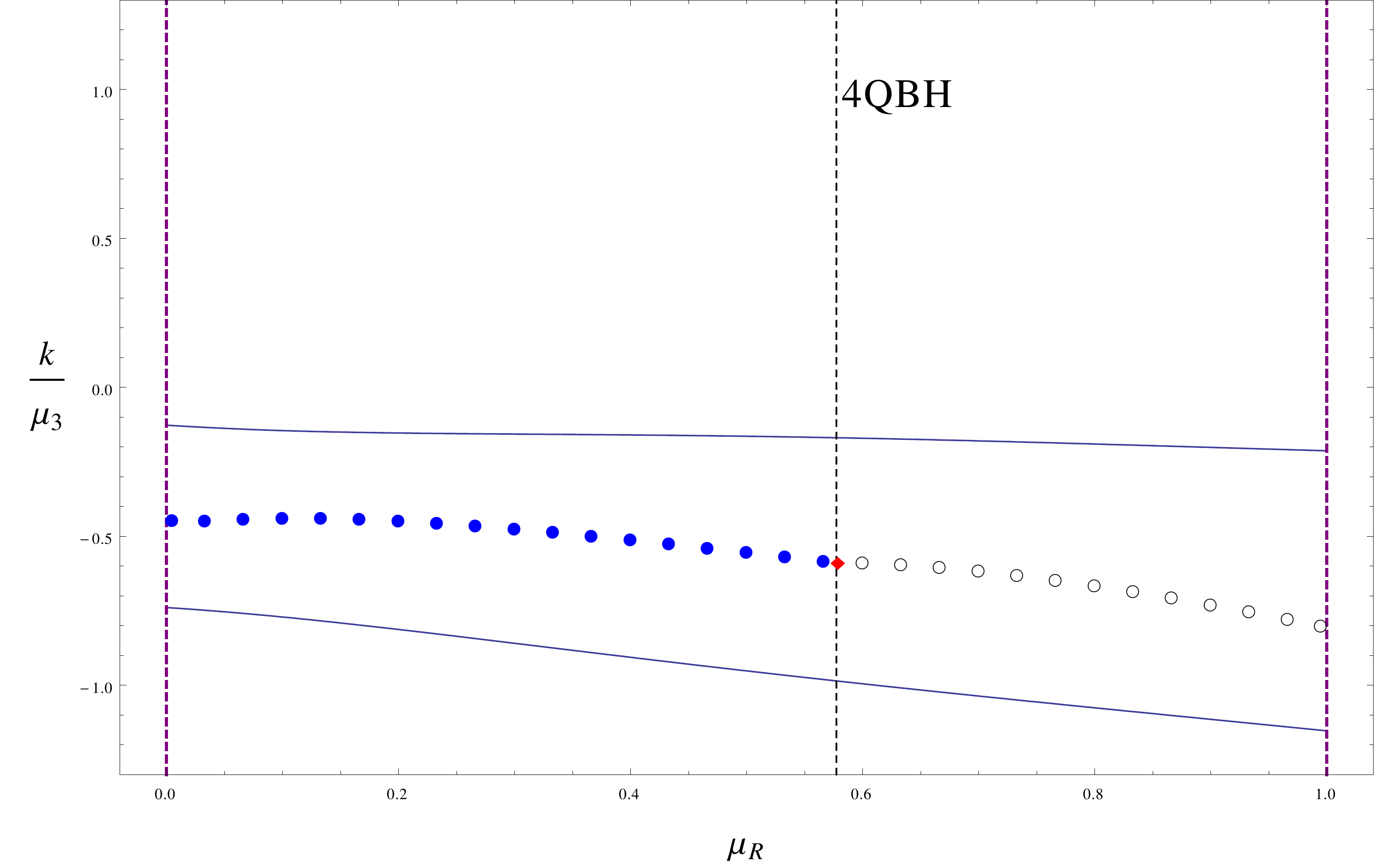}
\caption{Class 5 fermions for the (3+1)QBH. Unlike their net-charged brethren, there exists no oscillatory region for the net neutral modes, but a single ``oscillatory point" at the pole/zero transition.
\label{3plus1-5Fig}}
\end{center}
\end{figure}

The parameter $\nu$ \eno{Nu} is given by
\eqn{3-1QBH_nu}{
\nu^2 = {m^2 (1 - 3 \mu_R^2)^2 \over 48 (1 - \mu_R^4)} + {2 \over (1 + \mu_R^2)} {\tilde{k}^2 \over \mu_3^3} - {(q_3 (1 - \mu_R^2) + 2 \sqrt{3} q_1 \mu_R^3)^2 \over 72 (1 - \mu_R^2)(1 + \mu_R^2)^2} \,,
}
where the shifted momentum \eno{ktilde} is
\eqn{}{
\tilde{k} = k - {(-1)^{\alpha} \over 4} \left(p_1 \mu_1 + {p_3 \over \sqrt{3}} \mu_3 \right) \,.
}
We numerically obtained $\omega =0$ Green's functions as a function of $k$ for all five classes over the range $0 < \mu_R < 1$, imposing infalling boundary conditions 
by requiring $U$ to satisfy \eno{AdS2Bdy} with $\omega =0$. Fermi surface singularities are then identified as momenta $k=k_F$ for which the source is zero $A_+ = 0$; we also identify zeros as momenta $k = k_L$ for which the response vanishes $A_- = 0$. These results are plotted
 in figures \ref{3plus1-1Fig}-\ref{3plus1-5Fig}, with Fermi surface singularities given as blue dots, and zeros as open circles. 
 The plots show the $\alpha=2$ component of each spinor; $\alpha = 1$ modes are obtained simply by exchanging $k \to -k$.
 We also indicate oscillatory regions in green crosshatch, with their boundary $k = k_{\rm osc}$ determined by $\nu_{k_{\rm osc}} \equiv 0$. We additionally plot the lines of $k$ for which $\nu_k = 1/2$; this describes the boundary between the non-Fermi liquid behavior region (inside) and the Fermi liquid region (outside).

 Examining the results, the five classes fell into two distinct categories: the net-charged and net-neutral fermions behave rather differently.
For the net-charged fermions (classes 1, 2 and 3) one always finds an oscillatory region; for the first class this extends across the entire region, while in the other cases it begins on the left or right side respectively, but terminates some distance after crossing the four-charge black hole line. In general lines of poles (or zeros) either persist to the edge of the parameter space, or end on an oscillatory region. Each fermion has at least one Fermi surface singularity for any given value of $\mu_R$, with the exception of class 3 where the line of poles disappears into the oscillatory region just before $\mu_R = 1$.
Class 2 has two Fermi momenta $k_F$ with opposite sign for some small values of $\mu_R$; similar situations have been interpreted as a thick shell of occupied states between the two values of $|k_F|$ \cite{DeWolfe:2011aa}.
  The three classes match precisely at the four-charge point, as they must.

The net-neutral fermions (classes 4 and 5) look rather different. On one side of the four-charge point, there is a line of zeros; on the other side, a line of poles. Precisely at the four-charge point, one line turns into the other. Also unlike the net-charged case, there is no oscillatory region. However, one can determine that precisely at the four-charge point, there is a single point indicated by a red diamond where $k_{\rm osc} = -1/\sqrt{3}$ gives $\nu_{k_{\rm osc}} = 0$; this ``oscillatory point" is precisely where the line of poles turns into a line of zeros, respecting the pattern that a line of poles or zeros may terminate only at a momentum $k = k_{\rm osc}$. Precisely at this point --- which agrees between the two classes --- the Green's function is a nonzero, finite constant. We will comment more on this point at the end of the next subsection.

In all cases, both net-charged and net-neutral, the Fermi surface singularities stay within the non-Fermi liquid region. Thus continuing the pattern observed in the case of ${\cal N}=4$ Super-Yang-Mills theory, this strongly coupled maximally supersymmetric conformal field theory seems only to show non-Fermi liquid behavior, not Fermi liquid behavior. The same is not true for the zeros; in classes 2 and 3 the line of zeros extends into the $\nu > 1/2$ region. This implies that in the alternate quantization --- which is not dual to ABJM theory but in principle defines a dual CFT, as much as any bottom-up construction --- Fermi liquids would be present. It is very interesting that the top-down theories seem to avoid Fermi liquid behavior, when this is easy to obtain in a bottom-up construction; these results make this distinction sharper still.


\subsection{The 2+2-charge black hole}

Another interesting sector of the supergravity theory is made of the $2+2$-charge black holes (2+2QBH), which are defined by setting the charges equal in pairs,
\begin{eqnarray}
Q_2 \equiv Q_a = Q_b \,, \quad \quad \tilde{Q}_2 \equiv Q_c = Q_d \,,
\end{eqnarray}
corresponding to turning on the gauge fields
\begin{eqnarray}
B \equiv A_a = A_b \equiv \Phi_2(r) dt \,, \quad \quad
\tilde{B} \equiv A_c = A_d \equiv \tilde\Phi_2(r) dt \,,
\end{eqnarray}
with field strengths $G \equiv dB$, and $\tilde{G} \equiv d\tilde{B}$.

In addition to simplifying the gauge sector, this also sets two of the three scalars to zero, and we define
\begin{eqnarray}
\gamma \equiv \phi_1, \quad \quad \phi_2 = \phi_3 = 0 \,.
\end{eqnarray}
The Lagrangian then becomes
\begin{equation}
 e^{-1}\mathcal{L} = R-\frac{1}{2}(\partial \gamma)^2 + \frac{4}{L^2} + \frac{2}{L^2}\cosh\gamma - \frac{1}{2} e^{\gamma} G^2 - \frac{1}{2} e^{-\gamma} \tilde{G}^2 \,,
\end{equation}
and the black hole backgrounds are
\eqn{}{
& A(r)=-B(r)=\log\frac{r}{L}+\frac{1}{2}\left[\log\left(1+\frac{Q_2}{r}\right) + \log\left(1+\frac{\tilde{Q}_2}{r}\right) \right] \,, \cr
& h(r) = 1-\frac{r(r_H+Q_2)^2(r_H+\tilde{Q}_2)^2}{r_H(r+Q_2)^2(r+\tilde{Q}_2)^2} \,, \quad
 \gamma  = \log\left(1+\frac{Q_2}{r}\right)-\log\left(1+\frac{\tilde{Q}_2}{r}\right)\,, \cr
 &\Phi_2(r)= \frac{\eta_2}{L}\sqrt{\frac{Q_2}{r_H}}(r_H+\tilde{Q}_2)\left(1-\frac{r_H+Q_2}{r+Q_2}\right)  \,, \quad
 \tilde\Phi_2(r) = \frac{\tilde\eta_2}{L}\sqrt{\frac{\tilde{Q}_2}{r_H}}(r_H+Q_2)\left(1-\frac{r_H+\tilde{Q}_2}{r+\tilde{Q}_2}\right)\,.
}
The thermodynamic properties are
\begin{align}
 T=\frac{3r_H^2+(Q_2+\tilde{Q}_2)r_H-Q_2 \tilde{Q}_2}{4\pi L^2 r_H} \ \ , \ \ \ \ s = \frac{1}{4G L^2}(r_H+Q_2)(r_H+\tilde{Q}_2) \ .
\end{align}

\eqn{}{
\mu_2 = \frac{\sqrt{2}\, \eta_2}{L^2}\sqrt{\frac{Q_2}{r_H}}(r_H+\tilde{Q}_2) \,, \quad \quad
\tilde\mu_2 = \frac{\sqrt{2}\, \tilde\eta_2}{L^2}\sqrt{\frac{\tilde{Q}_2}{r_H}}(r_H+Q_2) \,.
}
\eqn{}{
\rho_2 = {\eta_2 \over 2 \pi} \sqrt{2 Q_2 \over r_H} s \,, \quad \quad
\tilde\rho_2 = {\tilde\eta_2 \over 2 \pi} \sqrt{2 \tilde{Q}_2 \over r_H} s\,.
}
where again the chemical potentials are defined with respect to canonically normalized gauge fields.
Extremality occurs for
\eqn{}{
3r_H^2+ (Q_2 + \tilde{Q}_2) r_H-Q_2 \tilde{Q}_2 =0  \quad\quad\quad {\rm (extremal \ (2+2)QBH)} \,.
}
Solving for $\tilde{Q}_2$, we have
\eqn{}{
\tilde{Q}_2 = {Q_2 r_H + 3 r_H^2 \over Q_2 - r_H}  \quad\quad\quad {\rm (extremal \ (2+2)QBH)} \,,
}
indicating $r_H \leq Q_2$. These extremal (2+2) charge solutions are  again regular black holes. There is a one-parameter space of such solutions, 
\eqn{}{
\tilde\mu_R \equiv {\mu_2 \over \tilde\mu_2} = 2 \sqrt{Q_2 r_H \over Q_2^2 + 2 Q_2 r_H - 3 r_H^2}
 \quad\quad\quad {\rm (extremal \ (2+2)QBH)}\,.
}
Here $\tilde\mu_R$ can take all positive values $0 \leq \mu_R \leq \infty$. The limit $\tilde\mu_R \to 0$ corresponds to the ``extremal" 2QBH, described in section~\ref{RGFlowSec}, and the point $\tilde\mu_R = 1$ is the 4QBH, connecting to the parameter space of the (3+1)QBHs.

Since the original charges are all equivalent, gauge invariance mandates a symmetry exchanging $Q_2$ and $\tilde{Q}_2$, which exchanges the corresponding gauge fields and changes the sign of the scalar $\gamma$:
\eqn{GaugeInvar}{
Q_2 \leftrightarrow \tilde{Q}_2 \,, \quad \quad \gamma \to - \gamma\,, \quad \quad \Phi_2 \leftrightarrow \tilde{\Phi}_2 \,,
}
which also exchanges $(\mu_2 , \rho_2)$ and $(\tilde\mu_2, \tilde\rho_2)$ and hence sends $\tilde\mu_R \to 1/\tilde\mu_R$. Thus the region $1 \leq \tilde\mu_R \leq \infty$ is equivalent to $0 \leq \tilde\mu_R \leq 1$.

The Dirac equation in the (2+2)QBH background is
\begin{equation}
\label{22Dirac}
 \Big[i\gamma^{\mu}\nabla_{\mu} +\frac{\tilde m}{4L}(e^{-\gamma/2}-e^{\gamma/2}) +\frac{q_2}{4L} \gamma^{\mu}B_{\mu} +\frac{\tilde{q}_2}{4L} \gamma^{\mu}\tilde B_{\mu}  +\frac{i}{8}\sigma^{\mu\nu}\Big(p_2 e^{\gamma/2} G_{\mu\nu} +\tilde{p}_2 e^{-\gamma/2} \tilde G_{\mu\nu} \Big) \Big] \chi  =0 \,,
\end{equation}
with parameters
\eqn{}{
\tilde{m} \equiv  m_a + m_b  &= - \left( m_c +m_d \right)\,, \cr
q_2 \equiv q_a + q_b \,, \quad \tilde{q}_2 \equiv q_c + q_d \,, \quad & \quad 
p_2 \equiv p_a + p_b \,, \quad \tilde{p}_2 \equiv p _c + p_d \,.
}
In this case, there are six distinct non-degenerate fermion eigenvectors, which we sort into classes I-VI:
\begin{center}
\begin{tabular}{|c|l|l|l|l|l|l|}\hline
Class &$\chi^{(q_a,q_b,q_c,q_d)}$  & $\tilde m$ & $q_2$ & $\tilde{q}_2$ & $p_2$ & $\tilde{p}_2$\\\hline
 
I& $\chi^{(-1,+1,+1,+3)}$, $\chi^{(+1,-1,+3,+1)}$, $\chi^{(-1,+1,+3,+1)}$, $\chi^{(+1,-1,+1,+3)}$ & $2$ & 0 & 4 & 0 & 0\\
 II &$\chi^{(+1,+3,-1,+1)}$, $\chi^{(+3,+1,+1,-1)}$, $\chi^{(+1,+3,+1,-1)}$, $\chi^{(+3,+1,-1,+1)}$ & $-2$ & 4 & 0 & 0 & 0\\
III & $\chi^{(+1,+1,+3,-1)}$, $\chi^{(+1,+1,-1,+3)}$ & $2$ & 2 & 2 & $2$ & $-2$\\
IV & $\chi^{(-1,+3,+1,+1)}$, $\chi^{(+3,-1,+1,+1)}$  & $-2$ & 2 & 2 & $-2$ & $2$\\
 \hline

V&$\chi^{(-1,-1,+3,-1)}$, $\chi^{(-1,-1,-1,+3)}$ & $2$ & $-2$ & 2 & $-2$ & $-2$\\
VI &$\chi^{(-1,+3,-1,-1)}$, $\chi^{(+3,-1,-1,-1)}$& $-2$ & 2 & $-2$ & $-2$ & $-2$\\
 \hline
\end{tabular}
\end{center}
\bigskip
\noindent
The first four classes are net-charged, and the last two net-neutral.

\begin{figure}
\begin{center}
\includegraphics[scale=0.25]{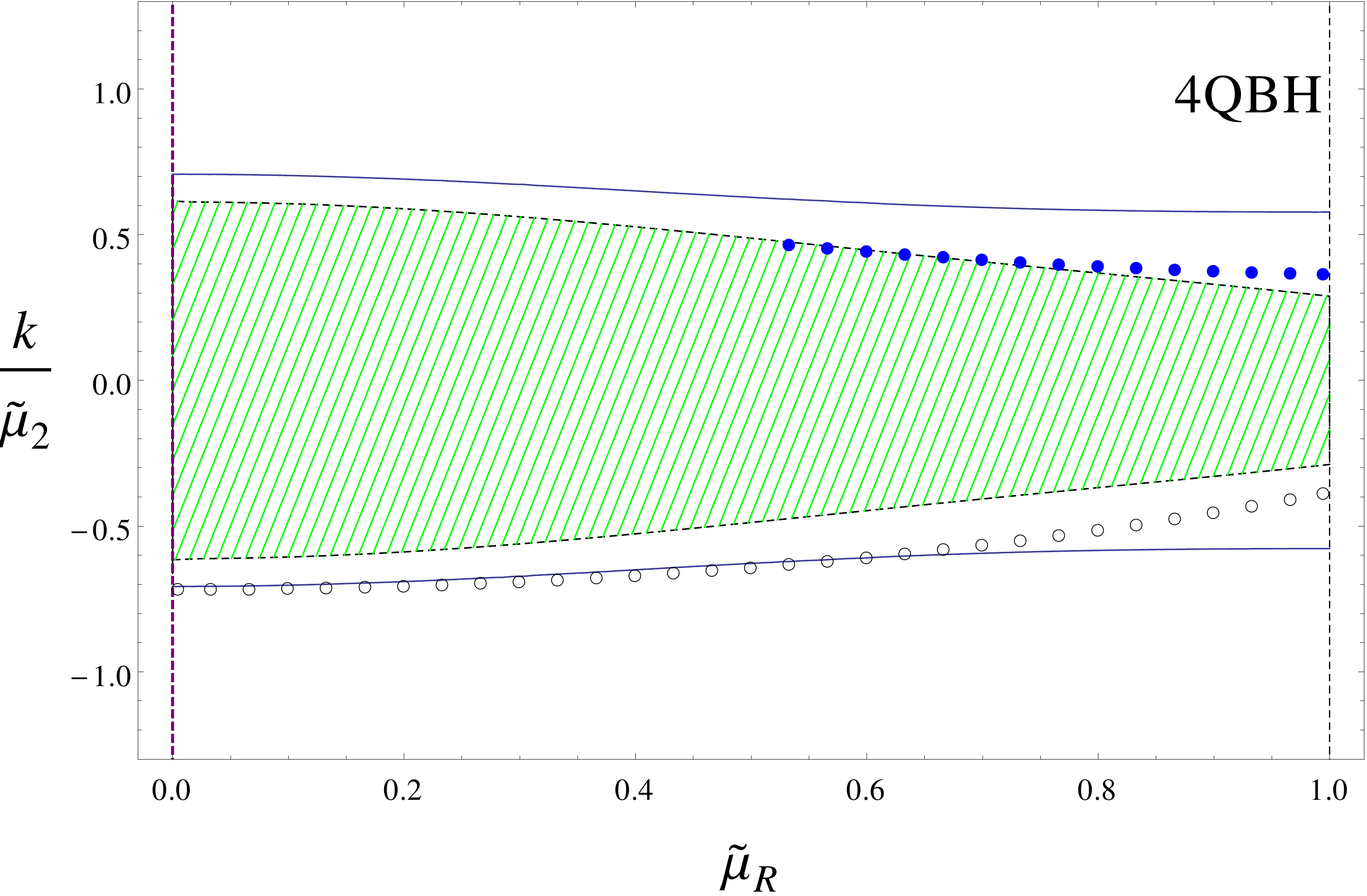}
\includegraphics[scale=0.25]{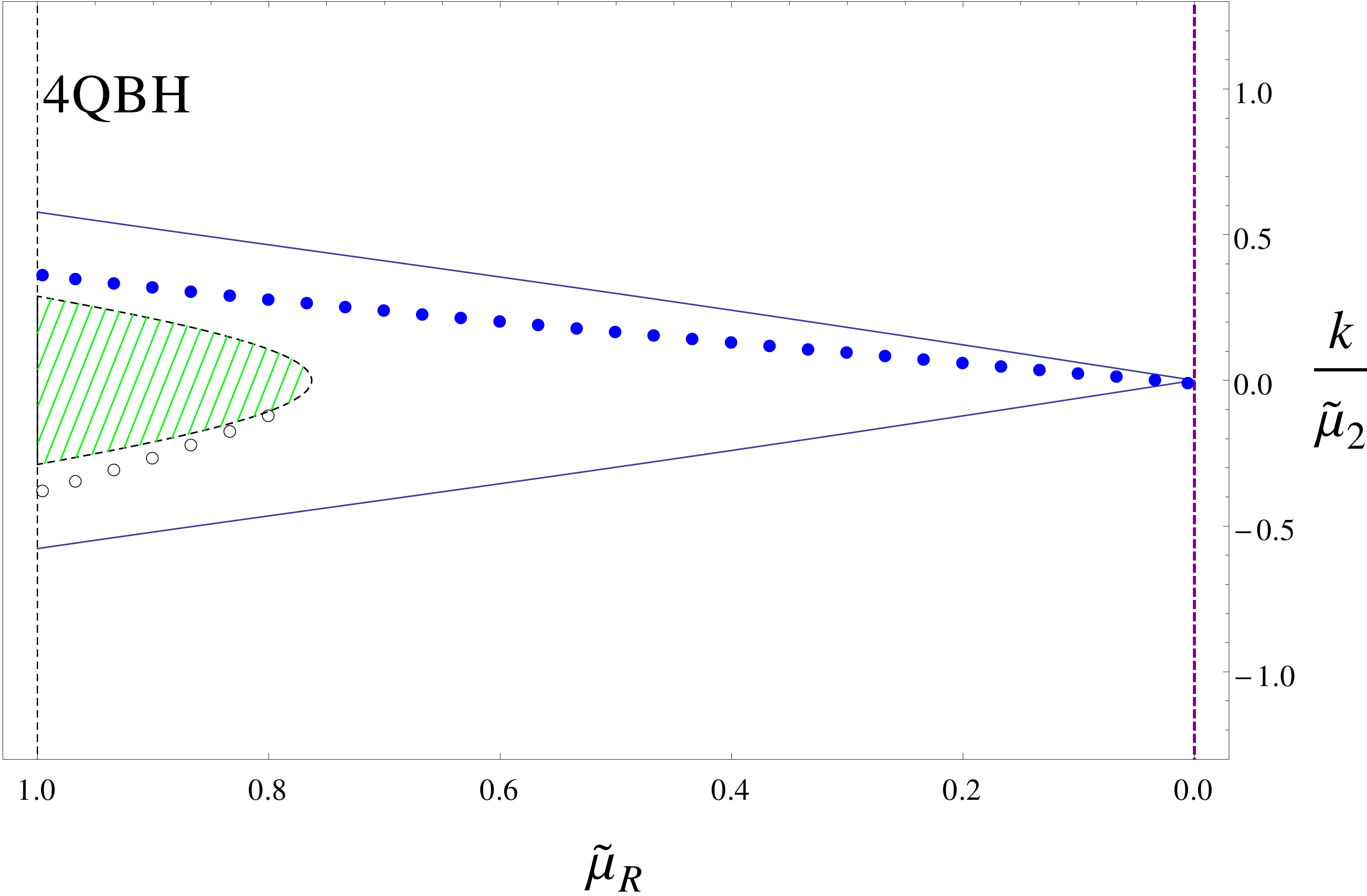}
\caption{Poles and zeros of the retarded Green's function for the class I fermion from $0 < \tilde\mu_R < 1$ (left) and for the class II fermion from $1 < \tilde\mu_R < 0$ (right). Viewed together the two plots depict the entire range $0 < \tilde\mu_R < \infty$ for class I or $\infty > \tilde\mu_R > 0$ for class II.
\label{2plus2-12Fig}}
\end{center}
\end{figure}

Gauge invariance requires that under the exchange \eno{GaugeInvar} the total physics of all fluctuations is invariant.
For this to be true, the form of the Dirac equation \eno{22Dirac} demands that the spectrum of fermions  be carried into itself under
\eqn{FermiPairs}{
\tilde{m} \to - \tilde{m} \,, \quad \quad q_2 \leftrightarrow \tilde{q}_2 \,, \quad \quad p_2 \leftrightarrow \tilde{p}_2 \,.
}
Glancing at the table we see this is indeed the case; the six classes form three partner pairs (I, II), (III, IV) and (V, VI) that are exchanged.


\begin{figure}
\begin{center}
\includegraphics[scale=0.25]{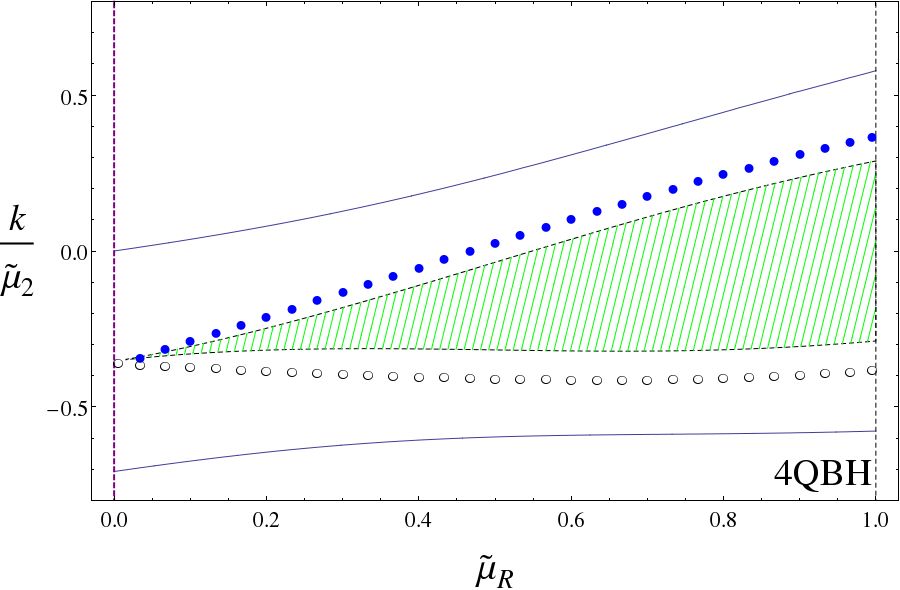}
\includegraphics[scale=0.25]{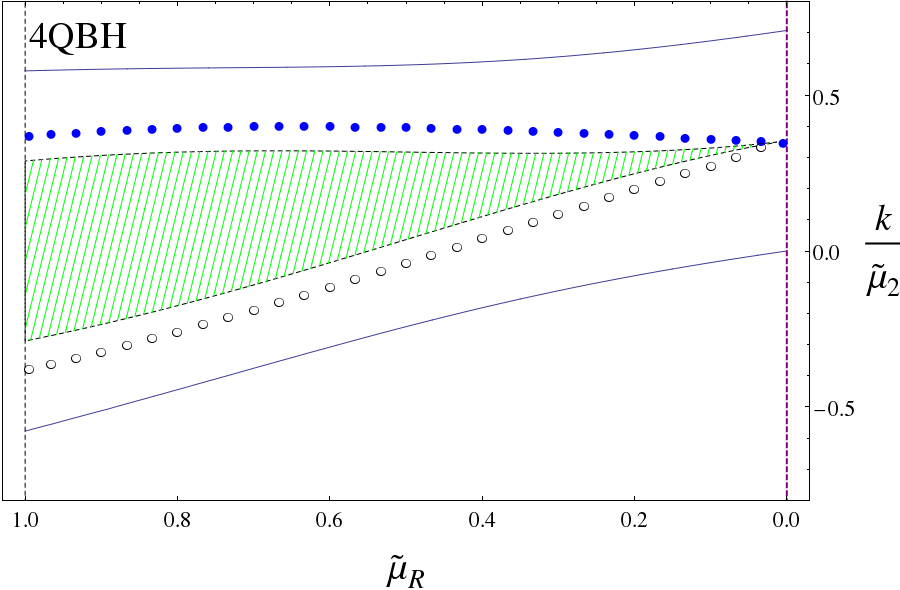}
\caption{Poles and zeros of the retarded Green's function for the class III fermion from $0 < \tilde\mu_R < 1$ (left) and for the class IV fermion from $1 < \tilde\mu_R < 0$ (right), or 
the entire range $0 < \tilde\mu_R < \infty$ for class III or $\infty > \tilde\mu_R > 0$ for class IV.
\label{2plus2-34Fig}}
\end{center}
\end{figure}

The $X$, $u$ and $v$ functions appearing in the Dirac equation are now
\eqn{}{
X&= \frac{\tilde{m}\, (e^{\gamma/2}-e^{-\gamma/2})\  e^B }{4L\sqrt{h}} = {\tilde{m}  e^B \sinh \gamma/2 \over 2 L \sqrt{h}} \cr
u&= \frac{1}{\sqrt{h}}\big[\omega  +\frac{q_2}{4L} \Phi_2+\frac{\tilde{q}_2}{4L} \tilde\Phi_2 \big]
\quad\text{,}\qquad
v= \frac{e^{-B}}{4}\big[p_2 \, e^{\gamma/2} \Phi_2' + \tilde{p}_2 \, e^{-\gamma/2} \tilde\Phi_2'  \big] \,,
}
and the $\nu$ parameter \eno{Nu} is
\eqn{}{
\nu^2= {-1 -\mu_R^2 + 2 {\cal S} \over 4 (1 + \mu_R^2)} \tilde{m}^2
+ {2 \over 1 +\mu_R^2} {\tilde{k}^2 \over \tilde\mu_2^2}
- { \left(q_2 \mu_R \left( -2 + \mu_R^2 + 2 {\cal S} \right)^{3/2} + \tilde{q}_2 \left( 1 + \mu_R^2 - {\cal S}\right)^{3/2} \sqrt{-1 + \mu_R^2 + {\cal S}} \right)^2\over 48 (1+ \mu_R^2)^2 (-2 + \mu_R^2 + 2 {\cal S} )^2} \,,
}
where we defined
\eqn{}{
{\cal S} \equiv \sqrt{1 - \mu_R^2 + \mu_R^4} \,,
}
and the shifted momentum \eno{ktilde} is
\eqn{}{
\tilde{k} \equiv k - {(-1)^\alpha \over 4 \sqrt{2}} (p_2 \mu_2 +\tilde{p}_2 \tilde{\mu}_2) \,. 
}
Due to the equivalence \eno{GaugeInvar}, \eno{FermiPairs}, it is not necessary to study all six classes of fermions over the entire range $0 < \tilde\mu_R < \infty$; one fermion for $0 < \tilde\mu_R \leq 1$ is equivalent to its partner fermion over $1 < \tilde\mu_R < \infty$. Thus we may either study three classes of fermions over the entire parameter space, or all six over half the parameter space. In practice we studied all six fermions over the range $0 < \tilde\mu_R \leq 1$, to avoid the complications of $\tilde\mu_R$ extending over an infinite range. However, we find it convenient to show the plots of classes II, IV, and VI with the horizontal axis reversed, placed next to their partner classes I, III, and V, respectively. Then one can alternately interpret each pair of figures as describing the behavior of a single fermion over the entire range $0 < \tilde\mu_R < \infty$; to do this one simply interprets one of the two plots as being the other fermion in the pair, with $\tilde\mu_R \to 1/\tilde \mu_R$ and  $k/\mu_2$ instead of $k/\tilde\mu_2$. Thus we can visualize the entire range of $\tilde\mu_R$ for all distinct fermions in a compact way.

\begin{figure}
\begin{center}
\includegraphics[scale=0.25]{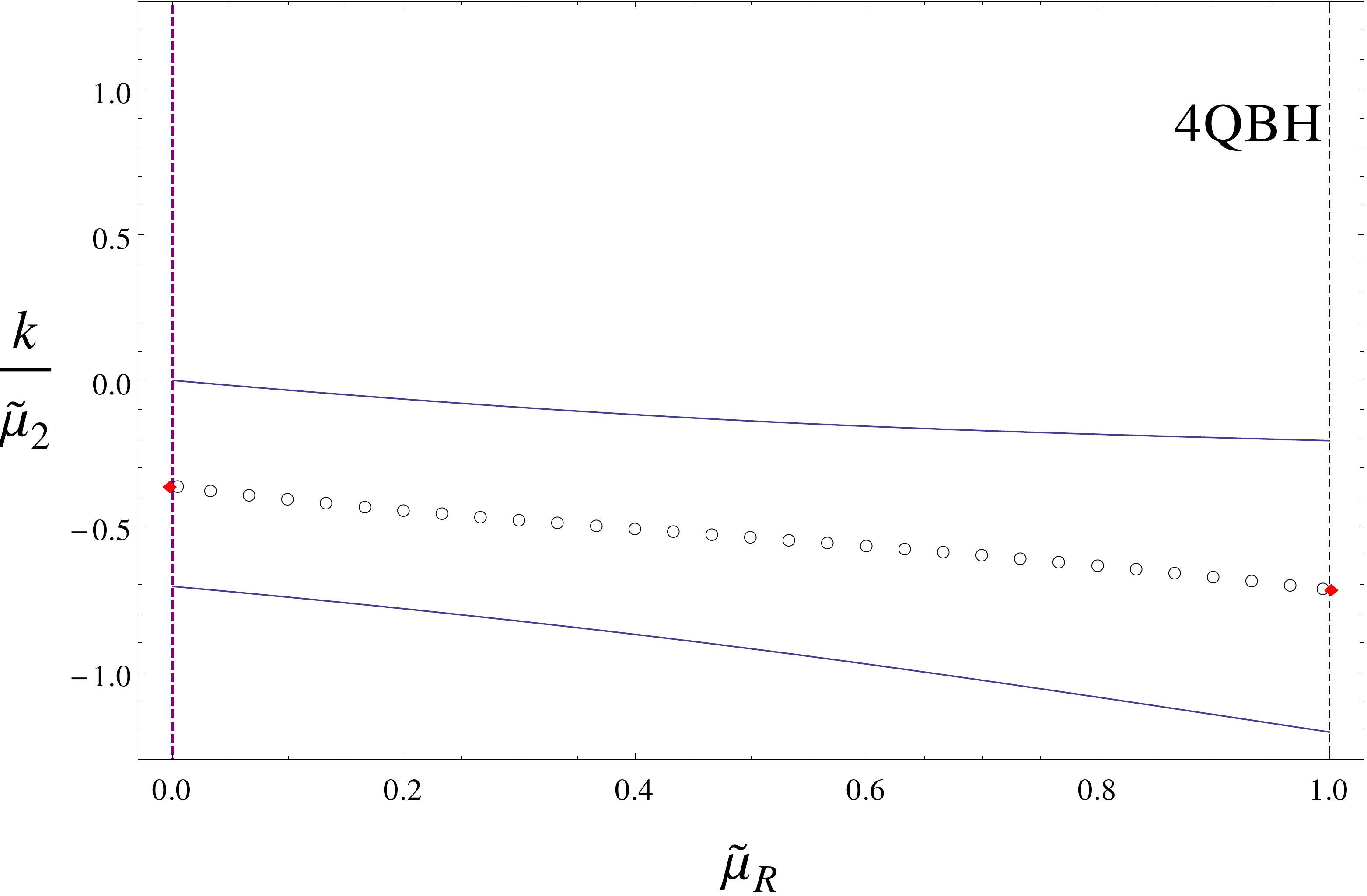}
\includegraphics[scale=0.25]{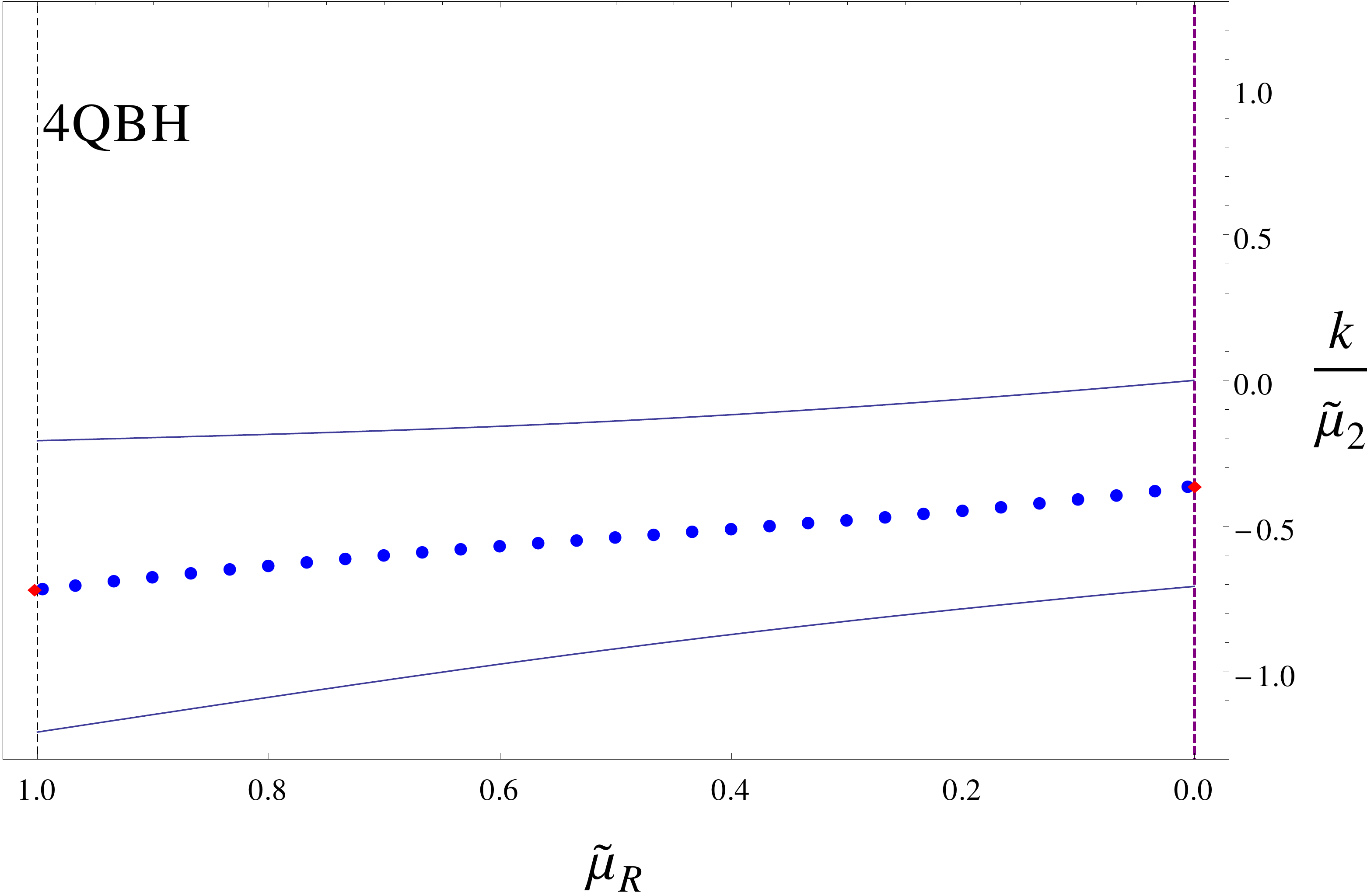}
\caption{Poles and zeros of the retarded Green's function for the class V fermion from $0 < \tilde\mu_R < 1$ (left) and for the class VI fermion from $1 < \tilde\mu_R < 0$ (right), or 
the entire range $0 < \tilde\mu_R < \infty$ for class V or $\infty > \tilde\mu_R > 0$ for class VI.
\label{2plus2-56Fig}}
\end{center}
\end{figure}

The (2+2)QBH results are qualitatively very similar to the (3+1)QBH results. Again the net-charged fermions have an oscillatory region as well as lines of poles and zeros that begin  either at the ends of the parameter space or at oscillatory regions. Again the poles stay within  $\nu_{k_F}  < 1/2$, indicating non-Fermi liquid behavior exclusively. The net-neutral fermions again have a line of poles turning into a line of zeros at a special ``oscillatory point" where $k = k_{\rm osc}$, that is, where $\nu_k = 0$. The (2+2)QBH results for the net-charged and net-neutral classes must match on to the (3+1)QBH results at the 4QBH, and indeed they do.

It is evident that the results shown in figures~\ref{2plus2-34Fig} and \ref{2plus2-56Fig} have rotational and reflectional symmetries, respectively, when combined with an exchange of zeros and poles.
One can show using the inversion equivalence \eno{GaugeInvar}, \eno{FermiPairs} in conjunction with charge conjugation
\eno{Conjugation} and the chirality flip \eno{ChiralityFlip} that 
a 180-degree rotation on the combined figures along with an exchange of poles and zeros is equivalent to $(q_2, p_2) \leftrightarrow (\tilde{q}_2, -\tilde{p}_2)$ on the fermions, while left-right parity along with an exchange of poles and zeros is equivalent to $(q_2, p_2) \leftrightarrow (-\tilde{q}_2, \tilde{p}_2)$; the invariance of classes (III, IV) under the former operation and of classes (V, VI) under the latter explains the symmetries of their respective figures. 
Note that one can argue that any fermion with $(q_2, p_2) = (-\tilde{q}_2, \tilde{p}_2)$ such as our class V and VI net-neutral fermions must be invariant under the inversion of the Green's function at the 4QBH, and thus cannot have a pole or zero there, but can have the transition between a pole and zero that we observe.

One fermion of particular interest is class II, for which the line of Fermi surfaces has $k/\tilde\mu_2 \to 0$ as $\tilde\mu_R \to 0$. This is the one case neutral under $\tilde{Q}_2$; as a result, one can renormalize to $k/\mu_2$ and one discovers $k/\mu_2 \to 0.354 \approx 1/\sqrt{8}$ which gives
\eqn{}{
\nu_{k_F, {\rm II}}(\tilde\mu_R \to 0) \to {1 \over 2} \,.
}
Hence while it never reaches the Fermi liquid region of $\nu_{k_F} > 1/2$ it asymptotically approaches a marginal Fermi liquid in the limit. An analogue of this MFL fermion exists in the five-dimensional case as well \cite{DeWolfe:2012uv}. We comment more on this case, and show the plot of $k_F/\mu_2$, in section~\ref{RGFlowSec}.

The abrupt variation in the spectrum at the 4QBH, where some net-neutral Fermi surfaces disappear and others appear, is reminiscent of a phase transition; since this occurs while varying parameters at zero temperature, it would constitute a quantum critical point. This interpretation is in accord with the interpretation \cite{Iqbal:2011in} of the inverse correlation length squared as the sum of mass and charge contributions to the $\nu^2$ parameter  \eno{CorrLength}; indeed at the four-charge black hole the masses of all fermions go to zero, while for the net-neutral fermions the charge contribution vanishes as well. (The lack of massless, chargeless fermions explains the failure of such an apparent critical point to appear when all charges were equal in the ${\cal N}=4$ SYM case.) The presence of the Pauli couplings shifting the momentum allows the pole-to-zero transition to occur at a nonzero $k_F$. We note that no sign of such a critical point is visible in the susceptibilities coming from the black hole thermodynamics \eno{TEqn}-\eno{RhoEqn}; it has been argued in \cite{Iizuka:2011hg} that such a discrepancy is a result of the large-N limit.

While we have restricted ourselves to considering the (3+1) and (2+2) charge black holes only, given a fermion with a pole on one side of the 4QBH but a zero on the other it is natural to ask what happens to the Green's function if one circles around the 4QBH in the full parameter space. A natural guess is that the critical point we observe may extend into a critical surface, separating pole and zero regions. Indeed, if one makes a small variation of the 
 chemical potentials near the  4QBH (which has $\mu_a = \mu_b = \mu_c = \mu_d$) and asks when a solution to $\nu_{k} =0$ exists, one finds that the fermion $\chi^{(+3, -1, -1, -1)}$ still has an oscillatory point at a single value of $k$ if the chemical potentials vary along the codimension-one surface
\eqn{VaryMu}{
\delta\mu_a = 0 \,, \quad \quad \delta\mu_b + \delta\mu_c + \delta\mu_d =0 \,,
}
while along other directions there are no oscillatory regions or points.
It is natural to surmise that the pole-zero transition again occurs at each oscillatory point, thus separating the parameter space into disjoint regions of poles and of zeros for this fermion. For other net-neutral fermions, the analogous permutation of \eno{VaryMu} holds; note that for different fermions, the transitions between poles and zeros occur in different places in the general parameter space.


\section{The extremal three-charge black hole and the gap}
\label{ThreeChargeSec}

We now turn to study the three-charge black hole (3QBH), the special case of the (3+1)QBH where the single charge $Q_1$ is set to zero. The background geometry is given by
\begin{align}
\label{3QBHSoln1}
& A(r)=-B(r)=\log\frac{r}{L}+\frac{3}{4}\log\left(1+\frac{Q_3}{r}\right)\,, \\
\label{3QBHSoln2}
 &h(r) = 1-\left(\frac{r_H+Q_3}{r+Q_3}\right)^3 \,, \quad\quad
 \phi = \frac{1}{2}\log\left(1+\frac{Q_3}{r}\right)\,, \\
\label{3qbh_gauge_pot}
 &\Phi_3(r)= \frac{\eta_3}{L}\sqrt{Q_3(r_H+Q_3)}\left(1-\frac{r_H+Q_3}{r+Q_3}\right) \,, \quad \quad
 \Phi_1(r) = 0 \,,
\end{align}
and the associated thermodynamic quantities are
\begin{align}
 &T=\frac{3}{4\pi L^2}\sqrt{r_H(r_H+Q_3)} \ \ , \ \ \ \ s = \frac{\sqrt{r_H}}{4G L^2}(r_H+Q_3)^{3/2} \,, \\
&\mu_3 = \frac{\eta_1}{L^2}\sqrt{3\, Q_3(r_H+Q_3)} \,, \quad \quad \mu_1 = 0 \,, \quad \quad
\rho_3 = {\eta_3 \over 2 \pi} \sqrt{3Q_3 \over r_H} s \,, \quad \quad \rho_1 =0\,.
\end{align}
This background describes a state in the M2-brane theory where three chemical potentials are set equal and the fourth one is set to zero. Zero temperature corresponds to the extremal limit  $r_H\to 0$. Unlike the 1QBH and the 2QBH discussed in the next section, the extremal solution remains a black hole.  The extremal geometry, however, is singular at the horizon $r_H = 0$. This singularity allows the horizon radius
to go to zero, so this background has zero entropy density at zero temperature. This situation is very similar to the 5D 2QBH geometry studied in \cite{DeWolfe:2012uv}, dual to a zero-entropy state in $\mathcal{N}=4$ super-Yang-Mills. One might expect that the fermions in our 3QBH background will behave similarily to the fermions in the 5D 2QBH background, a suspicion that will be confirmed throughout this section.

In appendix~\ref{app5Dlift}, we describe how the near-horizon limit of this solution lifts to a five-dimensional geometry of the form $AdS_3 \times \mathbb{R}^2$, analogous to the six-dimensional lift discussed in \cite{DeWolfe:2012uv}; $AdS_3$ regions in the near-horizon limits of zero-entropy extremal black holes are also discussed in \cite{Balasubramanian:2007bs, Fareghbal:2008ar, SheikhJabbaria:2011gc, Johnstone:2013eg}. This lift removes the singularity, showing it is harmless. The inactive gauge field $a$ is identified as the gravitphoton, and consistent reduction of any fermion requires that its parameters obey
\eqn{mConstraint}{
m = -q_1 p_1 \quad \quad \to \quad \quad |m| = |q_1|\,,
}
where the second equality follows since $p_1 = \pm1$;
due to \eno{pRatio}, \eno{31Charges} this is indeed satisfied by all fermions we consider.

The analysis of the 3QBH is done with the same method used in previous sections, analyzing 
the Dirac equation in the near-horizon limit to identify infalling boundary conditions, then solving the full Dirac equations numerically and identifying poles in the Green's function. The essential difference from the regular case is an interval of frequencies centered on $\omega=0$, bounded above and below by an energy scale $\Delta$ which we call the gap, inside which excitations have zero width and are thus stable. Outside this region we again recover the more familiar non-zero excitation widths observed for the regular black holes.  

A similar gap was also observed in \cite{DeWolfe:2013uba}, see section~1.2 and Fig.~1 therein. In that paper, a possible interpretation of the gap region was given as follows: One postulates a sector of the theory, additional to the fermion sector studied, that for generic cases has no mass gap, and thus is responsible for the generically non-zero ground state entropy. This sector could also couple to the fermions which are the subject of study, providing a channel for them to decay through. However, at the 3QBH the ground state entropy vanishes, which is interpreted as an indication that this sector becomes gapped. This gapping then also causes fermions with low enough energy to become stable, implying that their decay channel has been removed and that they cannot decay due to self-interactions; this could potentially be a large-$N$ effect.

\subsection{Near horizon analysis of the 3QBH}

Taking the extremal limit $r_H \to 0$ of the 3QBH background, and then expanding near the singular horizon $r \to 0$, we obtain the same equation for both $U_\pm$, or equivalently for $\psi_\pm$:
\begin{equation}\label{3QBH_near-horizon}
 \psi''+\left(\frac{3}{2r}+\cdots\right)\psi'+\left(\frac{L^4 (\omega^2-\Delta^2)}{9 Q_3 r^3} + \frac{\frac{1}{16}-\nu_{3Q}^2+\mathcal{O}(|\omega|-\Delta)}{r^2} + \cdots \right)\psi = 0 \ ;
\end{equation}
where the dots represent terms of  higher order in $r$ or in $\mathcal{O}(|\omega|-\Delta)$, and we have defined the scale $\Delta$,
\begin{equation}
\label{Gap}
 \Delta \equiv \frac{\sqrt{3} |m| Q}{4 L^2} = \frac{1}{4}|m| \mu \,,
\end{equation}
along with the parameter
\begin{equation}
\label{3QBH_nu}
 \nu_{3Q}^2 \equiv \Big(  \tilde{k}_{3Q}+ \frac{(-1)^\alpha}{4}\text{sgn}(\omega m)\Big)^2-\frac{m^2}{48} +\frac{ q_3\, m}{24\sqrt{3}}\text{sgn}(\omega m)  \,,
\end{equation}
which includes the shifted momentum
\eqn{ktilde3Q}{
\tilde{k}_{3Q} \equiv  {L^2k \over\sqrt{3} Q} + (-1)^\alpha {p_3 \over 4 \sqrt{3}} = {k \over |\mu_3|} + (-1)^\alpha {p_3 \over 4 \sqrt{3}} \,.
}
The quantity $\nu^2_{3Q}$ will play a similar role to $\nu^2$ for the regular black holes; in fact, we will show that  
\eno{3-1QBH_nu} coincides with \eno{3QBH_nu} in the $\mu_R \to 1$ limit for the fermions we consider.

Equation~(\ref{3QBH_near-horizon}) shows us that there is something special about the energy scale $\Delta$. For the regular extremal black holes, the leading order, no-derivative term was suppressed close to $\omega=0$, forcing one to consider inner and outer regions there. Here instead, as for the 5D 2QBH \cite{DeWolfe:2013uba}, that occurs around $\omega=\Delta$; in fact, equation~(\ref{3QBH_near-horizon}) has the same structure as  Eq.~(57) of \cite{DeWolfe:2013uba} with the correspondence $r_{\rm 4D} \leftrightarrow r_{\rm 5D}^2$, and thus the same analysis can be used.

When $|\omega|$ is not near $\Delta$, the $1/r^2$ term in \eno{3QBH_near-horizon} can be neglected. Then for $|\omega| > \Delta$ we have complex oscillatory solutions
\eqn{}{
\psi \sim \exp\left(\pm \frac{i 2 L^2\sqrt{\omega^2 - \Delta^2}}{3 \sqrt{Q_3 r}} \right) \,,
}
clearly representing infalling and outgoing waves. For $|\omega| < \Delta$ on the other hand, the equation is purely real and we get growing and dying exponentials,
\eqn{}{
\psi \sim \exp\left(\pm \frac{2 L^2\sqrt{\Delta^2 - \omega^2}}{3 \sqrt{Q_3 r}} \right) \,.
}
When the frequency is close to the gap energy, the two no-derivative terms in the second parenthesis of \eno{3QBH_near-horizon} can be of similar magnitude. Therefore, we will divide the problem into an outer region where $r$ is large enough to neglect the $r^{-3}$-term, and an inner region where we must take both terms into account. The outer region admits power law solutions,
\eqn{}{
\psi \sim r^{-\frac{1}{4}\pm\nu_{3Q}} \ .
}
The inner region equation can be solved by $r^{-1/4}$ times Bessel functions or modified Bessel functions for $|\omega|>\Delta$ and $|\omega|<\Delta$, respectively. After imposing infalling boundary conditions on the inner region solutions, we can study their near-boundary ($r \to \infty$) behavior, allowing us to determine the ``IR Green function'', $\mathcal{G}(\omega)_{3Q}$. This plays the same role near $|\omega|=\Delta$ as its cousin $\mathcal{G}(\omega)$ does near $\omega=0$ for the regular black hole solutions.
Near $|\omega| \approx \Delta$ we have
\eqn{CurlyG3Q}{
\mathcal{G}(\omega)_{3Q} \sim (|\omega|-\Delta)^{2\nu_{3Q}}\,.
}
Using \eno{CurlyG3Q} we can derive expressions for the fluctuations near $|\omega|=\Delta$ analogous to \eno{GreensFunction}. Let $k_\Delta$ be the momentum leading to a pole at $\omega = \Delta$.
Then for $|\omega|>\Delta$, we have a formula similar to \eno{GreensFunction},
\eqn{}{
G_R \sim \frac{h_1}{(k-k_{\Delta}) - \frac{|\omega|-\Delta}{v_F} + \cdots - h_2 e^{-2\pi i \nu_\Delta}(|\omega|-\Delta)^{2\nu_\Delta}} \,,
}
where $\nu_\Delta \equiv \nu_{3Q}(k_\Delta)$.
Similar to the regular case, there is an imaginary part which controls the width of the fluctuation, and if the value of $\nu_{3Q}$ at the pole is greater than $1/2$ the excitations behave like those of a Fermi liquid, while if $\nu < 1/2$ their behavior is non-Fermi liquid type. For the case $|\omega|<\Delta$, we instead obtain
\eqn{}{
G_R \sim \frac{h_1}{(k-k_{\Delta}) - \frac{|\omega|-\Delta}{v_F} + \cdots - h_2 (\Delta-|\omega|)^{2\nu_{\Delta}}} \ , 
}
which is the analytic continuation of the previous equation to negative $|\omega|-\Delta$ (and similarly the Bessel function solution regular at the horizon for $|\omega| < \Delta$ is the continuation of the infalling solution at $|\omega| > \Delta$). Importantly, the phase has disappeared and the Green's function is manifestly real. This  implies that the width of the fluctuations it describes are zero and they are thus stable. 

From the point of view of the five-dimensional lift, the relation \eno{mConstraint} allows one to interpret the gap \eno{Gap} as the momentum of the fermion in the fifth dimension. The appearance of $\omega^2 - \Delta^2$ in the Dirac equation can then be understood as the five-dimensional momentum-squared. Thus excitations inside the gap are spacelike from the higher-dimensional point of view, and beyond the gap they become timelike. Analogous behavior was seen in \cite{DeWolfe:2013uba}.


\subsection{Connection with extremal (3+1)-charge black holes}

The sequence of extremal (3+1)QBHs parameterized by $\mu_R \equiv \mu_1/\mu_3$ approaches the 3QBH extremal solution as  $\mu_R \to 1$. This limit is somewhat counterintuitive, as $\mu_1 = 0$ for the strict 3QBH; there is an associated discontinuity in $\mu_1/\mu_3$ as a function of $Q_1/Q_3$ (or equivalently $\rho_1/\rho_3$, as discussed for the analogous five-dimensional case in \cite{DeWolfe:2013uba}; see in particular figure~4). Associated to this subtlety is the failure to commute of the operations of going to the 3QBH and taking $T \to 0$; going to the 3QBH first leads to the solution presented in \eno{3QBHSoln1}-\eno{3qbh_gauge_pot} with $\Phi_1 = 0$, while taking the extremal limit of (3+1)QBHs first shifts the potential by a constant:
\eqn{3QBHshift}{
\Phi_1 \to {\sqrt{3}Q_3 \over L}\,.
}
Due to the structure of \eno{uandv}, the only effect in the Dirac equation is to shift the zero point of the energy between the two descriptions:
\eqn{}{
\omega_{3Q} = \omega_{(3+1)Q} + {q_1 \over 4L } {\sqrt{3} Q_3 \over L} 
= \omega_{(3+1)Q} + {1\over 4} q_1 \mu_3 \,, 
}
Using the relation \eno{mConstraint} between $|m|$ and $|q_1|$ and the definition \eno{Gap} of the gap $\Delta$, we can further write
\eqn{freq_matching}{
\omega_{3Q} = \omega_{(3+1)Q} + \sgn(q_1) \Delta \,,
}
indicating that the limit of a sequence of Fermi surface singularities at $\omega_{3+1} = 0$ as $\mu_R \to 1$ in the (3+1)QBHs will manifest itself as a singularity in the corresponding 3QBH at $\omega = \pm \Delta$. This is reasonable, since it is above $|\omega| = \Delta$ that the 3QBH has decaying fluctuations matching those at the Fermi surface in the regular case. In the next subsection we will see that the figures from the (3+1)QBH section do indeed match to the 3QBH in each case on the appropriate side of the gap region.

\begin{figure}
\begin{center}
\includegraphics[scale=0.5]{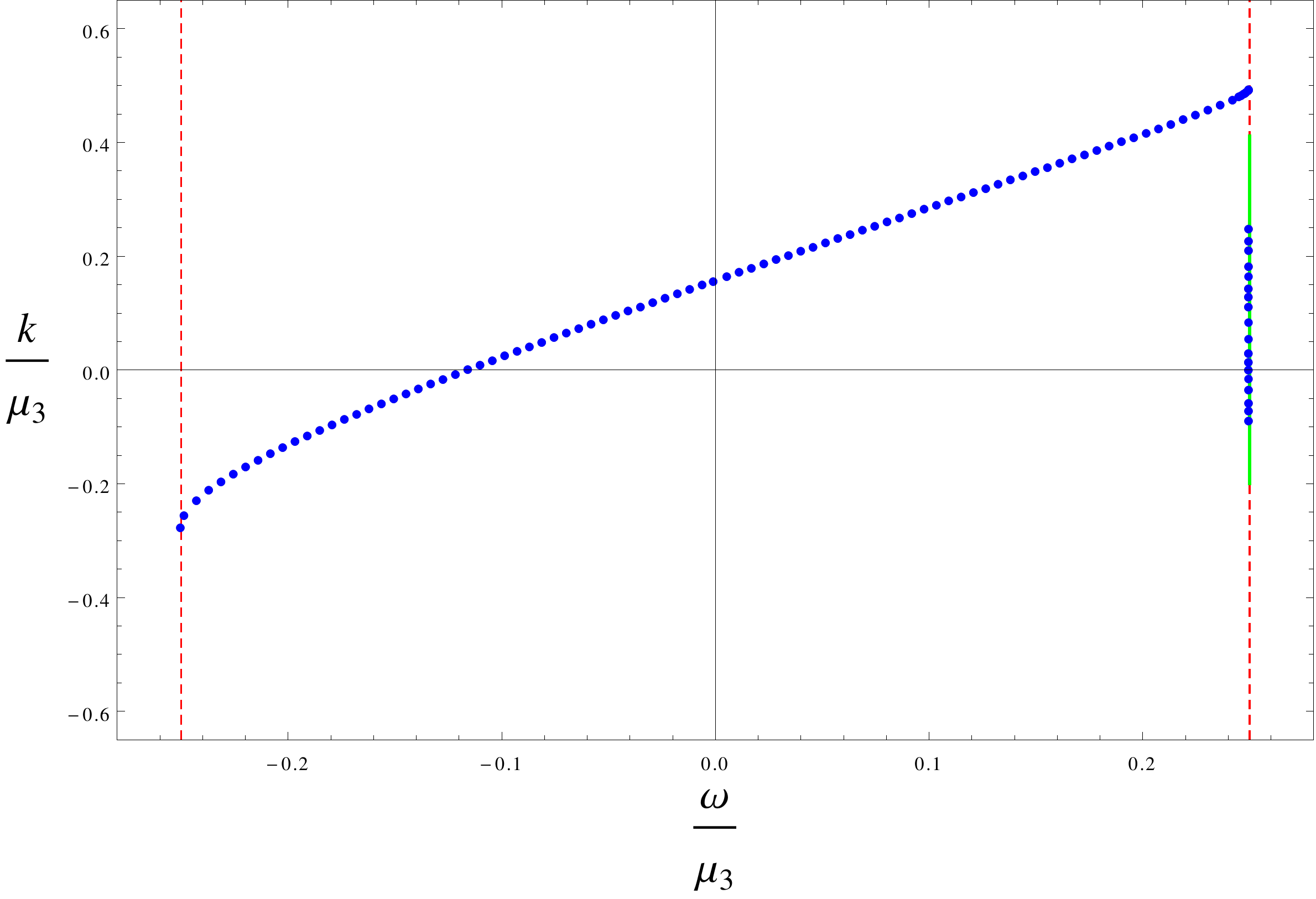}
\caption{ Class 1 fermions for the 3QBH. There is both a line of poles throughout the stable region $-\Delta < \omega < \Delta$, and a pair of poles nucleating very close to $\omega = \Delta$ before ending on the oscillatory region (green).
\label{3-1Fig}}
\end{center}
\end{figure}

These fluctuations are controlled by $\nu$ in the (3+1)QBH case, and $\nu_{3Q}$ in the 3QBH case; we would thus expect the two quantities to agree in the limit. Expanding \eno{3-1QBH_nu} around $\mu_R=1$ we find
\eqn{}{
\nu^2 = -\frac{m^2 - q_1^2}{48(\mu_R-1)} + \frac{1}{16}\left( \left( \tilde{k}_{3Q} - (-1)^\alpha p_1\right)^2 -\frac{3m^2}{2} -\frac{2q_1 q_3}{3\sqrt{3}} +\frac{7 q_1^2}{6} \right)  + \mathcal{O}(\mu_R-1) \ .
}
This would diverge in the limit $\mu_R \to 1$, but the relation \eno{mConstraint} sets the would-be diverging term to zero for all physical fermions. Moreover, the finite part  can be shown to agree exactly with \eno{3QBH_nu}, using the relationship \eno{mConstraint} together with  $\sgn(q_1)=\sgn(\omega)$, which follows from \eno{freq_matching}. It is interesting that the limits only coincide for fermions that can be lifted to five dimensions; this indicates that not just any fermion quantum numbers lead to consistent behavior throughout the parameter space, reinforcing the importance of a top-down description.

\begin{figure}
\begin{center}
\includegraphics[scale=0.5]{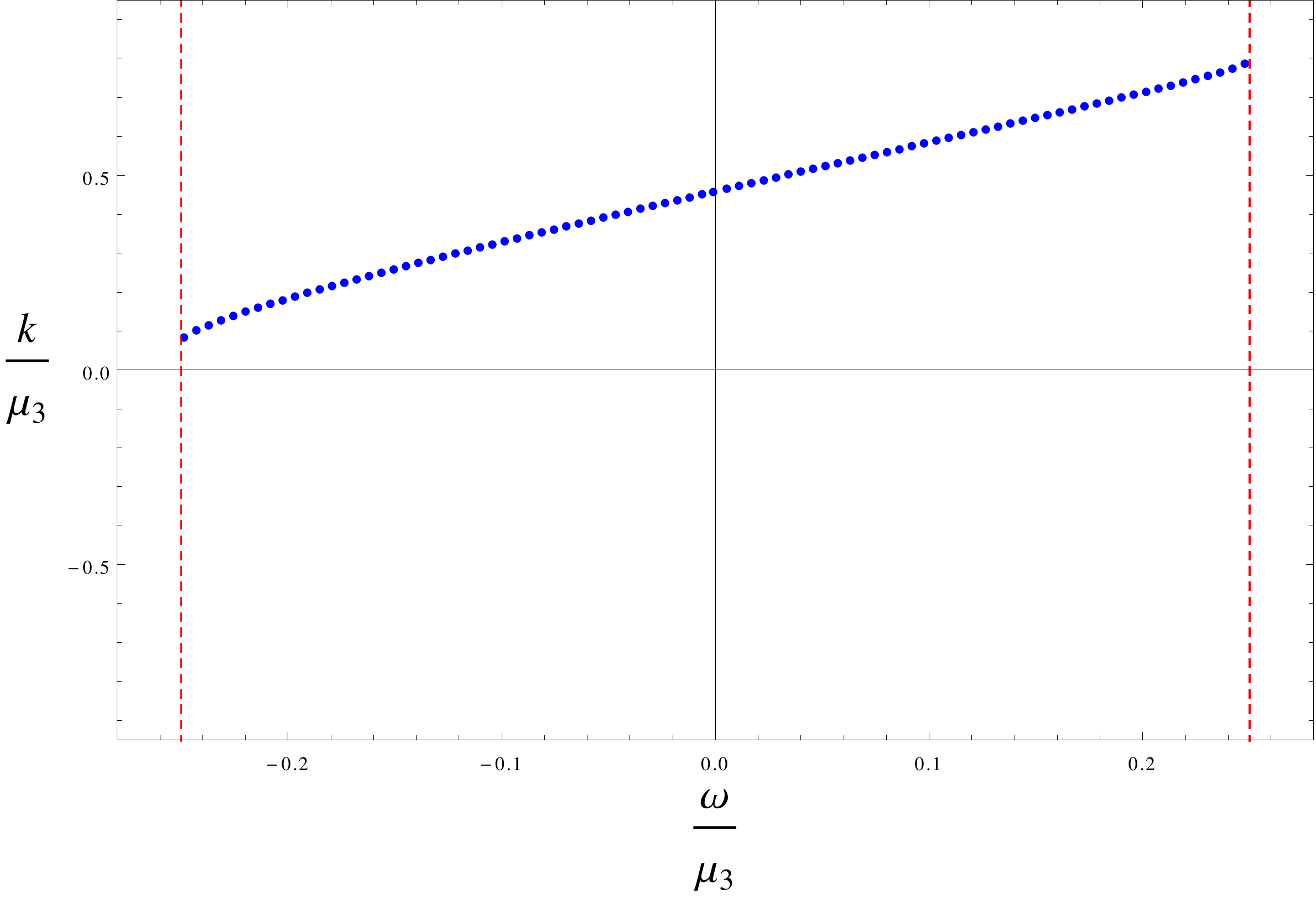}
\caption{ Class 2  fermions for the 3QBH. Here there is a line of poles only.
\label{3-2Fig}}
\end{center}
\end{figure}


\subsection{Fermion fluctuations and fermi surfaces}

\begin{figure}
\begin{center}
\includegraphics[scale=0.5]{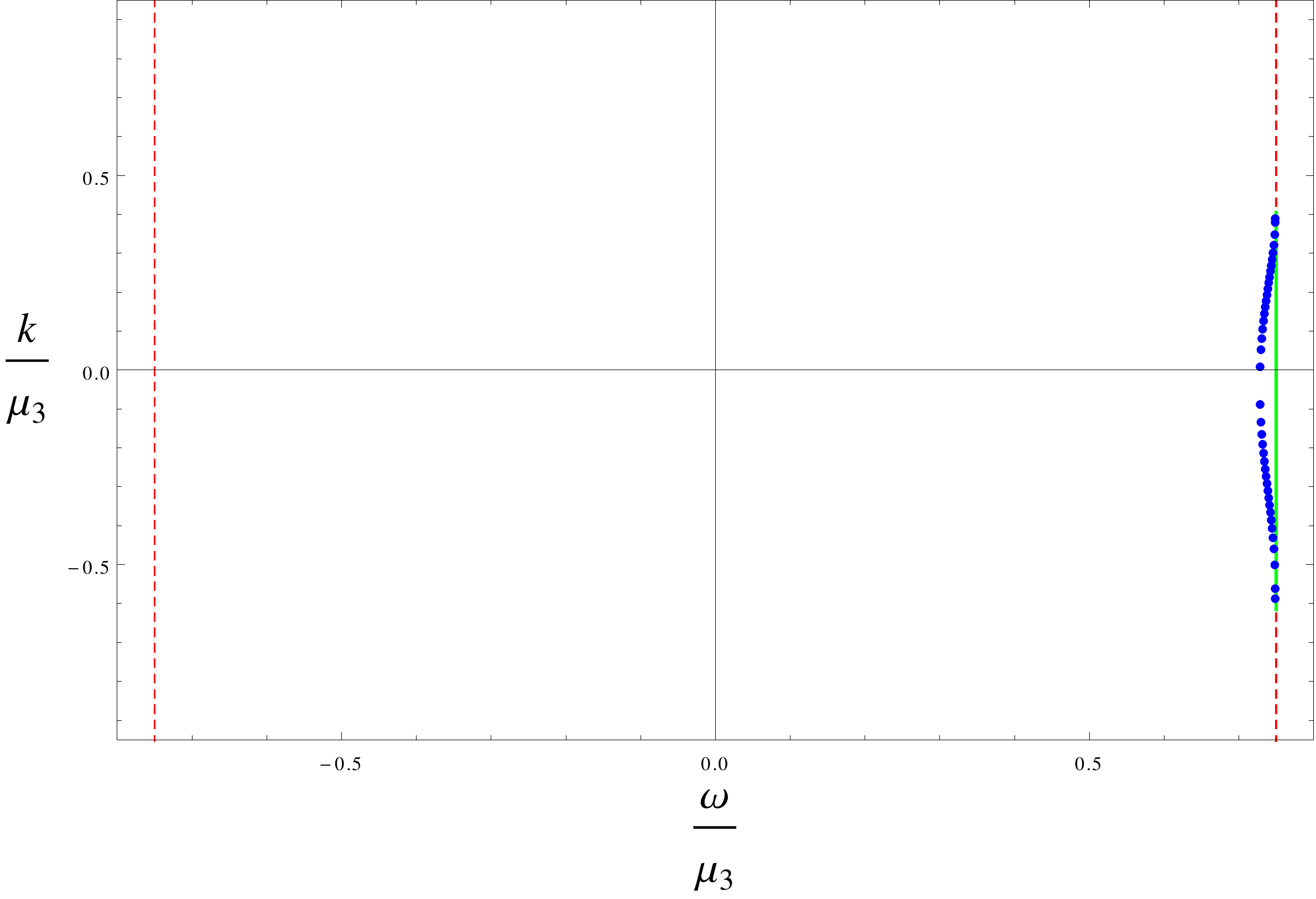}
\caption{ Class 3 fermions for the 3QBH. No line of poles through $\omega = 0$ exists, but a pair of poles nucleate near $\omega = \Delta$ and end on the oscillatory region.
\label{3-3Fig}}
\end{center}
\end{figure}

\begin{figure}
\begin{center}
\includegraphics[scale=0.5]{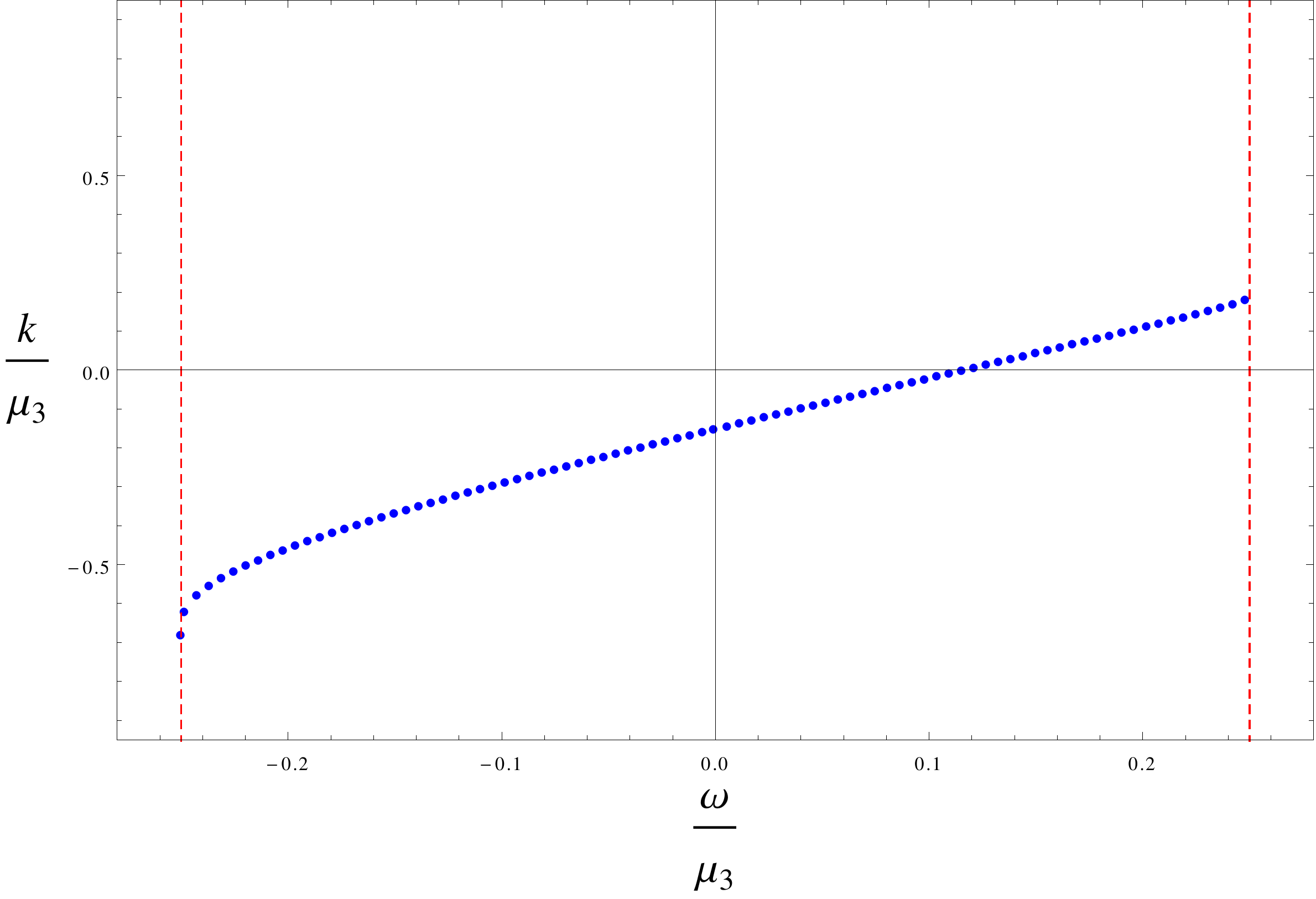}
\caption{Class 4  fermions for the 3QBH, with a line of poles only.
\label{3-4Fig}}
\end{center}
\end{figure}

\begin{figure}
\begin{center}
\includegraphics[scale=0.5]{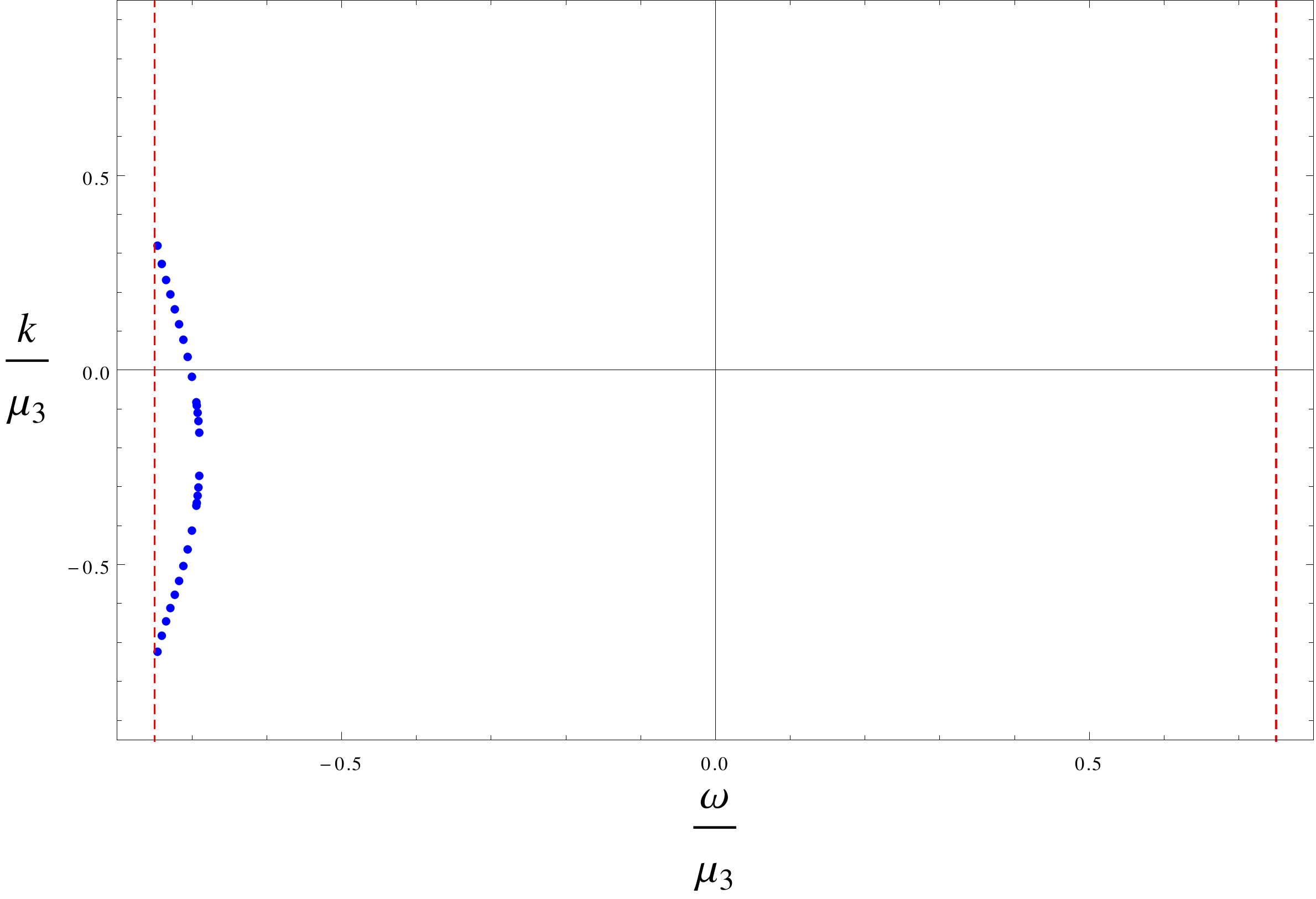}
\caption{ Class 5  fermions for the 3QBH, with a pair of poles nucleating near $\omega = - \Delta$.
\label{3-5Fig}}
\end{center}
\end{figure}

We now describe the results of numerically solving the Dirac equations for fermionic modes in the 3QBH. For the regular black holes, we were chiefly interested in whether a Fermi surface existed at a given $k=k_F$ at $\omega =0$, and we obtained the properties of nearby fluctuations. Here, we do more: we will look for poles in the Green's function as a function of $k$ for the entire stable region $-\Delta \leq \omega \leq \Delta$, and plot the results over the $\omega$-$k$ plane in Figs.~\ref{3-1Fig}-\ref{3-5Fig}; each of the five classes of fermion for the (3+1)QBH retains a distinct Dirac equation in the 3QBH limit. The location of the gap is marked with a vertical dotted red line. When applicable we also plot the extent of the oscillatory region at the gap in green.

For the fermions in class 1, 2 and 4, there is a Fermi singularity at zero frequency, part of a line of poles that stretches from one end of the stable region to the other. The dispersion near $\omega = 0$ is approximately linear in each case, which allows us to define a corresponding Fermi velocity $v_F$. Fermions of class 3 and 5, on the other hand, have no Fermi surface singularities near zero energy and hence are truly gapped.

As one moves away from $\omega = 0$ another phenomenon emerges: in some cases a new pair of poles nucleates at some nonzero $\omega$ and spreads apart as one approaches the gap. This is observed for fermions in class 1, 3 and 5. Poles that run into an oscillatory region at $\omega = \pm \Delta$ cease to exist.
 Note that for class 1 --- the only case with both a line of poles over the whole stable region as well as a nucleating pair --- it is difficult to see the poles nucleating on the right, since they appear very close to the gap and thus seem flattened along the horizontal direction. 


Poles for each fermion in the (3+1)QBH case in the $\mu_R \to 1$ limit should match onto either the left or right side of the gap in the 3QBH. Which side of the gap ought to match is decided by the sign of the fermions $q_1$ eigenvalue, with a positive (negative) $q_1$ meaning matching takes place at $\omega = \Delta(-\Delta)$. In all cases the appropriate matching occurs. For class 1, both of the nucleating poles run into the oscillatory region, but the pole at the right end of the long line matches with the one from Fig.~\ref{3plus1-1Fig}. For class 2, the pole on the left side of the gap matches with Fig.~\ref{3plus1-2Fig}. For class 3 both of the nucleating poles run into the oscillatory region, agreeing with Fig.~\ref{3plus1-3Fig} where the line of poles also hits the oscillatory region just before the edge. For class 4, the left side of the line of poles matches with the right side of Fig.~\ref{3plus1-4Fig}. And finally, for class 5 the right side where we do the matching displays no poles, agreeing with Fig.~\ref{3plus1-5Fig}.

We summarize our results in the table below, 
which lists the five classes of fermions and their $q_1$ values. When there is a pole at $\omega=0$, we list the corresponding $k_F/\mu_3$ and $v_F$. Furthermore, we list the values of $k_{\pm\Delta}/\mu_3$ for poles at $\omega = \pm \Delta$; in some cases there is more than one. If a pole momentum at the appropriate side of the gap matches with the (3+1)QBH, we underline it and provide the value of $\nu_{\pm\Delta}$ at that pole. 

\medskip
\begin{center}
\begin{tabular}{|c|l|l|l|l|l|l|l|}\hline
Class & $q_1$ & $k_F/\mu_3$ & $v_F$ & $k_{\Delta}/\mu_3$ & $\nu_{\Delta}$ & $k_{-\Delta}/\mu_3$ & $\nu_{-\Delta}$\\\hline
1& 1 & 0.163 & 0.769 & $-0.0832$; 0.254; \underline{0.499} & 0.248 & $-0.271$ & $\ \ \ \times$\\
2& $-1$ & 0.467 & 0.786 & 0.796 & $\ \ \times$ & \underline{0.0931} & 0.373\\
3& 3 & None & None & $-0.578$; 0.398 & $\ \ \times$ & None & None\\
 \hline
4& $-1$ & $-0.143$ & 0.748 & 0.189 & $\ \ \times$ & \underline{$-0.672$} & 0.0578\\
5& 3 & None & None & None & None & $-0.715$; 0.328 & $\ \ \ \times$\\
 \hline
\end{tabular}
\end{center}

\bigskip


\section{RG flow backgrounds: 2QBH and 1QBH}
\label{RGFlowSec}

We finally turn to the remaining boundaries of the parameter space: the two-charge black hole (2QBH), where two charges are set to zero and the other two are set equal, and the one-charge black hole (1QBH), where three charges are chosen to vanish. These cases have the property that it is not possible to take the extremal, zero-temperature limit without also setting the remaining chemical potential(s) to zero. In both cases one can take an $r_H \to 0$ limit with the appropriate $Q$ fixed, but one does not get a black hole: this limit removes the horizon, as well as shutting off the remaining gauge field, leaving a background with only the running scalar disturbing the metric, a so-called renormalization group (RG) flow geometry. The remaining parameter $Q$ no longer measures a charge, but instead controls the strength of the scalar perturbation.

The simplifications in the bosonic background render the Dirac equations fully solvable, and in the following subsections we will present the exact Green functions and briefly discuss the spectrum of the fermions. Although the backgrounds are nonthermodynamic, they are still worth commenting on. First, although this is outside the main thrust of our work, they are interesting as RG flow geometries and the fermionic Green's function reveals whether the spectrum is discrete or continuous, gapped or ungapped. For analogous five-dimensional cases, all fluctuations share the same spectrum due to the large supersymmetry \cite{Bianchi:2000sm}. Second, matching these  results to the endpoints of the regular cases can provide a check and analytic values for $k_F$ in the limit. Finally, these geometries are limits of nonzero-temperature backgrounds and may provide information about those thermodynamic cases.

\subsection{The one-charge black hole}

By setting $Q_3=0$ in the (3+1)QBH background we obtain the 1QBH:
\begin{align}\label{1qbh_background}
& A(r)=-B(r)=\log\frac{r}{L}+\frac{1}{4}\log\left(1+\frac{Q_1}{r}\right)\,,  \\
 &h(r) = 1-\frac{r_H^2(r_H+Q_1)}{r^2(r+Q_1)} \,, \quad \quad
 \phi = -\frac{1}{2}\log\left(1+\frac{Q_1}{r}\right)\,, \\
& \Phi_3(r)= 0 \,, \quad \quad
\label{1qbh_background2}
 \Phi_1(r) = \eta_1 \frac{r_H}{L}\sqrt{\frac{Q_1}{r_H+Q_1}}\left(1-\frac{r_H+Q_1}{r+Q_1}\right) \,.
\end{align}
The thermodynamic quantities are now
\begin{align}\label{1qbh_temp}
& T=\frac{3r_H+2 Q_1}{4\pi L^2}\sqrt{\frac{r_H}{r_H+Q_1}} \ \ , \ \ \ \ s = \frac{r_H^{3/2}}{4G L^2}(r_H+Q_1)^{1/2} \,, \\
&\mu_3 = 0 \,, \quad \quad \mu_1 = \eta_1 \frac{r_H}{L^2}\sqrt{\frac{Q_1}{r_H+Q_1}} \,, \quad \quad
\rho_3 = 0 \,, \quad \quad \rho_1 = {\eta_1 \over 2 \pi} \sqrt{Q_1 \over r_H} \,.
\end{align}
From (\ref{1qbh_temp}) we see that the extremal limit corresponds to taking $r_H \to 0$. This limit also causes all other thermodynamic quantities to vanish, since the horizon function $h(r) \to 1$, and the remaining gauge field $\Phi_1(r) \to 0$ as well. As promised we are then left with an RG-flow background consisting of AdS space deformed by an $r$-dependent scalar field. Unlike the  3QBH and the 2QBH discussed in the next subsection, there is no order-of-limits issue here, the same background is reached starting from the (3+1)QBH regardless of whether the 1QBH-limit or the extremal limit is taken first.

With the horizon function gone, fluctuations respect three-dimensional Lorentz invariance, and with no gauge fields the $q_i$ and $p_i$ parameters are irrelevant. 
Hence 
the uncoupled second order Dirac equation depends only on $\omega^2-k^2$, the mass parameter $m$ and the parameter $Q_1$. Defining a new variable $v = r/(r+Q_1)$, we obtain an exact solution for the spinor component infalling at the horizon in terms of Hankel functions of the first kind:
\eqn{}{
\chi_+ (r) = \left(\frac{\omega^2-k^2}{v}\right)^{1/4} H_{\frac{1-m}{2}}^{(1)}\left(\frac{2}{Q_1}\sqrt{\frac{\omega^2-k^2}{v}}\right) \,.
}
The Green function is obtained, as usual, from the asymptotic behavior of the spinor. The result is
\eqn{1QBHGreen}{
 G(\omega,k) = \frac{\sqrt{\omega^2-k^2}H_{\frac{-1-m}{2}}^{(1)}\left(\frac{2L^2 \sqrt{\omega^2-k^2}}{Q_1}\right)}{(\omega+k)H_{\frac{1-m}{2}}^{(1)}\left(\frac{2L^2\sqrt{\omega^2-k^2}}{Q_1}\right)} \,.
}
The imaginary part of the Green's function reveals a continuous spectrum; unlike the 2QBH
 in the next subsection, it does not display a mass gap.

The poles and zeros at $\omega =0 $ for the (3+1)QBHs plotted in figures~\ref{3plus1-1Fig}-\ref{3plus1-5Fig} all approach finite, nonzero values of $k/\mu_3$ as $\mu_R \to 0$, implying their values of $k/\mu_1$ are in all cases driven to zero; we can ask whether we see the corresponding poles or zeros at $\omega^2 - k^2 =0$ in the 1QBH Green's function. In fact \eno{1QBHGreen} has a zero for $m=-1$ and a pole for $m=3$, which matches correctly the classes 3, 4 and 5 with only a single pole or zero approaching $\mu_R \to 0$ limit. For classes 1 and 2 there are both poles and zeros as $\mu_R \to 0$; in both cases it is the zero, which has the larger value of $k$, that is reflected in the 1QBH Green's function.

\subsection{The two-charge black hole}

The 2QBH is the special case where $\tilde Q_2$ is set to zero in the (2+2)QBH background. The geometry is
\begin{align}\label{2qbh_background}
& A(r)=-B(r)=\log\frac{r}{L}+\frac{1}{2}\log\left(1+\frac{Q_2}{r}\right)\,, \\
& h(r) = 1-\frac{r_H(r_H+Q_2)^2}{r(r+Q_2)^2} \,, \quad \quad \gamma = \log\left(1+\frac{Q_2}{r}\right)\,, \\
& \Phi_2(r) = \eta_1\frac{\sqrt{Q_2 r_H}}{L}\left(1-\frac{r_H+Q_2}{r+Q_2}\right) dt \,, \quad \quad
\label{2qbh_background2}
 \tilde\Phi_2(r) = 0\,.
\end{align}
The thermodynamics are given by
\begin{align}\label{2qbh_temp}
 &T=\frac{3r_H+Q_2}{4 \pi L^2} \ \ , \ \ \ \ s = \frac{r_H}{4G L^2}(r_H+Q_2) \,, \\
&\mu_2 = \frac{\eta_2}{L^2}\sqrt{2 Q_2 r_H} \,, \quad \quad \tilde\mu_2 = 0 \,, \quad \quad
\rho_2 = {\eta_2 \over 2 \pi} \sqrt{2 Q_2 \over r_H} s \,, \quad \quad \tilde\rho_2 = 0 \,.
\end{align}
Looking at (\ref{2qbh_temp}), it is clear that in order to continuously tune the temperature down to zero, both $r_H$ and $Q_2$ must be taken to zero, leaving us with nothing but empty AdS space. The closest analogue to an extremal solution is to set $r_H = 0$ with $Q_2$ fixed and non-zero. In the limit $r_H \to 0$ the temperature approaches the value $Q_2/(4\pi L^2)$, but this limiting temperature does not strictly obtain since the horizon disappears at the endpoint. Similarly to the case for the 1QBH, this removes the last remaining gauge field and thus the chemical potential, leaving an RG flow geometry with a running scalar.

As was the case for the extremal 3QBH, there is an order-of-limits issue when this ``extremal'' 2QBH is approached from the (2+2)QBH, depending on whether $T \to 0$ or $\tilde Q_2 \to 0$ is imposed first. Taking the extremal limit first and then going to the 2QBH shifts the gauge potential $\tilde \Phi_2$ relative to \eno{2qbh_background2} by a constant,
\eqn{2qbh_order_of_limits}{
\tilde\Phi_2 \to {Q_2 \over L} \,.
}
The second-order Dirac equation depends only on $\omega^2 - k^2$, the mass parameter $\tilde{m}$ and the parameter $Q_2$, and as in the 1QBH case, it admits an analytic solution. Again we impose appropriate boundary conditions in the interior of the bulk. The solution for the spinor component $\chi_+$ is then
\begin{equation}
 \chi_+ (r) = \left(\frac{r}{r+Q_2}\right)^{-i \frac{\sqrt{16L^4(\omega^2-k^2)-\tilde m^2 Q_2^2}}{4Q_2}} \,,
\end{equation}
and the resulting Green function is 
\begin{equation}\label{2qbh_GF}
 G(\omega,k) = \frac{\tilde m Q_2 + i \sqrt{16L^4(\omega^2-k^2)-\tilde m^2 Q_2^2}}{4L^2(k+\omega)} \,.
\end{equation}
The imaginary part of the Green function displays a mass gap of $Q_2/2L^2$, then a continuum above the gap. This is very similar to the 1QBH in 5 dimensions studied in \cite{DeWolfe:2012uv}, dual to the Coulomb branch flow of $\mathcal{N}=4$ super-Yang-Mills, which also showed a continuous spectrum above a gap \cite{Freedman:1999gk, Bianchi:2000sm}.

Similarly to what we did for the 3QBH, we can match this to the $\tilde\mu_2 \to 0$ limit of the 
(2+2)QBH results, in the process deducing the precise values that the Fermi momenta and the momenta of zeros in the Green function should approach in this limit. Thanks to the shift \eno{2qbh_order_of_limits}, there is a relation between the energies analogous to \eno{freq_matching},
\eqn{}{
\omega_{2Q} = \omega_{(2+2)Q} + \frac{\tilde{q}_2 Q_2}{4 L^2 }\,.
}
The analytic solution (\ref{2qbh_GF}) tells us that fermions with negative (positive) $\tilde m$ have poles (zeros) at exactly $\omega_{2Q}^2 - k^2 =0$, meaning for $\omega_{(2+2)Q} =0$ there is a pole (zero) at
\eqn{}{
\left| k_{F(L)} \right| =   \frac{\tilde{q}_2 Q_2}{4 L^2 } \to \frac{\tilde q_2 \tilde \mu_2}{4\sqrt{2}} \,.
}
This indicates that class I should have a zero at $|k|=\tilde \mu_2/\sqrt{2}$, class II should have a pole at $k=0$, and class IV and VI (class III and V) should have a pole (zero) at $|k|=\tilde \mu_2/\sqrt{8}$. This agrees very well with our numerical results as seen in figures~\ref{2plus2-12Fig}-\ref{2plus2-56Fig}.
\begin{figure}
\begin{center}
\includegraphics[scale=0.4]{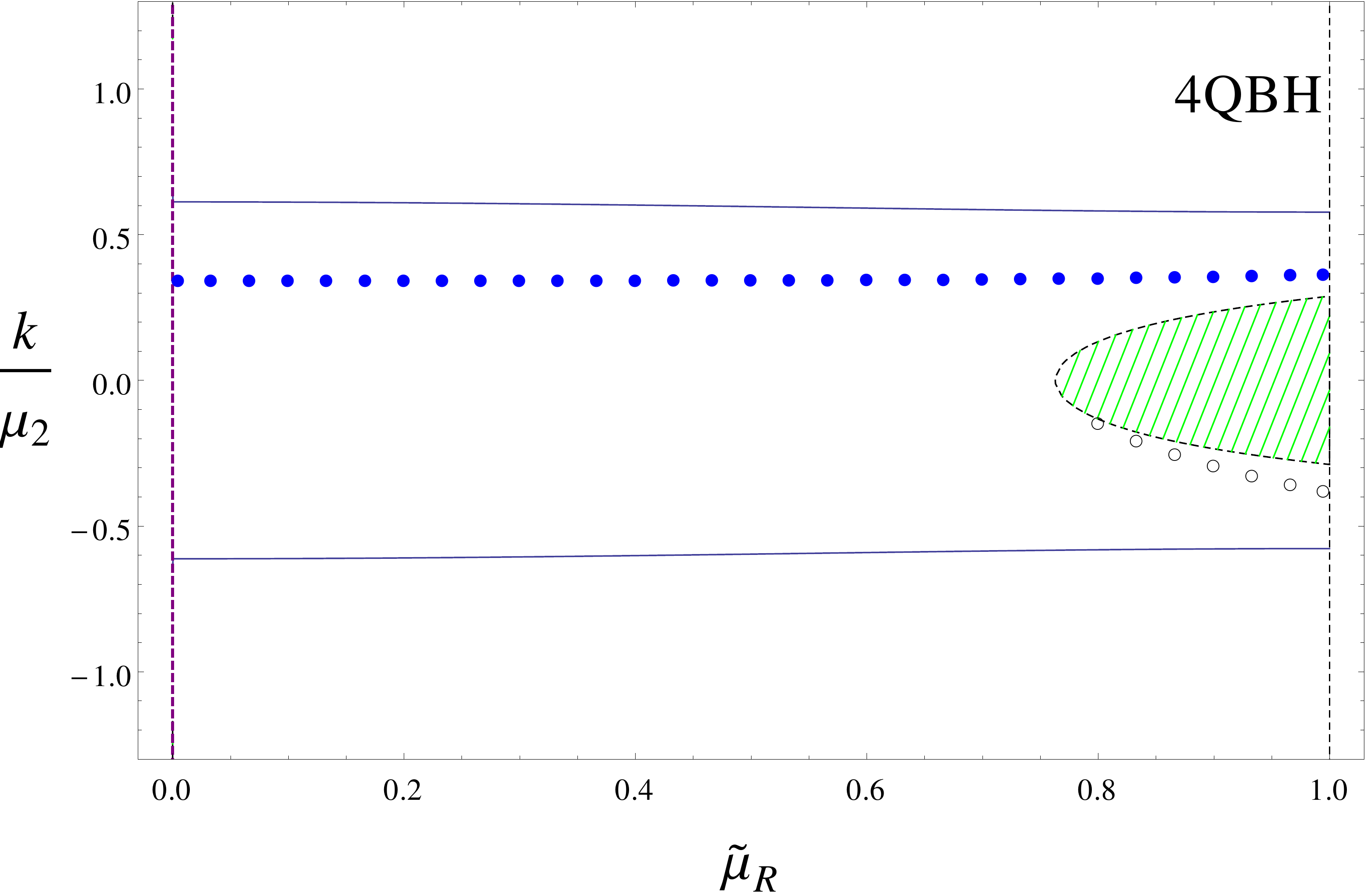}
\caption{Class II fermions for the (2+2)QBH, with $k$ normalized relative to $\mu_2$ instead of $\tilde\mu_2$. The Fermi surfaces all lie at $k_F\approx \mu_2/\sqrt{8}$.
\label{2plus2-2xFig}}
\end{center}
\end{figure}

The fermions in class II have a special behavior in the 2QBH limit since they are not charged under the gauge field $\tilde \Phi_2$. We refer to them as marginal Fermi liquid (MFL) fermions since they approach the value $\nu_k=0.5$ in this limit. For these fermions, we can normalize the Fermi momentum by $\mu_2$ instead of $\tilde \mu_2$ and still get a finite result, as is shown in Fig.~\ref{2plus2-2xFig}. For all the other fermions, the line of poles or zeros diverges in the $\mu_R \to 0$ limit with this normalization. Looking at this plot, it appears possible that the line of Fermi surfaces all lie at $k_F=\mu_2/\sqrt{8}$ independent of $\tilde \mu_R$, although we will not try to prove that here. This same behavior was observed for the MFL fermions in \cite{DeWolfe:2012uv}.

\section*{Acknowledgments}

We have enjoyed useful discussions and correspondence with Charles Cosnier-Horeau, Aristomenis Donos, Jerome Gauntlett, Steven S. Gubser, Victor Gurarie, Michael Hermele, Shamit Kachru, Giuseppe Policastro, Leo Radzihovsky, Dam Son and Jan Zaanen.   The work of O.D.\ and O.H.\ was supported by the Department of Energy under Grant No.~DE-FG02-91-ER-40672.  The work of C.R.\ was supported in part by European Union's Seventh Framework Programme under grant agreements (FP7-REGPOT-2012-2013-1) no 316165,
PIF-GA-2011-300984, the EU program ``Thales'' MIS 375734  and was also co-financed by the European Union (European Social Fund, ESF) and Greek national funds through the Operational Program ``Education and Lifelong Learning'' of the National Strategic Reference Framework (NSRF) under ``Funding of proposals that have received a positive evaluation in the 3rd and 4th Call of ERC Grant Schemes''.

\section*{Appendices}

\appendix
\section{Supergravity tensors}

In this appendix we present  explicit results for the supergravity tensors defined in section 2.2, evaluated for the scalar ansatz (\ref{ansatz}). Superscripts on two-index Levi-Civita symbols and generalized Kronecker deltas have the same meaning as described below (\ref{ansatz}) for four-index Levi-Civita symbols: the tensor is zero unless indices lie in the appropriate range, with $\alpha=1,...,4$ running over the index pairs $\{12,34,56,78\}$. For example, $\epsilon^{2}_{ij}=1 (-1)$ when $\{i,j\}$ is an even (odd) permutation of $\{3,4\}$, and zero otherwise, and $(\delta^1)^{kl}_{ij}$ is only non-zero when $\{i,j,k,l\}\in\{1,2\}$. 

The $u$- and $v$-tensors for the scalar ansatz are obtained by comparing (\ref{gauge_bein}) and (\ref{bein}). The result is 
\eqn{}{
 u_{ijkl}=\frac{1}{2}\cosh{\phi_1 \over 2} \sinh{\phi_2 \over 2} \sinh{\phi_3 \over 2} [\epsilon^1_{ij}\epsilon^2_{kl} + \epsilon^2_{ij}\epsilon^1_{kl} + \epsilon^3_{ij}\epsilon^4_{kl} + \epsilon^4_{ij}\epsilon^3_{kl}] \cr 
 + \frac{1}{2}\sinh{\phi_1 \over 2} \cosh{\phi_2 \over 2} \sinh{\phi_3 \over 2} [\epsilon^1_{ij}\epsilon^3_{kl} + \epsilon^3_{ij}\epsilon^1_{kl} + \epsilon^2_{ij}\epsilon^4_{kl} + \epsilon^4_{ij}\epsilon^2_{kl}] \cr 
 + \frac{1}{2}\sinh{\phi_1 \over 2} \sinh{\phi_2 \over 2} \cosh{\phi_3 \over 2} [\epsilon^1_{ij}\epsilon^4_{kl} + \epsilon^4_{ij}\epsilon^1_{kl} + \epsilon^2_{ij}\epsilon^3_{kl} + \epsilon^3_{ij}\epsilon^2_{kl}] \cr 
 + \cosh{\phi_1 \over 2} \cosh{\phi_2 \over 2} \cosh{\phi_3 \over 2} [\delta^1 + \delta^2 + \delta^3 + \delta^4]_{ij}^{kl}
 + \cosh{\phi_1 \over 2} [\delta^{12} + \delta^{34} - (\delta^1 + \delta^2 + \delta^3 + \delta^4)]_{ij}^{kl}  \cr
 +  \cosh{\phi_2 \over 2} [\delta^{13} + \delta^{24} - (\delta^1 + \delta^2 + \delta^3 + \delta^4)]_{ij}^{kl} 
 +\cosh{\phi_3 \over 2} [\delta^{14} + \delta^{23} - (\delta^1 + \delta^2 + \delta^3 + \delta^4)]_{ij}^{kl}\,,
}
\eqn{}{
 v_{ijkl}=-\frac{1}{2}\sinh{\phi_1 \over 2} \cosh{\phi_2 \over 2} \cosh{\phi_3 \over 2} [\epsilon^1_{ij}\epsilon^2_{kl} + \epsilon^2_{ij}\epsilon^1_{kl} + \epsilon^3_{ij}\epsilon^4_{kl} + \epsilon^4_{ij}\epsilon^3_{kl}] \cr 
 - \frac{1}{2}\cosh{\phi_1 \over 2} \sinh{\phi_2 \over 2} \cosh{\phi_3 \over 2} [\epsilon^1_{ij}\epsilon^3_{kl} + \epsilon^3_{ij}\epsilon^1_{kl} + \epsilon^2_{ij}\epsilon^4_{kl} + \epsilon^4_{ij}\epsilon^2_{kl}] \cr 
 - \frac{1}{2}\cosh{\phi_1 \over 2} \cosh{\phi_2 \over 2} \sinh{\phi_3 \over 2} [\epsilon^1_{ij}\epsilon^4_{kl} + \epsilon^4_{ij}\epsilon^1_{kl} + \epsilon^2_{ij}\epsilon^3_{kl} + \epsilon^3_{ij}\epsilon^2_{kl}] \cr 
 -  \sinh{\phi_1 \over 2} \sinh{\phi_2 \over 2} \sinh{\phi_3 \over 2} [\delta^1 + \delta^2 + \delta^3 + \delta^4]_{ij}^{kl} \cr
 - \frac{1}{2} \sinh{\phi_1 \over 2} [\epsilon^{12}_{ijkl} + \epsilon^{34}_{ijkl} - (\epsilon^1_{ij}\epsilon^2_{kl} + \epsilon^2_{ij}\epsilon^1_{kl} + \epsilon^3_{ij}\epsilon^4_{kl} + \epsilon^4_{ij}\epsilon^3_{kl})] \cr
 - \frac{1}{2} \sinh{\phi_2 \over 2} [\epsilon^{13}_{ijkl} + \epsilon^{24}_{ijkl} - (\epsilon^1_{ij}\epsilon^3_{kl} + \epsilon^3_{ij}\epsilon^1_{kl} + \epsilon^2_{ij}\epsilon^4_{kl} + \epsilon^4_{ij}\epsilon^2_{kl})] \cr
 -\frac{1}{2} \sinh{\phi_3 \over 2} [\epsilon^{14}_{ijkl} + \epsilon^{23}_{ijkl} - (\epsilon^1_{ij}\epsilon^4_{kl} + \epsilon^4_{ij}\epsilon^1_{kl} + \epsilon^2_{ij}\epsilon^3_{kl} + \epsilon^3_{ij}\epsilon^2_{kl})] \,.
}
From (\ref{t-tensor}) we obtain the T-tensor:
\eqn{}{
 T_{ijkl}=& \frac{3}{2}\left(\cosh{\phi_1 \over 2}\cosh{\phi_2 \over 2}\cosh{\phi_3 \over 2}-\sinh{\phi_1 \over 2}\sinh{\phi_2 \over 2}\sinh{\phi_3 \over 2}\right)\delta_{ij}^{kl}\ +\cr & \frac{3}{4}\left(\cosh{\phi_1 \over 2}\sinh{\phi_2 \over 2}\sinh{\phi_3 \over 2}-\sinh{\phi_1 \over 2}\cosh{\phi_2 \over 2}\cosh{\phi_3 \over 2}\right)[\epsilon^{12}+\epsilon^{34}]_{ijkl}\ +\cr
 & \frac{3}{4}\left(\sinh{\phi_1 \over 2}\cosh{\phi_2 \over 2}\sinh{\phi_3 \over 2}-\cosh{\phi_1 \over 2}\sinh{\phi_2 \over 2}\cosh{\phi_3 \over 2}\right)[\epsilon^{13}+\epsilon^{24}]_{ijkl}\ +\cr
 &\frac{3}{4}\left(\sinh{\phi_1 \over 2}\sinh{\phi_2 \over 2}\cosh{\phi_3 \over 2}-\cosh{\phi_1 \over 2}\cosh{\phi_2 \over 2}\sinh{\phi_3 \over 2}\right)[\epsilon^{14}+\epsilon^{23}]_{ijkl} \,.
}
The A-tensors are easily obtained from the T-tensor:
\begin{equation}
 A^1_{ij}= (\cosh{\phi_1 \over 2}\cosh{\phi_2 \over 2}\cosh{\phi_3 \over 2}-\sinh{\phi_1 \over 2}\sinh{\phi_2 \over 2}\sinh{\phi_3 \over 2})\ \delta_{ij}
\end{equation}
\eqn{}{
 A^2_{ijkl}=& -\left[\cosh{\phi_1 \over 2}\sinh{\phi_2 \over 2}\sinh{\phi_3 \over 2}-\sinh{\phi_1 \over 2}\cosh{\phi_2 \over 2}\cosh{\phi_3 \over 2}\right][\epsilon^{12}+\epsilon^{34}]_{ijkl} \cr
 & -\left[\sinh{\phi_1 \over 2}\cosh{\phi_2 \over 2}\sinh{\phi_3 \over 2}-\cosh{\phi_1 \over 2}\sinh{\phi_2 \over 2}\cosh{\phi_3 \over 2}\right][\epsilon^{13}+\epsilon^{24}]_{ijkl} \cr
 & -\left[\sinh{\phi_1 \over 2}\sinh{\phi_2 \over 2}\cosh{\phi_3 \over 2}-\cosh{\phi_1 \over 2}\cosh{\phi_2 \over 2}\sinh{\phi_3 \over 2}\right][\epsilon^{14}+\epsilon^{23}]_{ijkl} \,.
}
The squares of the A-tensors, which we need to determine the potential, are
\begin{align}
 |A^1_{ij}|^2=& 2\left(\cosh\phi_1+\cosh\phi_2+\cosh\phi_3 + \cosh\phi_1\cosh\phi_2\cosh\phi_3-\sinh\phi_1\sinh\phi_2\sinh\phi_3\right) \,,
\end{align}
and
\begin{align}
 |A^2_{ijkl}|^2= -12\left(\cosh\phi_1+\cosh\phi_2+\cosh\phi_3\right)  +36\left(\cosh\phi_1\cosh\phi_2\cosh\phi_3 - \sinh\phi_1\sinh\phi_2\sinh\phi_3\right) \,.
\end{align}
Finally, we can invert equation~(\ref{Stensor}) to obtain the S-tensor. The result  is
\eqn{}{
 S_{ijkl}= \frac{1}{2}\delta_{ij}^{kl}& + \frac{1}{2}[\cosh\phi_1\cosh\phi_2\cosh\phi_3+\sinh\phi_1\sinh\phi_2\sinh\phi_3][\delta^1+\delta^2+\delta^3+\delta^4]_{ij}^{kl}\cr
 & + \frac{1}{2}\cosh\phi_1 [\delta^{12} + \delta^{34} - (\delta^1 + \delta^2 + \delta^3 + \delta^4)]_{ij}^{kl} + \frac{1}{4}\sinh\phi_1 [\epsilon^{12}+\epsilon^{34}]_{ijkl}  \cr
 & + \frac{1}{2} \cosh\phi_2 [\delta^{13} + \delta^{24} - (\delta^1 + \delta^2 + \delta^3 + \delta^4)]_{ij}^{kl} + \frac{1}{4}\sinh\phi_2 [\epsilon^{13}+\epsilon^{24}]_{ijkl} \cr 
 & + \frac{1}{2} \cosh\phi_3 [\delta^{14} + \delta^{23} - (\delta^1 + \delta^2 + \delta^3 + \delta^4)]_{ij}^{kl} + \frac{1}{4}\sinh\phi_3 [\epsilon^{14}+\epsilon^{23}]_{ijkl}\cr
  +\frac{1}{4}& [ -\sinh\phi_1+\sinh\phi_1\cosh\phi_2\cosh\phi_3+\cosh\phi_1\sinh\phi_2\sinh\phi_3][\epsilon^1_{ij}\epsilon^2_{kl} + \epsilon^2_{ij}\epsilon^1_{kl} + \epsilon^3_{ij}\epsilon^4_{kl} + \epsilon^4_{ij}\epsilon^3_{kl}] \cr +\frac{1}{4}& [ -\sinh\phi_2+\cosh\phi_1\sinh\phi_2\cosh\phi_3+\sinh\phi_1\cosh\phi_2\sinh\phi_3]  [\epsilon^1_{ij}\epsilon^3_{kl} + \epsilon^3_{ij}\epsilon^1_{kl} + \epsilon^2_{ij}\epsilon^4_{kl} + \epsilon^4_{ij}\epsilon^2_{kl}] \cr +\frac{1}{4}& [ -\sinh\phi_3+\cosh\phi_1\cosh\phi_2\sinh\phi_3+\sinh\phi_1\sinh\phi_2\cosh\phi_3]  [\epsilon^1_{ij}\epsilon^4_{kl} + \epsilon^4_{ij}\epsilon^1_{kl} + \epsilon^2_{ij}\epsilon^3_{kl} + \epsilon^3_{ij}\epsilon^2_{kl}] \ .
}

\subsection{Specializing to the (3+1)QBH}

In the (3+1)QBH background the above quantities simplify, becoming:
\eqn{}{
u_{ijkl}=& -\frac{1}{2}\cosh{\phi\over 2} \sinh^2{\phi\over 2} [\epsilon^1_{ij}\epsilon^2_{kl} + \epsilon^2_{ij}\epsilon^1_{kl} + \epsilon^3_{ij}\epsilon^4_{kl} + \epsilon^4_{ij}\epsilon^3_{kl}+ \epsilon^1_{ij}\epsilon^3_{kl} + \epsilon^3_{ij}\epsilon^1_{kl} + \epsilon^2_{ij}\epsilon^4_{kl} + \epsilon^4_{ij}\epsilon^2_{kl} -\cr &\epsilon^1_{ij}\epsilon^4_{kl} - \epsilon^4_{ij}\epsilon^1_{kl} - \epsilon^2_{ij}\epsilon^3_{kl} - \epsilon^3_{ij}\epsilon^2_{kl}] 
 + \cosh^3{\phi\over 2} [\delta^1 + \delta^2 + \delta^3 + \delta^4]_{ij}^{kl} \cr &
 + \cosh{\phi\over 2} [\delta^{12} + \delta^{34} + \delta^{13} + \delta^{24} + \delta^{14} + \delta^{23} - 3(\delta^1 + \delta^2 + \delta^3 + \delta^4)]_{ij}^{kl} \,,
}
\eqn{}{
 v_{ijkl}=& -\frac{1}{2}\cosh^2{\phi\over 2}\sinh{\phi\over 2}\ [\epsilon^1_{ij}\epsilon^2_{kl} + \epsilon^2_{ij}\epsilon^1_{kl} + \epsilon^3_{ij}\epsilon^4_{kl} + \epsilon^4_{ij}\epsilon^3_{kl}+ \epsilon^1_{ij}\epsilon^3_{kl} + \epsilon^3_{ij}\epsilon^1_{kl} + \epsilon^2_{ij}\epsilon^4_{kl} + \epsilon^4_{ij}\epsilon^2_{kl} -\cr &\epsilon^1_{ij}\epsilon^4_{kl} - \epsilon^4_{ij}\epsilon^1_{kl} - \epsilon^2_{ij}\epsilon^3_{kl} - \epsilon^3_{ij}\epsilon^2_{kl}] +\sinh^3{\phi\over 2} [\delta^1 + \delta^2 + \delta^3 + \delta^4]_{ij}^{kl} \cr &
 -\frac{1}{2}\sinh{\phi\over 2} [\epsilon^{12}_{ijkl}+\epsilon^{34}_{ijkl}+\epsilon^{13}_{ijkl}+\epsilon^{24}_{ijkl}-\epsilon^{14}_{ijkl}-\epsilon^{23}_{ijkl}-\cr&(\epsilon^1_{ij}\epsilon^2_{kl} + \epsilon^2_{ij}\epsilon^1_{kl} + \epsilon^3_{ij}\epsilon^4_{kl} + \epsilon^4_{ij}\epsilon^3_{kl}+ \epsilon^1_{ij}\epsilon^3_{kl} + \epsilon^3_{ij}\epsilon^1_{kl} + \epsilon^2_{ij}\epsilon^4_{kl} + \epsilon^4_{ij}\epsilon^2_{kl} -\epsilon^1_{ij}\epsilon^4_{kl} - \epsilon^4_{ij}\epsilon^1_{kl} - \epsilon^2_{ij}\epsilon^3_{kl} - \epsilon^3_{ij}\epsilon^2_{kl})]
}
\begin{equation}
 T_{ijkl}= \frac{3}{2}[\cosh^3{\phi\over 2}+\sinh^3{\phi\over 2}]\ \delta_{ij}^{kl} - \frac{3}{4}e^{\phi/2}\cosh{\phi\over 2}\sinh{\phi\over 2}[\epsilon^{12}+\epsilon^{34}+\epsilon^{13}+\epsilon^{24}-\epsilon^{14}-\epsilon^{23}]_{ijkl} \ ,
\end{equation}
\begin{equation}
 A^1_{ij}= [\cosh^3{\phi\over 2}+\sinh^3{\phi\over 2}]\ \delta_{ij}
\end{equation}
\begin{equation}
 A^2_{ijkl}= e^{\phi/2}\cosh{\phi\over 2}\sinh{\phi\over 2}[\epsilon^{12}+\epsilon^{34}+\epsilon^{13}+\epsilon^{24}-\epsilon^{14}-\epsilon^{23}]_{ijkl} \ .
\end{equation}
\begin{equation}
 |A^1_{ij}|^2= 8[\cosh^3{\phi\over 2}+\sinh^3{\phi\over 2}]^2 \,, \quad \quad
 |A^2_{ijkl}|^2= 36\ e^{\phi}\sinh^2\phi \,.
\end{equation}
\eqn{}{
 S_{ijkl}=& \frac{1}{2}\delta_{ij}^{kl} +\frac{1}{2} [\cosh^3\phi-\sinh^3\phi] [\delta^1+\delta^2+\delta^3+\delta^4]_{ij}^{kl}\cr & + \frac{1}{2}\cosh\phi[\delta-(\delta^1+\delta^2+\delta^3+\delta^4)]_{ij}^{kl} +\frac{1}{4} \sinh\phi [\epsilon^{12}+\epsilon^{34}+\epsilon^{13}+\epsilon^{24}-\epsilon^{14}-\epsilon^{23}]_{ijkl}\cr &-\frac{1}{4} e^{-\phi}\sinh^2\phi [ \epsilon^1_{ij}\epsilon^2_{kl} + \epsilon^2_{ij}\epsilon^1_{kl} + \epsilon^3_{ij}\epsilon^4_{kl} + \epsilon^4_{ij}\epsilon^3_{kl}+ \epsilon^1_{ij}\epsilon^3_{kl} + \epsilon^3_{ij}\epsilon^1_{kl} + \epsilon^2_{ij}\epsilon^4_{kl} + \epsilon^4_{ij}\epsilon^2_{kl} -\cr &\epsilon^1_{ij}\epsilon^4_{kl} - \epsilon^4_{ij}\epsilon^1_{kl} - \epsilon^2_{ij}\epsilon^3_{kl} - \epsilon^3_{ij}\epsilon^2_{kl}] \,.
}


\subsection{Specializing to the (2+2)QBH}

The (2+2)QBH background admits even greater simpifications than the (3+1)QBH case:
\begin{equation}
 u_{ijkl}= [\cosh{\gamma\over 2}-1][\delta^{12} + \delta^{34}]_{ij}^{kl} + \delta_{ij}^{kl}\,,  \quad\quad
 v_{ijkl}= -\frac{1}{2}\sinh{\gamma\over 2}[\epsilon^{12}+\epsilon^{34}]_{ijkl} \,,
\end{equation}
\begin{equation}
 T_{ijkl}= \frac{3}{2}\cosh{\gamma\over 2}\ \delta_{ij}^{kl} - \frac{3}{4}\sinh{\gamma\over 2}[\epsilon^{12}+\epsilon^{34}]_{ijkl} \,,
\end{equation}
\begin{equation}
 A^1_{ij}= \cosh{\gamma\over 2}\ \delta_{ij} \,, \quad \quad
 A^2_{ijkl}= \sinh{\gamma\over 2}[\epsilon^{12}+\epsilon^{34}]_{ijkl} \,,
\end{equation}
\begin{equation}
 |A^1_{ij}|^2= 8\cosh^2{\gamma\over 2} \,, \quad \quad
 |A^2_{ijkl}|^2= 48\sinh^2{\gamma\over 2} \,,
\end{equation}
\eqn{}{
S_{ijkl}= \delta_{ij}^{kl} +\sinh^2{\gamma\over 2}[\delta^{12}+\delta^{34}]_{ij}^{kl} +\frac{1}{2}\cosh{\gamma\over 2}\sinh{\gamma\over 2}[\epsilon^{12}+\epsilon^{34}]_{ijkl} \,.
}


\section{Lifting the 3QBH to five dimensions}
\label{app5Dlift}

Here we summarize how the near-horizon region of the extremal 3QBH may be lifted to a non-singular five-dimensional $AdS_3 \times \mathbb{R}^2$ geometry, with the inert gauge field identified as the graviphoton. Additionally, the lift of a fermion action to five dimensions implies a relation between the fermion eigenvalues, which is satisfied by all the fermions we work with. This analysis is similar to what is done in \cite{DeWolfe:2013uba}, section 6, and we refer the reader there for additional detail.


The near-horizon ($r \to 0$) limit of the metric and scalar \eno{3QBHSoln1}, \eno{3QBHSoln2} in the extremal 3QBH  gives
\begin{equation}
\label{NearHorizon3QBH}
ds^2 = -\frac{3\sqrt{Q}r^{3/2}}{L^2}dt^2 + \frac{Q^{3/2}\sqrt{r}}{L^2}d\vec x^2 + \frac{L^2}{3\sqrt{Q}r^{3/2}}dr^2 \,,
\quad\quad e^{\phi} = \sqrt{\frac{Q}{r}} \ .
\end{equation}
A five-dimensional ansatz of the form
\begin{equation}
\label{metric_ansatz}
d \hat s^2 = e^{ \phi}ds^2 + e^{-2 \phi}(dz + \mathcal{A})^2 \ .
\end{equation}
dimensionally reduces to a four-dimensional action
\begin{equation}
\mathcal{L}_4 = \sqrt{-g}\left(R-\frac{3}{2}(\partial \phi)^2-\frac{1}{4}e^{-3\phi}\mathcal{F}^2 \right) \,,
\end{equation}
which matches the Einstein term, scalar kinetic term, and second gauge kinetic term in 
\eno{3plus1_lagrangian} if we identify the $a$ gauge field with the graviphoton ${\cal A}$. Then using this lift, the five-dimensional metric arising from \eno{NearHorizon3QBH} is
\begin{equation}
d \hat s^2 = -\frac{3Qr}{L^2}dt^2 +\frac{L^2}{3r^2}dr^2 +\frac{r}{Q}dz^2 +\frac{Q^2}{L^2}d \vec x^2 \,,
\end{equation}
which with the change of coordinates  $r\equiv\rho^2$ becomes
\begin{equation}
d \hat s^2 = -\frac{3Q\rho^2}{L^2}dt^2 +\frac{4 L^2}{3 \rho^2}d \rho^2 +\frac{\rho^2}{Q}dz^2 +\frac{Q^2}{L^2}d \vec x^2 \,,
\end{equation}
which is of the form $AdS_3 \times \mathbb{R}^2$, as promised. 
The leading order term in the four-dimensional potential in \eno{3plus1_lagrangian}  arises by dimensional reduction of a cosmological constant term,
\begin{equation}
\mathcal{L}_{5 \Lambda} = \sqrt{-\hat g} \hat \Lambda \to \mathcal{L}_{4 \Lambda} = \sqrt{-g}e^{\frac{\varphi}{\sqrt{3}}} \hat \Lambda \ ,
\end{equation}
with $\hat \Lambda = 3/L^2$.


Consider now reducing a massless, four-component Dirac fermion  $\lambda$ from five dimensions. Its action is
\begin{equation}
\mathcal{L}_{5 \lambda} = \sqrt{-\hat g} i \bar \lambda \gamma^{\underline{M}} \hat e^N_{\ \ \underline M} \hat \nabla_N \lambda \ ,
\end{equation}
where $\hat \nabla_N = \partial_N - \frac{1}{4}\hat \omega_N^{\ \ \underline{PQ}}\gamma_{\underline{PQ}}$.
We make an ansatz for this spinor in terms of a four-dimensional Dirac spinor $\chi$ 
\begin{equation}
\lambda(x^{\mu},z) = e^{i n z/R} e^{-\phi/4} e^{-\pi \gamma^{\underline 4}/4}\chi(x^\mu)\,,
\end{equation}
where $R$ is the radius of the compact $z$ coordinate.
Here we have rescaled by a power of $\phi$ to obtain canonical kinetic terms, and performed a chiral rotation to remove factors of $\gamma^{\underline 4}$ (which is $i$ times the four-dimensional chirality matrix) from the mass and Pauli terms.
We arrive at
\begin{equation}
e^{-1}\mathcal{L} = \bar\chi\left[i\gamma^{\mu}\nabla_{\mu} - \frac{n}{R}e^{3\phi/2} + \frac{n}{R}\gamma^{\mu}\mathcal{A}_{\mu} - \frac{i}{8}e^{-3\phi/2}\gamma^{\mu\nu}\mathcal{F}_{\mu\nu}\right]\chi \ .
\end{equation}
Comparing to the  3QBH fermion Lagrangian (\ref{3QBH_Lagrangian}) with the identification $a = {\cal A}$ we find agreement for the appropriate terms (including the term in the potential of leading order as $r \to 0$) given the identifications
\eqn{}{
 \frac{m}{4L} = \pm \frac{n}{R} \,, \quad \quad
 \frac{q_1}{4L} = \frac{n}{R} \,, \quad \quad
 p_1 = \mp 1 \,,
}
where the second choice of sign can be obtained by doing a
chirality flip \eno{ChiralityFlip}. Thus both the mass parameter $m$ and the charge $q_1$ are given by the momentum in the compact direction, and we have the relation
\begin{equation}
 m = - q_1 p_1 \,,
\end{equation}
as given in \eno{mConstraint}.
Checking the eigenvalue table for the (3+1)QBH in section \ref{3plus1_section}, we find perfect agreement with this constraint in all cases.

\bibliographystyle{JHEP}
\bibliography{4DFermi.bib}

\end{document}